\documentclass[%
 reprint,
superscriptaddress,
nofootinbib,
 amsmath,
 amssymb,
 aps,
]{revtex4-2}

\usepackage{mathtools} 
\usepackage{graphicx}
\usepackage{dcolumn}
\usepackage[super]{nth}
\usepackage{bm}

\usepackage{nicematrix}

\usepackage{color,soul}
\usepackage{xcolor,hyperref}
\hypersetup{
    colorlinks=true,
    linkcolor=[rgb]{0 0 0.5},
    urlcolor=[rgb]{0 0 0.5},
    citecolor= [rgb]{0 0 0.5}
}

\usepackage{dsfont} 
\usepackage{etoolbox}

\usepackage[capitalise]{cleveref}

\makeatletter
\patchcmd{\upbracefill}{\m@th}{\scriptstyle\m@th}{}{}
\patchcmd{\upbracefill}{$\braceld$}{$\scriptstyle\braceld$}{}{}
\patchcmd{\upbracefill}{\bracelu}{\bracelu\mkern-1mu}{}{}
\patchcmd{\upbracefill}{\hfill\braceru}{\hfill\mkern-1mu\braceru}{}{}
\makeatother

\makeatletter
\def\smallunderbrace#1{\mathop{\vtop{\m@th\ialign{##\crcr
   $\hfil\displaystyle{#1}\hfil$   \vspace{0.6ex}\crcr
   \noalign{\kern3\p@\nointerlineskip}%
\tiny\upbracefill\crcr\noalign{\kern3\p@}}}}\limits}
\makeatother

\renewcommand{\thesection}{\arabic{section}}
\renewcommand{\thesubsection}{\thesection.\arabic{subsection}}
\renewcommand{\thesubsubsection}{\thesubsection.\arabic{subsubsection}}

\makeatletter
\renewcommand{\p@subsection}{}
\renewcommand{\p@subsubsection}{}
\makeatother

\newcommand{\smashhat}[1]{\smash{\hat{#1}}}
\definecolor{lightgray}{rgb}{0.8, 0.8, 0.8} 

\newcommand{\bs}{\boldsymbol}

\newcommand{\ix}{\mathbf{x}}

\newcommand{\Y}{\mathbf{Y}}
\newcommand{\y}{\mathbf{y}}

\newcommand{\hatY}{\smash{\hat \Y}}

\newcommand{\tix}{\bs{\theta}}

\newcommand{\la}{\left\langle}
\newcommand{\ra}{\right\rangle}

\newcommand{\Exp}[2][]{\mathbb{E}_{#1}\left[#2\right]}

\newcommand{\ind}{\mathds{1}}

\newcommand{\pd}[2]{\frac{\partial #1}{\partial #2}}

\renewcommand{\d}{\mathrm{d}}
\newcommand{\ee}{\mathrm{e}}
\newcommand{\ii}{\mathrm{i}}
\newcommand{\partialt}{\frac{\partial}{\partial t}}
\renewcommand{\u}{{\mathbf u}}
\newcommand{\q}{\mathbf{q}}
\newcommand{\Q}{{\mathbf Q}}
\def\v{{\mathbf v}}
\newcommand{\vel}{{\bs v}}

\newcommand{\ptil}{\tilde{p}}

\renewcommand{\ss}{{\rm ss}}

\newcommand{\M}{{\rm M}}
\newcommand{\rS}{{\rm S}}

\newcommand{\bR}{\mathbf{R}}
\newcommand{\n}{{\bs n}}

\renewcommand{\r}{\mathbf {r}}

\newcommand{\F}{{\mathbf F}}
\newcommand{\C}{{\mathbf C}}

\newcommand{\J}{{\mathbf J}}

\newcommand{\I}{\mathcal{I}}
\newcommand{\rI}{{\rm I}}
\newcommand{\bxi}{{\bs \xi}}
\newcommand{\bsigma}{{\bs \sigma}}

\newcommand{\tr}[1]{#1^{\rm T}}
\newcommand{\T}{\mathsf{T}}

\newcommand{\Tiso}{\T^\text{(iso)}}

\newcommand{\Z}{\mathcal{Z}}
\newcommand{\A}{\mathbf{A}}

\newcommand{\kb}{{k}_\text{b}}

\newcommand{\tka}{\tilde{k}_\text{b}}
\newcommand{\zero}{\bs{0}}

\newcommand{\ku}{{k}_\text{u}}

\renewcommand{\k}{{{\bs k}^{(1)}_\text{b}}}
\newcommand{\kk}{\mathbf{K}^{(2)}_\text{b}}

\newcommand{\comment}[1]{}
\newcommand{\ab}{{\alpha\beta}}
\newcommand{\mn}{{\mu\nu}}
\newcommand{\abmn}{{\ab\mn}}

\newcommand{\avg}[1]{\la #1 \ra}

\newcommand{\unbound}{\text{\footnotesize unbound}}

\newcommand{\subeqn}[1]{\begin{subequations}%
    #1%
\end{subequations}}

\newcommand{\td}{\tilde}

\newcommand{\WK}{Wiener-Khinchin}

\let\originalleft\left
\let\originalright\right
\renewcommand{\left}{\mathopen{}\mathclose\bgroup\originalleft}
\renewcommand{\right}{\aftergroup\egroup\originalright}

\newcommand{\TODO}[1]{[{\textbf{\hl{TODO}}}:  \textit{#1}]}

\newcommand{\kBT}{{\rm k}_{\rm B}T}

\begin{document}

\preprint{APS/123-QED}

\title{Nonequilibrium noise emerging from broken detailed balance in active gels}

\author{Ashot Matevosyan}
 \email{ashmat@pks.mpg.de}
 \affiliation{Max Planck Institute for the Physics of Complex Systems, N\"{o}thnitzerst. 38, 01187 Dresden, Germany}
 \affiliation{Alikhanyan National Laboratory (Yerevan Physics Institute), 
 Alikhanian Brothers Street 2, Yerevan 375036, Armenia}
\author{Frank J\"ulicher}%
\affiliation{Max Planck Institute for the Physics of Complex Systems, N\"{o}thnitzerst. 38, 01187 Dresden, Germany}
\affiliation{Center for Systems Biology Dresden, Pfotenhauerst. 108, 01307 Dresden, Germany}
\affiliation{Cluster of Excellence Physics of Life, TU Dresden, 01062 Dresden, Germany}

\author{Ricard Alert}%
 \email{ralert@pks.mpg.de}
\affiliation{Max Planck Institute for the Physics of Complex Systems, N\"{o}thnitzerst. 38, 01187 Dresden, Germany}
\affiliation{Center for Systems Biology Dresden, Pfotenhauerst. 108, 01307 Dresden, Germany}
\affiliation{Cluster of Excellence Physics of Life, TU Dresden, 01062 Dresden, Germany}
\affiliation{Departament de F\'{i}sica de la Mat\`{e}ria Condensada, Universitat de Barcelona, Mart\'{i} i Franqu\`{e}s 1, 08028 Barcelona, Spain}
\affiliation{Universitat de Barcelona Institute of Complex Systems (UBICS), Barcelona, Spain}
\affiliation{Instituci\'{o} Catalana de Recerca i Estudis Avan\c{c}ats (ICREA), Barcelona, Spain}
%


\date{\today}

\begin{abstract}
In thermodynamic equilibrium, the fluctuation-dissipation theorem links thermal fluctuations and dissipation. Biological systems, however, are driven out of equilibrium by internal processes that produce additional, active fluctuations. Despite being relevant for biological functions such as intracellular transport, predicting the statistical properties of active fluctuations remains challenging. Here, we address this challenge in a minimal model of an active gel as a network of elastic elements connected by transient crosslinks. The crosslinkers' binding and unbinding rates break detailed balance, which drives the system out of equilibrium. Through coarse-graining, we derive fluctuating hydrodynamic equations including an active noise term, which emerges explicitly from the breaking of detailed balance.
Finally, we provide predictions for the stochastic motion of a tracer particle embedded in the active gel, which enables comparisons with microrheology experiments both in synthetic active gels and in cells. Overall, our work provides an explicit link between the statistical properties of active fluctuations and the molecular breaking of detailed balance. Thus, it paves the way toward complementing the fluctuation-dissipation theorem with a fluctuation-activity relation in active systems.

\end{abstract}

\maketitle


\section{Introduction}

Living matter is active: 
It is internally driven by irreversible processes such as the action of molecular motors. Over the last two decades, the theory of active matter has been developed to capture this internal activity.
The existing hydrodynamic theories for active materials have been fruitfully applied to both synthetic and biological systems \cite{ramaswamy2010mechanics,prost2015,marchetti2013,julicher2018,kruse2005generic}. However, these theories mostly focus on the \emph{average} effects of activity.
For example, the theory can predict how the collective action of molecular motors produces an average active stress and drive flows of the cell cortex \cite{mayer2010,prost2015}.

Yet, beyond these average effects, active processes generate prominent fluctuations, which are fundamentally different from equilibrium thermal fluctuations.
Active fluctuations can be stronger than thermal fluctuations, and they can impact biological processes such as intracellular transport.
For example, active fluctuations can enhance the diffusivity of macromolecules, vesicles, and biomolecular condensates \cite{brangwynne2009,parry2014,guo2014,fakhri2014,shu2024} and speed up motor-driven transport \cite{ezber2020,wolgemuth2020,ariga2021}. At the cellular scale, active fluctuations may enable cellular sensitivity to weak magnetic fields \cite{matevosyan2021nonequilibrium}, promote bleb nucleation \cite{alert2016,turlier2016,turlier2019}, and help to position the cell nucleus \cite{almonacid2015nucleus-positioning,rupprecht2018}.



Despite their relevance, active fluctuations have been largely overlooked in active matter theories so far. In part, this is due to the fact that, unlike in the case of thermal fluctuations, we lack a general theoretical framework to predict the statistical properties of active fluctuations. For example, there is no fluctuation-dissipation theorem to relate active fluctuations to dissipation \cite{gnesotto2018}.
Given this limitation, previous works took a phenomenological approach and proposed active noise terms with spatiotemporal correlations consistent with experimental data for particular systems
\cite{lau2003,basu2008thermal,ben-isaac2011,fodor2015,fodor2016,ahmed2018,bernheim2018,abbasi2023non}.

Here, we seek to relate the active noise to the underlying molecular processes. To this end, we start with a minimal model of an active gel as a network of elastic elements bound by molecular crosslinkers. At the subcellular level, these elastic elements can be cytoskeletal filaments, such as actin in the cell cortex or microtubules in the mitotic spindle. At the tissue level, the elastic elements can be cells crosslinked through cell-cell junctions. 
We introduce activity at the molecular level by breaking detailed balance in the binding kinetics of the crosslinkers.

Through explicit coarse-graining, we then derive the fluctuating hydrodynamic equations of the active gel. These equations include a nonequilibrium noise term consisting of active noise, which arises from the breaking of detailed balance, and passive noise, which encompasses both equilibrium thermal fluctuations and externally driven contributions such as an imposed shear. This latter contribution indicates that even externally driven passive systems can exhibit nonequilibrium noise within their hydrodynamic descriptions.

Overall, our results establish a fluctuation-activity relation that links the statistics of active noise at the mesoscopic scale to the breaking of detailed balance at the molecular scale. We also predict properties of the fluctuations of a tracer particle embedded in an active gel, which is often measured in microrheology experiments on cytoskeletal networks \cite{mizuno2007,brangwynne2008,toyota2011} and cellular cytoplasm \cite{lau2003,bursac2005,wilhelm2008,guo2014,fodor2015,fodor2016,nishizawa2017,ahmed2018,hurst2021,bacanu2023,umeda2023,muenker2024elife,muenker2024natmat}.

The paper is structured as follows: 
We begin by summarizing the minimal mesoscopic model of an active gel introduced in Ref. \cite{oriola2017}, which is presented in Section \ref{sec-model}. Starting from the nonequilibrium dynamics of molecular crosslinkers, Ref. \cite{oriola2017} derived the hydrodynamic equations of an active gel, which we summarize in Section \ref{sec-coarsen}. In Section \ref{sec-fluct}, we use the model to derive the fluctuating hydrodynamic equations, which allow us to predict the nonequilibrium noise that emerges in this minimal active gel. To understand the different sources of noise, in Section \ref{sec-ss} we study the nonequilibrium steady-state distribution of the system. In Section \ref{sec-noise-amplitude} we derive the explicit form of the active noise and demonstrate the effect of activity with a simple example.
In Section \ref{sec-observables} we study the fluctuations of a tracer particle embedded in an active gel, and thus get predictions for microrheology experiments. We conclude with a discussion of our results in Section \ref{sec-discussion}. 



\section{Model of an active gel as a transiently crosslinked network}
\label{sec-model}

\begin{figure*}
	\centering 
	\includegraphics[width=0.99\textwidth]{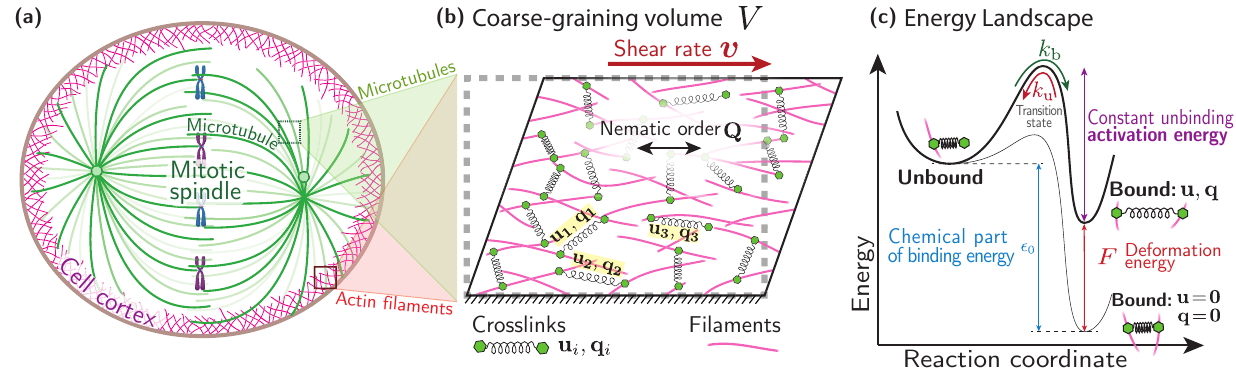}
	\caption{
    \textbf{(a)} Cell cortex and mitotic spindle, the relevant cellular structures where our model applies. Filaments are interconnected by active molecular motors/crosslinks.
\textbf{(b)} Coarse-graining volume $V$ containing the active gel model. The background filaments establish the nematic order $\Q$, while an external shear rate $\v$ is applied. Each bound molecular motor is characterized by a mesoscopic strain $\u$ and an orientation $\q$, independent of other motors.
\textbf{(c)} Energy landscape of the bound state. We illustrate the contributions to the detailed balance relation \eqref{kakd}, setting $\Omega=0$ for simplicity. In our model, the activation energy for unbinding is constant, making the unbinding rate $\ku$ independent of the bound state.
    }
    \label{fig-model}
\end{figure*}

Motivated by the cytoskeleton, the active gel is represented as a network of elastic elements, such as cytoskeletal filaments, connected by transient crosslinkers (\cref{fig-model}a,b). Each crosslinker is modeled as an elastic macromolecule, which can stochastically bind and unbind from the network. To reach a certain binding site, a linker generally has to stretch and reorient. Thus, a linker can be in multiple bound states, characterized by two symmetric and traceless tensors: its mesoscopic strain $\u$ and orientation $\q$ (\cref{fig-model}b).

To account for the dynamics of the linkers, we introduce $n(\u, \q, t) \, \d\u\, \d\q$ as the fraction of bound linkers with strain $[ \u, \u\! +\! \d\u ]$ and orientation $[ \q, \q \!+\! \d\q ]$ at time $t$.
The evolution of this distribution is described by the following advection-reaction equation \cite{oriola2017}:
\begin{align}\label{n-eqn}
    \pd{n}{t} +v_\ab \pd{n}{u_\ab} + \dot{Q}_\ab \pd{n}{q_\ab} =(1-\phi)\kb - n\,\ku.
\end{align}
This equation expresses that $n(\u, \q, t)$ evolves due to two processes: The left-hand side accounts for the change in the strain and orientation of bound linkers under externally-imposed deformations, including a uniform shear rate $v_\ab = \la \dot{u}_\ab \ra$ and change in global orientation $\dot Q_{\alpha\beta}$, where $Q_\ab = \la q_\ab \ra$. For example, a crosslink initially bound at state $(\u, \q)$ will deform to $(\u + \v t, \q + \dot{\Q} t)$ after time $t$, provided it remains bound and ignoring its diffusion.

The right-hand side of \cref{n-eqn} accounts for binding and unbinding events, which we assume to be independent Poisson processes.
Individual linkers stochastically bind and unbind with rates $\kb$ and $\ku$ respectively, which can depend on the strain $\u$ and orientation $\q$ of the linker in the bound state. The total binding rate is proportional to the fraction of unbound linkers, $1-\phi$, where 
\begin{align}\label{phi-def}
\phi(t) = \int n(\u, \q, t) \, \d\u\, \d\q
\end{align} 
is the fraction of bound linkers. Respectively, the total unbinding rate is proportional to the fraction of linkers bound to a specific state (c.f.~last term in \cref{n-eqn}).
The integrals $\int \dots \d\u\,\d\q$ are over all possible bound states; see \cref{app-integral} for the  definition and evaluation of these integrals.



At thermodynamic equilibrium, the binding and unbinding rates of individual linkers must fulfill detailed balance. In active systems, however, detailed balance is locally broken, which can be generically expressed as \cite{oriola2017,chen2020,julicher1997}
\begin{align}
\label{kakd}
\frac{\kb}{\ku} = \ee^{\beta \epsilon_0} \left( \ee^{-\beta F} + \Omega \right)\,.
\end{align}
Here, $\beta \equiv (\kBT)^{-1}$ is the inverse temperature, $\epsilon_0$ is the chemical part of the binding energy, while $F(\u,\q)$ represents the mechanical free energy difference per linker between the bound state $(\u,\q)$ and the unbound state due to stretching and reorientation of the linker (\cref{fig-model}c). The term $\Omega$ quantifies the departure from detailed balance, which we refer to as molecular activity. It is an a priori unknown function of the system parameters, the coordinates $\u$ and $\q$ of the bound state, and the chemical potential difference $\Delta \mu$ of ATP hydrolysis, with $\Omega \propto \Delta \mu$ near equilibrium \cite{julicher1997}.

At this stage, for each specific system, one should introduce the appropriate dependence on the unbinding rate $\ku$ on the crosslinker's bound state $\u,\q$. For simplicity, here we assume a state-independent unbinding rate. This assumption implies that the activation energy of the unbinding process is the same for all bound states (see \cref{fig-model}c) \cite{walcott2010}, which is a good approximation if the binding energy is dominated by the chemical contribution $\epsilon_0$ rather than the mechanical contribution $F$. Additionally, for convenience, we incorporate the factor with the chemical energy $\epsilon_0$ in \cref{kakd} into $\ku$. Thus, we have:
\begin{equation}
    \label{kakd-new}
    \kb(\u,\q) = \ku\left(\ee^{-\beta F(\u,\q)}+\Omega(\u,\q)\right).   
\end{equation} 
Furthermore, we consider that the system has an isotropic linear elastic response, both to strain and to nematic ordering \cite{lubensky2002}.
We now switch to a continuum description for a system with a density $\rho$ of linkers with free energy density $f(\u,\q) = \rho F(\u,\q)$ given by
\begin{equation}
    \label{free-energy} 
f(\u,\q) = \frac{\mu}{2} u_\ab u_\ab + D u_\ab q_\ab + \frac{\chi}{2} q_\ab q_\ab,
\end{equation}
where Greek indicates denote spatial coordinates, $\mu$ is the shear elastic modulus, $D$ is the elastonematic coefficient, and $\chi$ is the inverse nematic susceptibility \cite{lubensky2002}. 
With these choices, the model is fully specified.

Hereafter, we work with dimensionless quantities by rescaling energy by $\beta^{-1}$, time by $\ku^{-1}$, length by $(\beta\mu)^{-1/3}$. Therefore, pressure, stress and energy density are all in units of $\mu$.

\section{Constitutive relation for the average stress}
\label{sec-coarsen}
In continuum descriptions, a constitutive relation specifies how thermodynamic forces -- here the stress tensor -- depend on the relevant thermodynamic variables and their gradients \cite{mazur,landau-stat}. It closes the balance equations by encoding the material-specific response. In passive viscoelastic media this yields, for example, a viscous stress proportional to the strain-rate and an elastic stress derived from a free energy. In our case, detailed balance is broken and nematic order provides additional symmetry-allowed couplings, so the stress contains both dissipative terms and active, non-potential contributions such as a term proportional to the nematic order parameter $Q_{\ab}$ \cite{ramaswamy2010mechanics,prost2015,marchetti2013,julicher2018,kruse2005generic}. While such constitutive laws are typically derived phenomenologically, here we derive them from an explicit microscopic model of stochastic binding-unbinding linkers, thereby expressing the effective transport coefficients in terms of microscopic rates and elastic couplings \cite{oriola2017}.

Our goal is to obtain an equation for the stress tensor of the active gel, which fluctuates due to the stochastic linker dynamics.
The stress exerted by a bound linker in a coarse-graining region of volume $V$ is given by
\begin{equation}
\label{sigma-1-def}
    \sigma_\ab^\text{linker}(\u,\q)
    =
    \frac{1}{V}\pd{F(\u,\q)}{u_\ab}\equiv \frac{1}{N} \pd{f(\u,\q)}{u_\ab},
\end{equation}
where $N = \rho \, V$ is the number of linkers in volume $V$.
Summing over linkers, the total stress in this volume is:
\begin{equation}
\label{sigma-1}
    \hat\sigma_\ab(t) = \sum_{ i=1}^{N} 
    \sigma_\ab^\text{linker}(\hat\u_i,\hat\q_i)\; \ind_i(t),
\end{equation}
where the hat indicates fluctuating quantities, and $\ind_i(t)$ is the indicator function
\begin{equation}
    \label{def-ind}
    \ind_i(t) = \begin{cases}
    1 & \text{if linker $i$ is bound at time $t$} \\
    0 & \text{otherwise}
    \end{cases}
\end{equation}
that counts only the bound linkers in \cref{sigma-1}. When $\ind_i(t) = 1$, $(\hat\u_i,\hat\q_i)$ specifies the bound state of linker $i$.

In this section, we derive the constitutive relation for the average stress, defined as
\begin{equation}
\label{average-stress-def-0}
\begin{aligned}
    \sigma_\ab(t) =&\,
    \la \, \hat\sigma_\ab(t)\, \ra_{\sim n}
    \\
    =&  N \int n(\u,\q, t) \sigma^\text{linker}_\ab(\u,\q) \,\d\u\, \d\q
    \\
    =& \int n(\u,\q,t) \pd{f(\u,\q)}{u_\ab} \,\d\u\, \d\q,
\end{aligned}
\end{equation}

where $\la~\ra_{\sim n}$ denotes the average over the distribution $n(\u,\q, t)$. We consider identical linkers; therefore, the bound state of each linker follows the same distribution, with $\hat\u_i,\hat\q_i\sim n$. Note that \cref{average-stress-def-0} accounts for the stress due to bound linkers, with $\int n \,\d\u\,\d\q = \phi$ being their fraction. The unbound linkers exert no stress, and hence do not contribute to the stress.




To derive a hydrodynamic equation for the average stress, we multiply \cref{n-eqn} by $\pd{f}{u_\ab}$ and integrate over all bound states. Using that $\pd{f(\u,\q)}{u_\ab} = \mu u_\ab + D q_\ab$, this procedure yields the following constitutive relation (see Ref. \cite{oriola2017} and \cref{app-oriola-derivation}):
\begin{align}\label{sigma-average}
    \left(1+\frac{1}{\ku} \frac{\d}{\d t}\right)
    & \sigma_\ab= 2 \eta \,v_\ab-\nu \,\dot{Q}_\ab+\zeta Q_\ab.
\end{align}
This result shows that the system behaves like an active nematic viscoelastic fluid with relaxation time $\ku^{-1}$. The shear viscosity $\eta$, flow-alignment coefficient $\nu$, and
active stress coefficient $\zeta$ are obtained in terms of the model parameters:
\begin{equation}\label{coupling}
    \eta= \frac{\mu \phi}{2 \ku},
    \quad
    \nu=-\frac{D \phi}{\ku},
    \quad
    \zeta=(1-\phi)\left(\mu\Omega_u+D\Omega_q\right)
\end{equation}
where $\phi$ is the fraction of bound linkers, defined in \cref{phi-def}.
The scalars $\Omega_u$ and $\Omega_q$ are defined by
\begin{equation}
    \label{omega-uq-def}
\begin{aligned}
    \textstyle
    \int u_\ab \,\Omega(\u,\q) \;\d\u\, \d\q= \Omega_u Q_\ab,
    \\
    \textstyle
    \int q_\ab \,\Omega(\u,\q) \;\d\u\, \d\q = \Omega_q Q_\ab.
\end{aligned}
\end{equation}
These quantities arise from the breaking of detailed balance and give rise to the active stress term $\zeta Q_\ab$ in \cref{sigma-average}. 
As the integrands are tensors, the integral is proportional to $Q_\ab$, since it is the only symmetry-breaking tensor in the absence of external driving. In the presence of driving, the system could also have a macroscopic strain $U_{\alpha\beta} = \la u_{\alpha\beta} \ra$, which is also a symmetry-breaking tensor and should thus appear on the right-hand side of \cref{omega-uq-def}. Here, we consider weak external driving and hence we neglect this contribution.

The coupling coefficients \eqref{coupling} depend on $\phi(t)$, which is time-dependent. To specify its dynamics, we integrate \cref{n-eqn} over all bound states and obtain
\begin{align}\label{phi-average}
    \frac{1}{\ku}\frac{\d}{\d t} \phi = - (1+\Z) \phi +  \Z,
\end{align}
where $\Z$ is the duty ratio defined as 
\begin{align}
    \label{Z-def}
    \Z = \frac{1}{\ku}\int \kb(\u,\q)\,\d\u\, \d\q = \Z_\text{th}+\Omega_0.
\end{align}
Here, we decomposed the duty ratio into the thermal and active contributions (c.f.~\cref{kakd-new}):
\begin{align}
    \label{omega-0-def}
    \Z_\text{th} =\int \ee^{-\beta f/\rho}\,\d\u\, \d\q,
    \qquad
    \Omega_0 = \int \Omega\,\d\u\, \d\q.
\end{align}
The first integral is evaluated in \cref{Zth-res}.

As a useful approximation, we can use the steady-state fraction of bound linkers $\phi_\ss$ instead of $\phi(t)$ in \cref{coupling}. Setting the time derivative to zero in \cref{phi-average} yields
\begin{equation} \label{eq-phiss}
    \phi_\ss = \frac{\Z}{1+\Z} = \frac{\Z_\text{th}+\Omega_0}{1+\Z_\text{th}+\Omega_0}.
\end{equation}
where we clearly see that activity affects $\phi_\ss$ and hence all the coupling coefficients in \cref{coupling}.

\Crefrange{sigma-average}{Z-def} are the constitutive equations of the active gel defined by our mesoscopic model. 


\section{Constitutive relation with fluctuations}
\label{sec-fluct}
In this section, we go beyond the average stress and derive a stochastic differential equation (SDE) governing the temporal fluctuations of the stress tensor $\hat\bsigma$, as introduced in \cref{sigma-1}.
To this end, we start by writing \cref{sigma-average} for fluctuating quantities and incorporating a noise term $\hat{\bxi}^\sigma$:
\begin{align}\label{sigma-fluct}
    \textstyle
    \left(1+\frac{1}{\ku} \frac{\d}{\d t}\right) \hat\sigma_\ab= 
    2 \hat\eta\; v_\ab
    -
    \hat\nu\; \dot{Q}_\ab
    +
    \hat\zeta\; Q_\ab 
    + \hat\xi^\sigma_\ab.
\end{align}
Here, the external drivings $\v$ and $\dot\Q$ are imposed externally, thus not fluctuating. Above we defined the fluctuating coefficients
\begin{align}\label{transport-fluct}
    &\hat\eta \!=\! \frac{\mu }{2 \ku} \hat\phi,
    \quad
    \hat\nu \!=\! -\frac{D}{\ku}\hat\phi,
    \quad
    \hat\zeta \!=\! (1\!-\!\hat\phi)\left(\mu\Omega_u\!+\!D\Omega_q\right),
\end{align}
which fluctuate because of fluctuations in the fraction of bound linkers, $\hat\phi$. This quantity is defined similarly to the fluctuating stress in \cref{sigma-1}:
\begin{align}
\label{phi-1}
    \hat \phi(t) = \frac{1}{N}\sum_{i=1}^N \ind_i(t)\,, \quad \langle \hat\phi \rangle = \phi(t),
\end{align}
where the indicator function $\ind_i(t)$ is given by \cref{def-ind}, and $\la ~\ra$ denotes averaging over the distribution function $n(\u,\q, t)$. 
To capture the fluctuations in $\hat\phi$, we write the fluctuating version of \cref{phi-average}:
\begin{align}\label{phi-fluct}
    \frac{1}{\ku}\frac{\d}{\d t} \hat\phi = - (1+\Z) \hat\phi +  \Z + \hat\xi^{\phi},
\end{align}
which now includes the noise $\hat\xi^\phi$. After averaging over the fluctuations, \cref{sigma-fluct,transport-fluct,phi-fluct}  reduce to \cref{sigma-average,coupling,phi-average}, respectively.

Our objective now is to derive the statistical properties of the noise terms in \cref{sigma-fluct,phi-fluct} so that the stress fluctuations resulting from \cref{sigma-fluct} match those arising from the binding-unbinding dynamics in our mesoscopic model introduced in \cref{sec-model}.
Essentially, \cref{sigma-fluct,phi-fluct} define the noise;
substituting \cref{sigma-1,phi-1} into them yields
\begin{subequations}
\label{noise-def}
\begin{align}
    \label{phi-noise-def}
    &\hat\xi^\phi(t) = \frac{1}{N} \!\sum_{i=1}^N\!\Bigg\{
    \frac{1}{\ku}\frac{\d}{\d t} \ind_i(t) + (1\!+\!\Z)\ind_i(t)- \Z \Bigg\},
    \\
    \label{sigma-noise-def}
    &\hat\bxi^\sigma(t) = \frac{1}{N} \sum_{i=1}^N\Bigg\{
    (\mu \hat\u_i \!+\! D\hat\q_i)\left(\ind_i(t) \!+\! \frac{1}{\ku}\frac{\d}{\d t} \ind_i(t) \right)
    \\
    &-\frac{1}{\ku}(\mu \v\!+\! D \dot\Q)\ind_i(t)
    -(\mu \Omega_u\!+\!D \Omega_q)\Big(\!1-\ind_i(t)\!\Big) \!\Bigg\}. \nonumber
\end{align}
\end{subequations}
The fluctuating stress, fraction of bound linkers and noises defined in \cref{sigma-1,phi-1,phi-noise-def,sigma-noise-def} depend on the mesoscopic state of the system, specified by whether a linker is bound, $\ind_i(t)$, and to which state, ($\hat\u_i,\hat\q_i)$. In the Supplementary Material (SM), we solve the mesoscopic stochastic dynamics, obtaining the statistics of the stochastic process $\ind_i(t)$. This approach is direct and robust against model modifications, but it is technical and lacks intuitive clarity. Here, instead, we present a simpler and clearer method to obtain the properties of fluctuating hydrodynamic quantities.

First, the noises have zero mean, even in a non-stationary state, characterized by a non-stationary distribution of bound linkers $n(\u,\q, t)$. By averaging \cref{sigma-fluct}, and interchanging the time derivative and the average as $\la \tfrac{\d}{\d t}\hat\bsigma(t) \ra = \tfrac{\d}{\d t}\la \hat\bsigma(t)\ra=\tfrac{\d}{\d t}\bsigma(t)$, we get $\langle \hat\bxi^\sigma\rangle = 0$. Proceeding similarly from \cref{phi-fluct}, we get $\langle \hat\xi^\phi\rangle = 0$.
The zero mean of the noise can also be shown directly from \cref{noise-def} (see SM~\ref{sm-with-noise}).

We next obtain the second moment of the stress fluctuations, i.e. the autocorrelation function, defined as
\begin{align}
\label{C-sigma}
    C^{(\sigma)}_\abmn(t-t') \!=\! \la \,\hat\sigma_\ab(t)\hat\sigma_\mn(t')\,\ra_\ss \!- \la \hat\sigma_\ab \ra_\ss \la \hat\sigma_\mn \ra_\ss,
\end{align}
where $\la~\ra_\ss$ denotes the average over the steady-state distribution, which we obtain in \cref{sec-ss}. In the following, we calculate this correlation function and then find the spectrum of the noise \eqref{noise-def} that produces obtained stress autocorrelation in \cref{sigma-fluct}.

To simplify the notation, we combine the fluctuating variables into a ($d^2+1$)-dimensional vector $\hatY_t = (\hat\bsigma(t), \hat\phi(t))\equiv (\hat\sigma_{11}, \dots, \hat\sigma_{dd}, \hat\phi)$, where $d$ is the system's dimensionality. Here, in $\hatY_t$, time argument is indicated as the subscript $t$ to make the notation more compact. Then, \cref{sigma-fluct,phi-fluct} can be compactly written as
\begin{align}\label{Y-fluct-compact}
    \frac{1}{\ku} \frac{\d}{\d t} \hatY_t = -\frac{1}{\ku} \M (\hatY_t-\Y_\ss) + \hat\bxi_t,
\end{align}
where $\hat\bxi_t = (\hat\bxi^\sigma(t), \hat\xi^\phi(t))$ is the noise vector, $\Y_\ss$ is the steady-state average of $\hat\Y$, and $\M$ is a $(d^2+1) \times (d^2+1)$ block matrix defined as
\begin{align}
\label{M-def}
   \M=\begin{pNiceArray}{c|c}
        \ku & -(\mu \v +D \dot \Q) +\ku (\mu\Omega_u\!+\!D\Omega_q)\Q
        \\[1ex]  \hline
        0 & \ku (1+\Z)
\end{pNiceArray}.
\end{align}   
Here, the upper-left block is the $d^2$ dimensional identity matrix multiplied by $\ku$, lower-right block is the scalar $\ku(1+\Z)$.
Taking the Fourier transform of \cref{Y-fluct-compact} and using 
the \WK{} theorem (\cref{app-wiener}), we obtain a relation between the spectral density of the noise and the spectral density of $\hatY$:
\begin{align}\label{Sxi-SY}
    {\rm  S}_\bxi(\omega) =& \frac{1}{\ku^2}(\M + \ii\omega\rI)\,\rS_\Y(\omega)\,
    (\tr{\M} - \ii\omega\rI)
\end{align}
where $\rI$ is the identity matrix. The spectral density $\rS_\Y$ is the Fourier transform of the autocorrelation function of $\hatY_t$. We use the following Fourier transform convention:
\begin{align}
    \label{ft-convention}
    \rS_\Y(\omega) = \int_{-\infty}^{\infty} {\rm C}_\Y(t) \ee^{-\ii \omega t} \d t,
\end{align}
where the autocorrelation function is defined as
\begin{equation}
    \label{CY-line-2}
\begin{aligned}
    {\rm C}_\Y(t)=&\,\Exp{(\hatY_t-\Y_\ss)(\hatY_0-\Y_\ss)\tr{}}
    \\
    =&\,\Exp{~\Exp{(\hatY_t-\Y_\ss) \big| \hatY_0}~(\hatY_0-\Y_\ss)\tr{}}
    \\
    =&\,\Exp{~\left(\Exp{\hatY_t \big| \hatY_0}-\Y_\ss\right)~(\hatY_0-\Y_\ss)\tr{}}.
\end{aligned} 
\end{equation}
Here, all expectations are evaluated with respect to the steady-state distribution. In obtaining the second line, we consider $t>0$ and apply the tower rule for conditional expectations \cite{grimmett2020}.

We now compute the conditional expectation in \cref{CY-line-2}.
The expression $\Exp{\hatY_t \big| \hatY_0=\y}$ denotes the mean of the steady-state trajectories $\hat\Y_t$ that go through $\hatY_0=\y$.
Thus, we compute the conditional expectation by using the tower rule once more: First, we calculate the average of $\hatY_t$ for some specific initial configuration of the linkers $S_0$ at time $t=0$, and then we take the average over all such configurations that have a certain stress and fraction of bound linkers, given as $\hatY_0=\y$:
\begin{align}
    \Exp{\hatY_t \mid\y} = \Exp{~\Exp{\hatY_t \mid S_0(\y)}\mid \y}
    \nonumber
\end{align}
where $S_0(\y)$ is the initial configuration of the linkers that yields $\hatY_0=\y$. For example, 
if at $t=0$ only $K\le N$ linkers are bound, then $S_0(\y)$ can be written as:
\begin{equation}
    \begin{gathered}
    S_0(\y) = \{ \hat\u_{i} = \u_{i}, \hat\q_{i} = \q_{i} \text{ for } i=1,\dots,K \}
    \nonumber
    \end{gathered}
\end{equation}
such that   $\left(\sum_{i=1}^K \bsigma^\text{linker}(\u_i,\q_i), \frac{K}{N}\right) = \y\equiv(\bsigma(0), \phi(0))$.

Because the linkers are indistinguishable, the state of the system is fully described by the distribution of linkers $n( \u, \q, t)$.
Requiring an initial state $S_0(\y)$ is equivalent to requiring an initial distribution $n^\ast(\u,\q,{t\!=\!0})$ 
\begin{align}
    n^\ast(\u, \q, 0) = \frac{1}{N} \sum_{i=1}^K \delta(\u-\u_{i})\delta(\q-\q_{i})
    \nonumber
\end{align}
Then, the state of the system at time $t$ is given by the distribution of linkers $n^\ast( \u, \q,t)$, which evolved from the initial distribution $n^\ast( \u, \q,0)$
according to \cref{n-eqn}.
Therefore, we can write
\begin{align}
    \Exp{\hatY_t \mid S_0(\y)} = \la \hatY_t \ra_{\sim n^\ast} = (\bsigma(t), \phi(t)) \equiv \Y_t,
    \label{Y-important}
\end{align}
where for the second equality we used \cref{average-stress-def-0,phi-1}. To obtain how $\Y_t$ evolves from the initial condition, we use \cref{sigma-average,phi-average} for the average stress and fraction of bound linkers. They can be written in a compact form as
\begin{align}\label{Y-average}
    \frac{1}{\ku}\frac{\d}{\d t} \Y_t = -\frac{1}{\ku}\M (\Y_t-\Y_\ss),
\end{align}
where the matrix $\M$ is the same as in \cref{M-def}. The initial distribution $n^\ast( \u, \q,0)$  sets an initial condition $\Y_0 = \y$ for \cref{Y-average}. Thus
\begin{align}
    \Exp{\hatY_t \mid S_0(\y)} = \Y_t = \Y_\ss + \ee^{-\M  t}(\y - \Y_\ss) 
    \nonumber
\end{align}
The right-hand side depends on the initial state only through $\y$. The specific state of the linkers, denoted by $S_0$, is not relevant; only the initial stress and fraction of bound linkers matters.
Using these results, the conditional expectation in \cref{CY-line-2} evaluates to 
\begin{align} 
    \Exp{(\hatY_t-\Y_\ss) \big| \hatY_0} = \ee^{-\M  t} (\hatY_0-\Y_\ss).
    \nonumber
\end{align} 
Finally, replacing this result into \cref{CY-line-2}, we obtain the autocorrelation function:
\begin{align}
\label{CY}
    {\rm C}_\Y(t) =&\, \ee^{-\M  t} \Sigma \qquad \text{for} ~~t\ge0,
    \\
    \label{Ycov}
    \text{where}\quad \Sigma =&\, \Exp{(\hatY - \Y_\ss)(\hatY - \Y_\ss)\tr{}}
\end{align}
is the covariance matrix of $\hatY$.

We now compute the spectral density in \cref{Sxi-SY}. Hence, we need the autocorrelation function both for positive and negative times. Using the time-translation invariance of the steady-state, we have
\begin{align}
    \label{CY-time-translation}
    {\rm C}_\Y(t) = \tr{ {\rm C}_\Y(-t) } \qquad  \text{for} ~~t<0.
\end{align}
Then, using the result in \cref{CY} and the identity $\int \ee^{-{\rm A} t} \d t = -{\rm A}^{-1} \ee^{-{\rm A} t}+c$ for a non-singular matrix ${\rm A}$, we obtain
\begin{align}
\label{SY}
    \rS_\Y(\omega) =  \left(\M + \ii\omega \rI\right)^{-1}\Sigma + \Sigma\,\left(\tr{\M} - \ii\omega\rI\right)^{-1}.
\end{align}
Substituting this result into \cref{Sxi-SY} gives:
\begin{align*}
    \ku^2\,\rS_\bxi(\omega) =& 
    \left(\M + \ii\omega\rI\right)
    \left(\M + \ii\omega\rI\right)^{-1}\Sigma
    \left(\tr{\M} - \ii\omega\rI\right)
    \\
    &
     \!\!+ \left(\M + \ii\omega\rI\right)\Sigma\,\left(\tr{\M} - \ii\omega\rI\right)^{-1}
    \!\left(\tr{\M} - \ii\omega\rI\right)
    \\
    =&\,
    \Sigma
    \left(\tr{\M} - \ii\omega\rI\right)
     + \left(\M + \ii\omega\rI\right)\Sigma
    \\
    =& \Sigma \,\tr{\M}+\M\,\Sigma.
\end{align*}
This means that the noise spectrum is independent of $\omega$, corresponding to white noise:
\begin{align}
    \label{white-noise}
    \langle \hat\bxi(t) \hat\bxi(t')\tr{} \rangle = \frac{1}{\ku^2}(\Sigma\, \tr{\M} + \M \,\Sigma) \;\delta(t - t').
\end{align}


\noindent We close this section with a few remarks:
\begin{enumerate}
\setlength{\leftskip}{12pt} 

\item[(i)] We obtained that the noise on the stress and the fraction of bound linkers, in \cref{sigma-fluct,phi-fluct}, is white. This property comes from the fact that the linker dynamics is Markovian, i.e. memoryless. 
Yet, as the time derivative of $\hatY$ appears in \cref{Y-fluct-compact}, the dynamics of the stress and the fraction of bound linkers is not memoryless. Rather, their correlations decay exponentially as shown in \cref{CY}.
\item[(ii)] Our results imply that, whenever an averaged constitutive relation \eqref{Y-average} also holds for the conditional average $\Y_t = \Exp{\hatY_t | \Y_0}$, then the noise is white. In \cref{app-colored-noise}, 
we demonstrate how this condition breaks down in the presence of colored noise (time-correlated noise). Specifically, we start with colored noise in \cref{Y-fluct-compact} and show that the conditional average does not satisfy \cref{Y-average}.
\item[(iii)] Note that \cref{white-noise} does not impose a specific distribution of the noise; $\hat\bxi(t)$ is not necessarily Gaussian and $\hat\bxi$ is not a Wiener process \cite{grimmett2020}. In fact, in our system with linkers, the noise $\bxi^\sigma$ is a stochastic process with jumps, as can be seen from \cref{phi-noise-def,sigma-noise-def}.
\end{enumerate}

Having determined the noise spectrum, we now turn to its amplitude, given by $(\Sigma\tr{\M} + \M\Sigma)$, where $\Sigma$ is the covariance matrix of $\hat\Y$ (see \cref{Ycov}). Since $\Sigma$ characterizes the steady state of the system, our next step is to determine the steady-state distribution $n_\ss(\u,\q)$ in order to calculate $\Sigma$.

\section{Steady state distribution}
\label{sec-ss}

In this section, we derive the steady-state distribution $n_\ss(\u,\q)$, which gives the steady-state fraction of linkers bound at each state $(\u,\q)$. Using this distribution, in the next section we calculate the covariance matrix $\Sigma$ of $\hatY$ defined in \cref{Ycov}, which is then used to calculate the noise amplitude in \eqref{white-noise}.

To find the steady-state distribution, we set the time derivative in \cref{n-eqn} to zero. Solving the resulting equation with the method of integrating factors (see \cref{app-ss}), we obtain
\begin{align}\label{ss-10}
    n_\ss(\u,\q) = \frac{1}{1\!+\!\Z}\int_0^\infty\!\!\!\kb(\u\!-\!\v \tau,\; \q\!-\!\dot\Q \tau) \;\ee^{-\ku \tau} \;\d\tau.
\end{align}
The steady-state distribution deviates from the equilibrium distribution because of (i) the external driving, given by the shear and rotation rates $\v$ and $\dot\Q$, and (ii) the activity $\Omega$ in the binding rate $\kb$ (see \cref{kakd-new}). Setting both of these effects to zero, \cref{ss-10} reduces to the Boltzmann distribution $n_\text{eq} = (1+\Z)^{-1} \ee^{-\beta F}$ for an equilibrium system.


We now provide an interpretation of \cref{ss-10}, which describes how external drives ($\v$, $\dot\Q$) distort the steady-state distribution away from equilibrium. 
Consider a linker that is bound to state $(\u,\q)$. If the linker has been bound for a time $\tau$, it must have initially bound to state $(\u_0, \q_0) = (\u - \v \tau, \q - \dot\Q \tau)$, and subsequently drifted to $(\u, \q)$ due to external driving. This explains the argument of the binding rate $\kb(\u_0,\q_0)$ in \cref{ss-10}. The probability that a linker binds at $(\u_0, \q_0)$ is $\kb(\u_0,\q_0)/\int \kb(\u,\q)\d\u\,\d\q = \kb(\u_0,\q_0)/(\Z\ku)$, where we used \cref{Z-def}.
Thus, the probability of finding a linker bound at $(\u,\q)$ conditioned on being bound for a time $\tau$ is equal to the fraction of bound linkers $\phi_\ss$ times $\kb(\u_0,\q_0)/(\Z\ku)$.
Averaging this result over $\tau$ yields
$n_\ss(\u,\q)=\phi_\ss \langle\; \kb(\u - \v \tau, \q - \dot\Q \tau)/(\ku \Z) \;\rangle_\tau$.
In \cref{sm-ss-prob}, we show that $\tau$ is exponentially distributed with rate $\ku$, from which \cref{ss-10} follows.

\begin{figure}
    \centering
    \includegraphics[width=\linewidth]{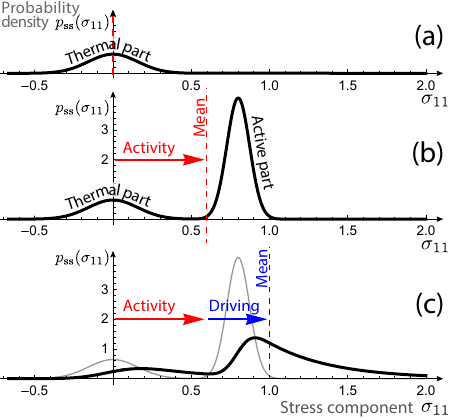}
	\caption{\textbf{Activity and external drivings shift the stress probability distribution.}
Steady-state distribution of the $\sigma_{11}$ component of the stress tensor in equilibrium \textbf{(a)}, in the presence of activity \textbf{(b)}, and of activity and external driving \textbf{(c)}. The stress distribution results from the distribution of bound linkers given in \cref{ss-10} (see text). Each bound linker contributes a stress $\bsigma^{\text{linker}}(\u,\q)$ as defined in \cref{sigma-1-def}.
\textbf{(a)} Isotropic stress distribution in thermal equilibrium ($\Omega=0$ in \cref{kakd-new}). The distribution is symmetric around $\sigma_{11}=0$ with vanishing mean stress.
\textbf{(b)} Effect of activity on the stress distribution. The distribution has contributions from thermal and active fluctuations; the latter arise from the breaking of detailed balance in \cref{kakd}.
For the active part,
we consider the molecular activity $\Omega(\u,\q)$ presented in \cref{sec-specific}, which promotes binding at certain orientation $\q=\Q$. We take $Q_{11}=1$. Because we also choose a positive elastonematic coupling $D>0$ (which couples $\u$ and $\q$, see free energy \cref{free-energy}), the linker is biased to binding at states with positive $u_{11}$. Thus, the stress $\sigma_{11}\sim \mu u_{11}+Dq_{11}$ is also biased towards positive values. Overall, the mean stress becomes positive because of the symmetry-breaking molecular activity.
\textbf{(c)} Shearing the system at a rate $v_{11}>0$ additionally shifts the distribution in the direction of $v_{11}$, as encoded in \cref{ss-10}.
Parameter values, chosen for illustration purposes, are given in \cref{params-fig1d}. See \cref{app-plots} for more details, and the distribution function's visualisation in two dimensions.
}
	\label{fig-sigmadist-1d} 
\end{figure} 

We now discuss the effects of activity and external driving on the steady-state distribution for the stress $\bsigma$.
The stress that a linker bound at $(\u,\q)$ exerts on the network is given by \cref{sigma-1-def}. Hence, we use \cref{sigma-1-def} to change the variables of $n_\ss(\u,\q)$ from $\u,\q$ to $\bsigma$, and thus obtain the steady-state distribution for the stress tensor (see \cref{app-plots} for details).
From it, we calculate the average stress (see \cref{app-sigma-evaluation})
\begin{align}
\label{sigma-ss}
	\bsigma_\ss = \frac{\phi}{\ku}\left(\mu \v + D \dot \Q\right) + \zeta \Q,
\end{align}
which matches the long-time limit of the constitutive relation in \cref{sigma-average}. \Cref{fig-sigmadist-1d} illustrates how \textit{activity} and \textit{external driving} shift the stress distribution $p_\ss(\bsigma)$.
Although the contributions from \textit{activity} and \textit{external driving} are additive in the average stress in \cref{sigma-ss}, their effects on the steady-state distribution $n_\ss$ are not additive.
One reason is that these two effects do not commute: 
Consider first turning on the activity (going from \cref{fig-sigmadist-1d}a to \cref{fig-sigmadist-1d}b). This change affects the population of bound linkers, which are then stretched and rotated by the external driving (going from \cref{fig-sigmadist-1d}b to \cref{fig-sigmadist-1d}c). 
Now, consider first shifting the distribution of bound linkers due to driving. This shift would affect the states available for binding, and hence, subsequently turning on activity would have a different effect on the distribution of bound linkers.
Whereas it does not impact the average stress (\cref{sigma-ss}), this interdependence of the effects of \textit{activity} and \textit{external driving} on the steady-state distribution of the linkers gives rise to cross-coupling terms in 
the variance of the stress tensor $\la \hat\sigma_\ab \hat\sigma_\mn\ra_\ss - \la \hat\sigma_\ab \ra_\ss\la\hat\sigma_\mn\ra_\ss$, calculated in \cref{sigma-var-result-sm} in appendix. As a result, it also impacts the noise amplitude, which we calculate in the following section.




\section{Amplitude of the noise}
\label{sec-noise-amplitude}
Having established the steady-state distribution of bound linkers, \cref{ss-10}, we are now ready to calculate the covariance matrix $\Sigma$, introduced in \cref{Ycov}, and hence the amplitude of the noise given in \cref{white-noise}. Specifically, for the stress covariance, we evaluate integrals like 
\begin{align}
    \la \hat\sigma_\ab \hat\sigma_\mn\ra_\ss 
    =\frac{1}{N}\int \pd{f}{u_\ab} \pd{f}{u_\mn} n_\ss(\u,\q)\,\d \u \d \q,
    \nonumber
\end{align} 
where we have used \cref{sigma-1-def,sigma-1} and assumed independent linkers. This integral, along with the other components of $\Sigma$, can be directly evaluated as shown in \cref{app-sigma-evaluation}. Then, using \cref{white-noise}, we determine the amplitude of the noise in the constitutive relation \cref{sigma-fluct}:
\begin{align}\label{sigma-noise-correlation}
    \langle \hat\xi^\sigma_\ab (t)\hat\xi^\sigma_\mn (t')\rangle = \Lambda^\sigma_\abmn \frac{1}{V}\delta(t-t'),
\end{align}
where the amplitude $\Lambda^\sigma_\abmn$ decomposes into four parts:
\begin{align}\label{Sigma-decomposed}
    \Lambda^{\sigma}_\abmn =& \,
    \Lambda^{\text{thermal}}_\abmn
    + \Lambda^{\text{driven}}_\abmn
    + \Lambda^{\text{active}}_\abmn
    + \Lambda^{\text{cross}}_\abmn,
\end{align}
which we discuss below.

\subsection{Thermal, driven, and active contributions}

In thermodynamic equilibrium, where both the activity and the external drivings vanish ($\Omega = 0$, $\v = \dot\Q = 0$), only the thermal contribution remains, and it is given by
\begin{align}\label{Sigma-thermal}
    \Lambda^{\text{thermal}}_\abmn=    4\eta \kBT ~\Tiso_\abmn,
\end{align}
where $\Tiso_\abmn =\frac{1}{2}(\delta_{\alpha\nu}\delta_{\beta\mu}+\delta_{\alpha\mu}\delta_{\beta\nu})-\frac{1}{d} \delta_\ab \delta_{\mu\nu}$ is the \nth{4}-rank isotropic tensor, $\kBT$ is the thermal energy, and $\eta$ is the equilibrium viscosity given by \cref{coupling} with $\phi=\phi_\text{eq}=\phi_\ss(\Omega=0)$. In equilibrium, the noise and dissipative coefficient $\eta$ must satisfy the Fluctuation-Dissipation theorem (FDT) \cite{landau-stat,mazur}. In \cref{app-FDT-passive}, we verify that \cref{Sigma-thermal} indeed satisfies the FDT. Hence, our model reproduces equilibrium thermal noise when detailed balance is satisfied ($\Omega = 0$).

The external driving puts the system out of equilibrium while preserving detailed balance at the mesoscopic scale, so that the system remains passive, but driven. 
In our model, this corresponds to $\Omega=0$ and $\v,\, \dot\Q \ne 0$.
Interestingly, this external driving leads to additional noise with amplitude
\begin{align}\label{Sigma-driven}\Lambda^{\text{driven}}_\abmn=&\frac{2\phi_\ss}{\rho \ku^3}
    (\mu v_\ab +D \dot Q_\ab)
    (\mu v_\mn +D \dot Q_\mn)
\end{align}
Thus, our model predicts a nonequilibrium contribution to the noise in passive driven systems. While the external drivings are held constant in our model, \cref{Sigma-driven} suggests that this noise contribution would be multiplicative if $\v$ and $\dot\Q$ were treated as dynamical variables. 
The driven contribution $\Lambda^{\text{driven}}$ depends on temperature only indirectly through the fraction of bound linkers $\phi_\ss$ (see \cref{eq-phiss}). Moreover, since $\Lambda^{\text{driven}}$ is of second order in the drivings $\v$ and $\dot\Q$, it is often neglected in linear approximations \cite{mazur, basu2008}.

When activity is introduced ($\Omega\neq 0$), two additional contributions to the noise amplitude emerge: a purely active contribution and a cross contribution from the combined effects of activity and external driving. The purely active contribution is given by
\begin{align}
\label{sigma-active}
\Lambda^{\text{active}}_\abmn
=&\,
-4\eta\kBT \; \Tiso_\abmn\;\frac{\Omega_0}{\Z}
\\
&\hspace{-6ex} 
+\frac{2}{\rho \ku(1+\Z)}\!\int \!
\left(\mu u_\ab\!+\!D q_\ab \right)
\left(\mu u_\mn\!+\!D q_\mn \right)\Omega \,\d\u\, \d\q,
\nonumber
\end{align}
which depends explicitly on the function $\Omega$ that quantifies the breaking of detailed balance; $\Omega_0$ is the zeroth moment of $\Omega$ defined in \cref{omega-0-def}, and the second moments of $\Omega$ appear in the second line. This result provides a fluctuation-activity relation, as it connects the amplitude of the active noise with the microscopic activity, namely the breaking of detailed balance specified by $\Omega$. Whereas the first term in \cref{sigma-active} is an isotropic contribution that can be subsumed in an effective temperature, the second contribution is anisotropic and cannot.

Finally, the cross contribution is
\begin{align}
\label{Sigma-cross}
\Lambda^{\text{cross}}_\abmn=&\frac{2\zeta}{\rho \ku^2}
    \Big[
     (\mu v_\ab \!+\!D \dot Q_\ab) Q_\mn
     \!+\!
    (\mu v_\mn \!+\!D \dot Q_\mn) Q_\ab \Big],
\end{align}
which depends directly both on the drivings $\v$ and $\dot\Q$, and on activity through the active stress coefficient $\zeta$ (see \cref{coupling}). The origin of this cross term was discussed in the last paragraph of \cref{sec-ss}.

Notice that the decomposition into passive and active noise contributions in \cref{Sigma-decomposed} is not perfect:
The ``thermal'' part $\Lambda^\text{thermal}_\abmn$ depends on activity through viscosity $\eta$ (see \cref{coupling}). Moreover, the duty ratio $\Z$, defined in \cref{Z-def}, depends on both activity $\Omega$ and temperature. Therefore, the active contribution $\Lambda^\text{active}_\abmn$ depends on temperature too. In the following section, we study the active part of the noise amplitude.

\subsection{Symmetries of the noise amplitude tensor}
\label{sec-reduced-repr}
The noise amplitude $\Lambda^\sigma_\abmn$ in \cref{sigma-noise-correlation} is a fourth-rank tensor with 81 entries. However, it has symmetries that strongly reduce its number of degrees of freedom. These symmetries arise from two facts: (i) that $\Lambda^\sigma_\abmn$ is defined in \cref{sigma-noise-correlation} as the product of the noise tensor $\hat \xi_\ab$, which is itself a second-rank traceless and symmetric tensor; 
and (ii) that we assume a uniaxial nematic, such that the system is invariant under rotations around the axis of nematic order.

The rank-4 tensors $\T$ that satisfy above symmetries have the following properties:
\begin{itemize}
    \setlength{\leftskip}{12pt} 
    \setlength\itemsep{0pt}

    \item Traceless with respect to the first and last pairs of indices: $\T_{{\xi\xi}\mu\nu}=0
    ,~\T_{\alpha\beta{\xi\xi}}=0$,
    \item Symmetric with respect to the first and last pairs of indices: $\T_{{\alpha\beta}\mu\nu}=\T_{{\beta\alpha}\mu\nu}
    ,~
    \T_{\alpha\beta{\mu\nu}}=\T_{\alpha\beta{\nu\mu}}$,
    \item Symmetric under the exchange of the first and last pairs of indices: $\T_{{\alpha\beta}\mu\nu}=\T_{\mu\nu{\alpha\beta}}$,
    \item Rotationally invariant around the nematic director $\n$: $
        R_{\alpha\alpha'}R_{\beta\beta'}R_{\gamma\gamma'}R_{\nu\nu'}\T_{\alpha'\beta'\mu'\nu'}=\T_\abmn$,
    where $R$ is an arbitrary rotation matrix 
    around the nematic director axis $\n$,
\end{itemize}
where the first three properties follow from (i) and the last property follows from (ii).

The linearly independent tensors that satisfy these requirements are (see \cref{app-axis-invariant} for derivation)
\begin{subequations}
\label{Tm-def}
\begin{align}
    \T^\text{(0)}_\abmn
    =& \frac{1}{2}(\delta_{\alpha\nu}\delta_{\beta\mu}\!+\!\delta_{\alpha\mu}\delta_{\beta\nu})\!-\!\frac{1}{3} \delta_\ab \delta_{\mu\nu}
    \equiv \Tiso_{\abmn},
    \label{Tm-0}
    \\
    \T^\text{(1)}_\abmn
    =& \frac{1}{4}\left[
        \left(\delta_{\alpha\mu} N_{\beta\nu} + \delta_{\beta\nu} N_{\alpha\mu}\right)
        +
        \left(\delta_{\alpha\nu} N_{\beta\mu}+\delta_{\beta\mu} N_{\alpha\nu}\right)
        \right] \nonumber
    \\
    &- \frac{1}{3} \left(\delta_\ab  N_{\mu\nu} + \delta_{\mu\nu} N_\ab  \right),
    \label{Tm-1}
    \\
    \T^\text{(2)}_\abmn
    =& N_\ab N_{\mu\nu}.
\end{align}
\end{subequations}
where $N_\ab = n_\alpha n_\beta - \frac{1}{3}\delta_\ab$ is the orientation tensor, with $\n$ being the nematic director with $|\n|=1$. 
For a uniaxial nematic, it relates to the full nematic order-parameter tensor as $Q_\ab = |\Q| N_\ab$ where $|\Q|$ is the strength of the nematic order.
The superscripts denote the order in $\Q$ of each tensor. The first tensor $\Tiso$ is the (totally symmetric) isotropic rank-4 tensor, which has already appeared in \cref{Sigma-thermal}.

All the contributions to the noise amplitude in \cref{Sigma-decomposed} can be expressed as a linear combination of the tensors in \cref{Tm-def}. We now do this for the active noise amplitude \cref{sigma-active}. Similar to \cref{omega-uq-def}, we decompose the integrals in \cref{sigma-active} as\
\begin{subequations}
\label{omega-2}
\begin{align}
    \textstyle\int u_\ab u_\mn \Omega\,\d\u\, \d\q = \sum_{i=0}^{2} \Omega_{uu}^{(i)} \T^{(i)}_\abmn,
    \\
    \textstyle\int q_\ab q_\mn \Omega\,\d\u\, \d\q =  \sum_{i=0}^{2} \Omega_{qq}^{(i)} \T^{(i)}_\abmn,
    \\
    \textstyle\int \tfrac{1}{2}(u_\ab q_\mn \!+\! q_\ab u_\mn) \Omega\,\d\u\, \d\q = \sum_{i=0}^{2} \Omega_{uq}^{(i)} \T^{(i)}_\abmn.
\end{align}
\end{subequations}
These relations uniquely define nine scalar quantities, since the tensors in \eqref{Tm-def} are linearly independent. These quantities -- $\Omega_{uu}^{(i)}$, $\Omega_{uq}^{(i)}$, and $\Omega_{qq}^{(i)}$ for $i=0,1,2$ -- characterize how the breaking of detailed balance generates active fluctuations. In terms of them, the active noise amplitude can be expressed as
\begin{multline}
\Lambda^{\text{active}}_\abmn
=-4\eta\kBT \; \Tiso_\abmn\;\frac{\Omega_0}{\Z}  \\
+ \sum_{i=0}^{2}\left(
    \mu^2 \Omega_{uu}^{(i)} + 2 \mu D\,\Omega_{uq}^{(i)} + D^2 \,\Omega_{qq}^{(i)}
\right) \T^{(i)}_\abmn.
\end{multline}
This compact representation shows how the nonequilibrium noise statistics can be expressed in terms of the 12 activity scalar quantities ($\Omega_0$, two first moments, 9 second moments) introduced above.

\subsection{Orientation-selective activity: a minimal illustrative case}
\label{sec-specific}

Here, we illustrate the properties of the active noise by taking a simple specific form of the molecular activity:
\begin{equation}\label{activity-specific}
    \Omega(\u,\q)=\Omega(\u)\; \delta(\q-\bs{Q}).
\end{equation}
This choice corresponds to activity that selectively promotes binding to a prescribed alignment $\Q$, enforced here through a Dirac delta function.
In practice, however, the delta function should be understood as an idealization. Real systems exhibit a finite width in alignment selectivity. This is particularly important when the activity is negative ($\Omega(\u,\q)<0$), corresponding to suppressed binding: in that case, the binding rate $\kb$ in \cref{kakd} must remain positive, which can only be satisfied by a sharply peaked but finite-width function instead of $\delta(\q-\bs{Q})$. For clarity and analytical simplicity, we will continue to use the delta function representation in what follows.

Additionally, we assume that the strain part $\Omega(\u)$ in \cref{activity-specific} is invariant under any rotation. 
With this symmetry, the first moments of the activity defined in \cref{omega-uq-def} evaluate to
\begin{align}\label{Omega-uq-specific} 
    \Omega_u = 0,
    \qquad
    \Omega_q = \Omega_0.
\end{align}
These quantities measure the departure from detailed balance, and they appear as factors in the active stress coefficient $\zeta$ in \cref{coupling}.

Respectively, here we found that the second moments of the molecular activity determine the active noise strength. Using their definition in \cref{omega-2}, we obtain
\begin{equation}
\label{Omega-uuqq-specific}
    \Omega^{(0)}_{uu}= A \Omega_0, 
    \quad
    \Omega^{(2)}_{qq}= \Omega_0 |\Q|^2.
\end{equation}
with all other coefficients being zero (note that $Q_\ab Q_\mn=|\Q|^2 \T^{(2)}_\abmn$). Here we introduced a standard-deviation-like parameter $A>0$, which characterizes the width of the function $\Omega(\u)$.

Ignoring contributions from the external driving in \cref{Sigma-decomposed}, the total noise magnitude for this simple choice of activity is given by
\begin{multline}
\label{lambda-simple}
    \Lambda^{\sigma}_{\ab\mn} =
    4\eta \kBT \frac{1}{1+\Omega_0/\Z_\text{th}} ~\Tiso_\abmn 
    \\
    \frac{2 \Omega_0}{\rho \ku (1\!+\!\Z_\text{th}+\Omega_0)} \left(A\mu^2\Tiso_{\ab\mn}
    +
    D^2 Q_\ab Q_\mn
    \right)
\end{multline}
In the activity-dominated regime, where $\Omega_0 \gg \Z_\text{th}$, the thermal contribution becomes negligible. Beyond this limit, two distinct active contributions remain. The first one is a direct analogue of the thermal term. In the activity-dominated regime, the viscosity takes the form
$\eta=\frac{\mu\phi}{2\ku}=\frac{\mu \Omega_0}{2\ku (1+\Omega_0)}$,
and the parameter $A$, defined in \cref{Omega-uuqq-specific}, plays the role analogous to $\frac{\rho\kBT}{\mu}$ in the Boltzmann factor of \cref{kakd-new}. Thus, with effective temperature $T_\text{eff}=\frac{A\mu}{\rho k_{\rm B}}$, the first term in the second line of \cref{lambda-simple} can therefore be interpreted as an equilibrium-like contribution, equivalent to $4\eta k_{\rm B} T_\text{eff}$.

The second term, proportional to $Q_\ab Q_\mn$, represents a genuinely nonequilibrium active contribution that has no analogue within the equilibrium thermodynamic framework. This term arises because molecular activity $\Omega$ biases crosslink binding toward specific orientations characterized by $\Q$. Owing to the elastonematic coupling coefficient $D$, crosslinks preferentially bind with local strain fields $\u \sim D\Q$. Consequently, this mechanism introduces anisotropy in both the stress fluctuations and the corresponding noise amplitude.

Overall, the specific choice of molecular activity in \cref{activity-specific} significantly simplifies the results, reducing the number of independent activity parameters from twelve to only two. In the following section, we investigate the dynamics of a tracer particle embedded in an active gel described by this choise of molecular activity.


\section{Fluctuations of a tracer particle}
\label{sec-observables}
In the previous sections, we derived the noise term in the constitutive relation for the stress tensor of an active gel. Now, we discuss how this noise affects a tracer particle.
Specifically, we calculate the fluctuation spectrum and the response function of a tracer particle immersed in the active gel. In experiments, these quantities are measured by passive and active microrheology, respectively, and comparing them can reveal departures from the FDT that indicate the presence of active fluctuations.

If the tracer particle experiences a displacement $\mathbf{R}(t)$ when subject to an external force $\F^\text{ext}(t)$, the response function $\chi^R$ is defined, in Fourier space, by
\begin{align}\label{chi-def}
    \tilde{R}_\alpha(\omega) = \chi^R_\ab(\omega) \tilde{F}^\text{ext}_\beta(\omega).
\end{align}
Respectively, the fluctuation spectrum is defined as the Fourier transform of the autocorrelation function of the particle's position:
\begin{align}\label{S-def}
    \rS^R_\ab(\omega) = \int_{-\infty}^\infty \la \hat R_\alpha(t) \hat R_\beta(0)\ra \ee^{-\ii \omega t} \d t.
\end{align}
In thermodynamic equilibrium, these two quantities are related by the fluctuation-dissipation theorem (FDT):
\begin{align}\label{FDT}
    \rS^{R,\text{eq}}_\ab(\omega) = -\frac{2\kBT}{\omega}  \chi^{\prime\prime}_\ab(\omega),
\end{align}
where $\chi^{\prime\prime}_\ab$ is the imaginary part of the response function.
Below, we calculate $\chi^R$ and $\rS^{R}$ for a tracer particle in an active gel, and demonstrate the FDT violation due to broken detailed balance in the active gel.

To this end, we first write the fluctuating hydrodynamic equations for an incompressible active viscoelastic fluid that follows the constitutive relation \cref{sigma-fluct}. Here, we focus on a situation with a constant and uniform nematic orientation $\Q$, and in the low-Reynolds limit characteristic of intracellular flows. Thus, the fluctuating hydrodynamic equations are
\subeqn{
    \label{fd-time}
\begin{align}
    &\partial_\beta  \hat\sigma^\text{tot}_\ab=0\,, 
    \quad \hat\sigma^\text{tot}_\ab = \hat\sigma_\ab \! -\! \delta_\ab\, \hat p \,, 
    \quad \partial_\alpha \hat v_\alpha \!=\! 0\,,
    \\
    &\left(1+\frac{1}{\ku} \pd{}{t}\right)\hat \sigma_\ab = 2\hat\eta\,  \hat v_\ab -\hat \zeta\, Q_\ab+ \hat\xi^\sigma_\ab.
    \label{fd-time-2}
\end{align}
}
The total stress tensor $\hat\sigma^\text{tot}$ includes the deviatoric stress $\hat\sigma$ and the pressure $\hat p$. 
The strain rate tensor is defined as $\hat v_\ab = \frac{1}{2}(\partial_\alpha \hat v_\beta + \partial_\beta \hat v_\alpha)$, where $\hat v_\alpha$ is the velocity field.
As before, the symbol `$\hat~$' denotes fluctuating quantities.

\Cref{fd-time-2} has three main differences with respect to \cref{sigma-fluct}:
\begin{itemize}
\setlength{\leftskip}{12pt} 

\item[(i)] The flow-alignment coupling term is absent since we took the nematic order to be constant, i.e., $\dot\Q=0$.

\item[(ii)] Rather than taking an imposed shear $v_{\alpha\beta}$, the velocity field $\hat \vel$ is now a dynamical variable and thus exhibits fluctuations. This means that, unlike in \cref{sigma-fluct}, the strain rate tensor $\hat v_\ab$ is also fluctuating. Whereas the 
driven and cross contributions to the noise ($\Lambda^{\text{driven}}$ and $\Lambda^{\text{cross}}$ in \cref{Sigma-driven,Sigma-cross}) 
depend on $v_\ab$, they are of higher order in the strain rate than the thermal and active contributions ($\Lambda^{\text{thermal}}$ and $\Lambda^{\text{active}}$),
as argued in \cref{sec-noise-amplitude}. Thus, we neglect $\Lambda^{\text{driven}}$ and $\Lambda^{\text{cross}}$ by assuming weak driving, and we focus on the effects of activity.

\item[(iii)] \cref{fd-time-2} includes a partial time derivative because the equations are formulated for space-dependent fields $\sigma_\ab(\r,t)$ and $v_\alpha(\r,t)$.  
In contrast, \cref{sigma-fluct,sigma-noise-correlation} were written for a small coarse-graining volume $V$. Here, we decompose the space into such small volumes and assign stress, strain rate, noise, etc., to each volume element.
Additionally, we assume that the noise is uncorrelated across different coarse-graining volumes. Hence, for sufficiently small $V$, noise is delta-correlated in space:
\begin{align}
    \langle \hat\xi^\sigma_\ab (\r, t)\hat\xi^\sigma_\mn (\r',t')\rangle = \Lambda^\sigma_\abmn \delta(\r-\r')\delta(t-t').
\end{align}

\end{itemize}
  

We now consider a spherical tracer particle of radius $a$ immersed in the active gel, and we calculate the tracer's positional fluctuations following Refs. \cite{bedeaux1974,bedeaux1977}.
To this end, we impose no-slip boundary conditions for the particle, meaning the fluid velocity at the particle's surface matches the particle's velocity $U_\alpha$:  
\begin{align}\label{bc}
    \hat v_\alpha(\r, t) = \hat U_\alpha(t), \quad  \r \in \hat S(t)
\end{align} 
where $\hat S(t)$ is the surface of the tracer particle defined by points satisfying $|\r - \hat{\bR}(t)| = a$.

To obtain the response function, we neglect inertia of the particle and consider the balance of an external force on the particle with the drag force exerted by the gel:
\begin{align}\label{tracer-balance}
    &0=\hat F^\text{gel}_\alpha(t) + \hat F^\text{ext}_\alpha(t) =  \oint\limits_{\hat S(t)}
    \hat \sigma^\text{tot}_\ab \,\d S_\beta
    + \hat F^\text{ext}_\alpha(t).
\end{align}
To evaluate the drag force, we solve \cref{fd-time,bc} and
obtain a generalisation of the Stokes law for a viscoelastic medium (in \cref{app-response-function}, we linearize the boundary condition \eqref{bc} assuming small tracer displacements and solve with \cref{fd-time}):
\begin{align}\label{stokes-time}
    F^\text{gel}_\alpha(t) = -6\pi a\eta \int_{0}^{t} \ku \ee^{-\ku (t-t')}  U_\alpha(t') \d t',
\end{align}
where $\eta=\la \hat \eta\ra_\ss$ is the steady-state viscosity of the gel. 
Transforming this result to Fourier space, we obtain the response function as defined in \cref{chi-def} (see \cref{app-response-function} for details):
\begin{align}
    \label{chi-final}
    \chi^R_\ab(\omega) = \chi^R(\omega)\delta_\ab,
    \quad
    \chi^R(\omega) = \frac{(\omega/\ku) - \ii}{6\pi a \eta \omega}.
\end{align}
 
We now compute the position fluctuation spectrum $S^{R}_\ab$ of the free tracer particle. This means $\hat F^\text{ext}=0$ and the tracer undergoes Brownian motion. The fluctuation spectrum can be written as (see \cref{app-position-fluct})
\begin{align} 
    \label{spectrum-reduction}
    \rS^R_{\theta\phi}(\omega) \!=\! \tfrac{1+\omega^2/\ku^2}{(6\pi a \eta \omega)^2}\!\! \!\int\limits_{|\r|>a}\!\!\!\!
     \d^3 \r\,
    P_{\alpha \theta,\beta}(\r)P_{\mu \phi,\nu}(\r)\,
    \rS^\text{partial}_{\ab\mn}(\omega).
\end{align}
Here, $P_{\alpha\gamma,\beta}(\r) = \frac{\partial}{\partial r_\beta }P_{\alpha\gamma}(\r)$, and
\begin{equation}
    P_{\alpha\gamma}(\r)=\frac{3 a}{4 r}\left(1+\frac{a^2}{3r^2} \right) \delta_{\alpha \gamma}
    +\frac{3 a}{4 r^3}\left(1-\frac{a^2}{r^2}\right)\,r_\alpha r_\gamma
    \label{P-def}
\end{equation}
is a generalization of the Oseen tensor to describe the flow field around a spherical particle of size $a$ as \cite{landau-fluid,kim2013microhydrodynamics} 
\begin{align}
    v_\alpha(\r-\bR, t) = P_{\alpha\gamma}(\r-\bR)\;U_\gamma(t).
    \nonumber
\end{align}
Respectively, $\rS^\text{partial}_{\ab\mn}(\omega)$ is the partial stress spectrum, defined as the stress fluctuation spectrum at fixed strain rate $v_\ab=0$, so that it only captures the fluctuations arising from the noise $\hat\xi^\sigma_\ab$ and from the active stress coefficient $\hat\zeta$.
Recall that the full stress fluctuation spectrum is given by \cref{SY} and corresponds to the $d^2 \times d^2$ upper-left block of $\rS_\Y(\omega)$.


\begin{figure}[t]
    \centering
    \includegraphics[width=\linewidth]{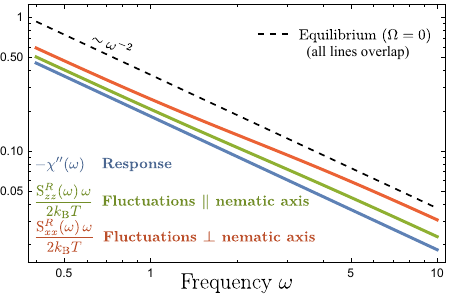}
    \caption{Visualisation of the response function $\chi(\omega)$ from \cref{chi-final} and the fluctuation spectrum $\rS^R(\omega)$ from \cref{SR-final}. The latter is scaled using the FDT \eqref{FDT} to highlight the departure from equilibrium. The parameters are $\ku=\kBT=\mu=\chi=\Omega_0=a=1$, $D=0.8$, $\rho=A=0.15$ and $|\Q|=1$ with $z$-axis of nematic order.
    The dashed line indicates the equilibrium case $\Omega=0$, when all three curves coincide. Note that the dashed line is above the response function (blue solid line), because the viscosity $\eta$ also depends on activity.}
    \label{fig-microrheology}
\end{figure}

To evaluate the integral in \cref{spectrum-reduction}, we decompose the partial stress spectrum into invariant tensors (c.f.~\cref{sec-reduced-repr} and \cref{Tm-def}):
\begin{align} 
\label{decompose}
    \rS^\text{partial}_{\ab\mn}(\omega) = \sum_{i=0,1,2} S^{(i)}(\omega) \T^{(i)}_\abmn,
\end{align}
with scalar functions $S^{(i)}(\omega)$ as coefficients.

Evaluating the integral \eqref{spectrum-reduction} for each $\T^{(i)}$ tensor, which form the basis of decomposition, we obtain (see \cref{app-position-fluct}): 
\begin{multline}
     \rS^{R}_{\theta\phi}(\omega) = \frac{1+\omega^2/\ku^2}{(6\pi a \eta \,\omega)^2} 
     \pi a
     \left[
     \left(
     3 S^{(0)}(\omega)+\tfrac{2}{5}S^{(2)}(\omega) \right)\delta_{\theta\phi}\right.
     \\
     \left.+\left(
     \tfrac{3}{4} S^{(1)}(\omega)+\tfrac{3}{7}S^{(2)}(\omega) \right) N_{\theta\phi}\right],
     \nonumber
\end{multline}
where we recall that the orientation tensor $N_{\alpha\beta}$ is defined by $Q_{\alpha\beta} = |\Q| N_{\alpha\beta}$, with $|\Q|$ being the strength of the nematic order. 
Finally, comparing 
\cref{decompose} with our previous result for the partial stress spectrum
$\rS_{\ab\mn}(\omega)$, we get the analytical from of the $S^{(i)}$ coefficients (see \cref{app-tracer-special}):
\begin{multline}
    \label{SR-final}
    \textstyle
\rS^{R}_{\theta\phi}(\omega) =\frac{2\kBT}{\omega} \frac{1}{6\pi a \eta \,\omega}
     \left[
        \left(1+\frac{\Omega_0}{\Z}\left(\frac{A \mu}{\rho\kBT}-1\right)\right)\delta_{\theta\phi}+\right.
        \\
        \textstyle
        \left.
            +\frac{D}{\rho\kBT}\frac{D}{\mu} \frac{\Omega_0}{\Z} |\Q|^2
        \left(\!1\!-\!\frac{\Omega_0}{\Z}\frac{(1+\Z)^2-1}{(1+\Z)^2+\omega^2/\ku^2}\!\right)\left(\tfrac{2}{15}\delta_{\theta\phi} \!+\! \tfrac{1}{7} N_{\theta\phi}\right)
    \right].
\end{multline}

When detailed balance is satisfied, with $\Omega=0$ in \cref{kakd}, we have $\Omega_0=0$, and \cref{SR-final} reduces to the \cref{FDT}, thus retrieving the fluctuation-dissipation theorem. When $\Omega_0\neq 0$, \cref{SR-final} provides the contributions of activity to the fluctuation spectrum, which thus departs from equilibrium. Thus, \cref{SR-final} provides a fluctuation-dissipation-activity relation for a tracer particle. This relation generalizes the fluctuation-dissipation theorem to include active fluctuations that emerge from the breaking of detailed balance ($\Omega_0\neq 0$). In addition, in the presence of activity, the spectrum also becomes anisotropic due to the nematic order $\Q$. On \cref{fig-microrheology}, we visualize the spectrum \eqref{SR-final} and response function \eqref{chi-final}; the difference between fluctuations and response quantifies the departure from equilibrium, and the difference between parallel and perpendicular fluctuations showcases the anisotropy of the fluctuations.



\section{Discussion and conclusion}
\label{sec-discussion}

In summary, we derived the properties of nonequilibrium fluctuations of an active gel, which we modeled as a network of elastic elements connected by transient crosslinks. Our results show explicitly how active noise emerges from nonequilibrium, irreversible processes at the molecular scale. In particular, we obtained the fluctuation spectra of both the stress tensor of the gel and the position of a tracer particle. The results show how the breaking of detailed balance for the crosslinker binding kinetics leads to departures from the fluctuation-dissipation theorem (FDT). Thus, our work extends the FDT to active systems by providing a fluctuation-dissipation-activity relation, where the activity is given in terms of the function $\Omega$ that characterizes the breaking of detailed balance.

Our model predicts the active fluctuation spectrum for a tracer particle, which can be measured in microrheology experiments performed either in reconstituted cytoskeletal networks or in living cells \cite{gnesotto2018,lau2003,mizuno2007,toyota2011,wilhelm2008,guo2014,fodor2015,fodor2016,nishizawa2017,ahmed2018,bacanu2023,umeda2023,muenker2024elife,muenker2024natmat}. We predict that active fluctuations are anisotropic, which could be directly tested in such experiments. Regarding the frequency dependence of the fluctuations, our model is too simple for a direct comparison to experiments. To attempt a quantitative match, our simple model could be generalized to account for (i) active forces exerted while a motor is bound \cite{mackintosh2008,levine2009,fodor2015,fodor2016,ahmed2018,mura2019,umeda2023}, (ii) the complex power-law rheology of the cytoskeleton \cite{lau2003,guo2014,ahmed2018,hurst2021,ebata2023,abbasi2023non,muenker2024natmat}, and (iii) consider diffusion of the crosslinkers \cite{mulla2022}. 
Whereas here we predicted active white noise in the stress tensor, these generalizations could enable the emergence of colored noise.

Looking forward, our results could be potentially applied to cytoskeletal networks such as the mitotic spindle \cite{brugues2014physical,smith2015} and actomyosin \cite{mizuno2007,toyota2011}, to the cell interior \cite{lau2003,guo2014,fodor2015,nishizawa2017,ahmed2018,hurst2021,bacanu2023,muenker2024elife,muenker2024natmat,narinder2026}, or even to DNA gels \cite{bertrand2012} and the contents of the cell nucleus \cite{weber2012,smith2015,chu2017,das2024}. In addition, our picture of nonequilibrium crosslinked networks also applies at cellular interfaces \cite{oriola2017,janes2022}, where molecular adhesions based on E-cadherin and integrin bind cells to other cells and to the extracellular matrix, respectively. Thus, our theoretical framework could be used to study fluctuations in cellular traction forces \cite{plotnikov2012,hennig2020,moriel2023}. 

From a theoretical perspective, our approach to active fluctuations complements works that characterize the properties and consequences of correlated noise arising from active baths such as bacterial suspensions \cite{kanazawa2020,maes2020,seyforth2022,massana-cid2024,paoluzzi2024,majhi2025,vilallobos-concha2025,frechette2025}. It also contributes to the ongoing effort to extend fluctuation-dissipation relations and thermodynamic inequalities to nonequilibrium states \cite{fodor2016how,nardini2017,dalcengio2019,horowitz2020,dalcengio2021,caprini2021,seara2021,han2021,solon2022,sorkin2024,johnsrud2025,kirkpatrick2025}. In particular, it would be interesting to relate the mechanical fluctuations studied here with the associated heat fluctuations arising from the chemical reactions that power active molecular processes \cite{mabillard2023}. Overall, our results pave the way towards a stochastic hydrodynamics of active gels \cite{basu2008,Basu2012}.

\section*{Acknowledgments}

We thank Jaume Casademunt, David Oriola, Ahandeep Manna and Sagnik Garai for discussions, and Joseph Hutchinson for related Gillespie simulations of our model. R.A. acknowledges funding from the European Union through the ERC Starting Grant “Living\_Fluctuations” (No. 101114584).

{
\appendix
\renewcommand{\thesubsection}{\thesection.\arabic{subsection}}

\allowdisplaybreaks
\newenvironment{nodisplaybreaks}
  {\begingroup\allowdisplaybreaks[0]}
  {\endgroup}

\section{Definition and evaluation of integrals involving tensors}

\label{app-integral}
In the main text, we encounter integrals over all possible binding states, characterized by traceless and symmetric tensors $\u$ and $\q$. In $d$ dimensions, each tensor has $d^2$ entries, but the constraints of tracelessness and symmetry reduce the number of independent entries to $d(d+1)/2 - 1$. Therefore, the space of all possible binding states has a dimension of $d(d+1) - 2$.

Working in Cartesian space simplifies the interpretation of the integral over all $2d^2$ entries of the tensors $\u$ and $\q$. To this end, we define a density of states $g(\u, \q)$, which is non-zero only when $\u$ and $\q$ are symmetric and traceless. Specifically, we choose:
\begin{multline}
    \textstyle
    g(\u,\q) =
    \delta\left(\sum_{i=1}^{d} u_{ii}\right) \delta\left(\sum_{i=1}^{d} q_{ii}\right)
    \\
    \textstyle
    \times \prod_{1\le i<j \le d} \delta(u_{ij} -u_{ji})
    \delta(q_{ij} -q_{ji}).
    \label{density-of-states}
\end{multline}
The first two delta functions ensure that $\u$ and $\q$ are traceless, while the remaining delta functions enforce their symmetry.

To ensure that binding events occur only in valid states, we modify the binding rate in \cref{kakd-new} as follows:
\begin{align}
    \kb(\u,\q) =  \ku\;g(\u,\q)
    \left[ \ee^{ \frac{\beta}{\rho} f(\u,\q)} + \Omega(\u,\q) \right].
\end{align}
This modification guarantees that linkers are bound exclusively to valid states at any given time.

\subsection{Evaluation of the thermal component of the duty ratio}
In this section, we evaluate the thermal component of the duty ratio, denoted by:
\begin{align} \label{Zth-1}
    \Z_\text{th} = \int g(\u,\q) \,\ee^{-\frac{\beta}{\rho} f(\u,\q)} \,\d \u \d \q,
\end{align}
where the free energy density $f(\u,\q)$ is defined in \cref{free-energy}.

This integral represents the ratio of the binding rate to any bound state, $\int \kb(\u,\q) g(\u,\q) \d\u\, \d\q$, to the unbinding rate $\ku$ in a passive system, where $\Omega \equiv 0$.

Using the rotational symmetry of the free energy, the integral \cref{Zth-1} can be factorized as follows:
\begin{equation}
    \Z_\text{th}= \;\mathcal{I}_1^{\;d(d-1)/2}\;\mathcal{I}_2,    
    \label{Zth-factors}
\end{equation}
where
\begin{multline}
    \I_1=
    \int   \d u_{12}\d u_{21}\d q_{12}\d q_{21}
    \;\delta(u_{12}\!-\!u_{21})\delta(q_{12}\!-\!q_{21})
    \\ 
    \times 
    \ee^{-\frac{\beta}{\rho}\left(\frac{\mu}{2}(u_{12}^2+ u_{21}^2) +D (u_{12}  q_{12}+u_{21}  q_{21})  + \frac{\chi}{2}(q_{12}^2+ q_{21}^2) \right)},
    \label{I1-def}
\end{multline}
\begin{multline}
    \I_2=\int  \prod_{i=1}^{d} \d u_{ii}\d q_{ii}\;
    \delta\Big(\sum_{i=1}^{d} u_{ii}\Big)\delta\Big(\sum_{i=1}^{d} q_{ii}\Big)
    \\
    \times
    \ee^{-\frac{\beta}{\rho}\left(\frac{\mu}{2}\left(\sum_{i=1}^{d} u_{ii}^2\right) +D \left(\sum_{i=1}^{d} u_{ii}q_{ii}\right)  + \frac{\chi}{2}\left(\sum_{i=1}^{d} q_{ii}^2\right) \right)}.
    \label{I2-def}
\end{multline}
Here, $\I_1$ is the contribution from off-diagonal components, while $\I_2$ is the contribution from the diagonal.
We will evaluate each of these integrals separately. For reference, consider the following simple Gaussian integral:
\begin{equation}\label{gaussian-simple}
\int_{-\infty}^{\infty} \ee^{-\frac{\beta}{\rho}\left(\frac{\mu}{2}u^2 +D u  q  + \frac{\chi}{2}q^2 \right)} \d u\,\d q = \frac{2\pi\rho}{\beta\sqrt{\mu\chi-D^2}}.
\end{equation}

For the first integral $\I_1$ in \cref{I1-def}, after integrating over $u_{21}$ and $q_{21}$, we obtain:
\begin{align}
    \I_1\!=&\!\!
    \int \textstyle  \!\!\d u_{12}\d q_{12}
    \exp\left[{-\frac{2\beta}{\rho}\left(\frac{\mu}{2}u_{12}^2\! +\!D u_{12}  q_{12}\! +\! \frac{\chi}{2}q_{12}^2 \right)}\right]
    \nonumber
    \\
    =&
    \frac{\pi\rho}{\beta\sqrt{\mu\chi-D^2}} \label{I1-res},
\end{align}
which is exactly the Gaussian integral \eqref{gaussian-simple} with $\beta$ replaced by $2\beta$.

We evaluate the second integral $\I_2$ in \cref{I2-def} by applying a change of variables through a rotational transformation $R$ to the diagonal components. Specifically, we transform the variables as follows:
\begin{equation}
    \label{diag-transform}
    \begin{aligned}
        (u_{11}, u_{22}, \dots, u_{dd}) &\rightarrow (u_{(1)}, u_{(2)}, \dots, u_{(d)}),
        \\
        (q_{11}, q_{22}, \dots, q_{dd}) &\rightarrow (q_{(1)}, q_{(2)}, \dots, q_{(d)}).
    \end{aligned}
\end{equation}
The transformation $R$ is chosen such that:
\begin{equation}
    \label{u1-q1}
\begin{aligned}
    u_{(1)} &= \tfrac{1}{\sqrt{d}} \left( u_{11} + u_{22} + \dots + u_{dd} \right),
    \\
    q_{(1)} &= \tfrac{1}{\sqrt{d}} \left( q_{11} + q_{22} + \dots + q_{dd} \right).
\end{aligned}
\end{equation}
This implies that the new basis vector corresponding to $u_{(1)}$ (and $q_{(1)}$) is $\frac{1}{\sqrt{d}}(1, 1, \dots, 1)$. The choice of the remaining basis vectors is not unique, but the final result is independent of this choice. Using \cref{u1-q1}, the delta functions can be expressed as:
\begin{align}
    \delta\Big(\sum_{i=1}^{d} u_{ii}\Big)\delta\Big(\sum_{i=1}^{d} q_{ii}\Big) = \frac{1}{d} \,\delta(u_{(1)}) \delta(q_{(1)}).
\end{align}

The rotational transformation \cref{diag-transform} preserves the scalar product: in the exponent of \cref{I2-def}, we only have the scalar products $(u \cdot u)$, $(u \cdot q)$, and $(q \cdot q)$, where $u = (u_{11}, u_{22}, \dots, u_{dd})$ and $q = (q_{11}, q_{22}, \dots, q_{dd})$ are $d$-dimensional vectors: 
\begin{equation}
\begin{gathered}
    \sum_{i=1}^{d} u_{ii}^2 =     \sum_{i=1}^{d} u_{(i)}^2, 
    \quad     
    \sum_{i=1}^{d} q_{ii}^2 = \sum_{i=1}^{d} q_{(i)}^2,
    \\
    \sum_{i=1}^{d} u_{ii}q_{ii} =     \sum_{i=1}^{d} u_{(i)}q_{(i)}.
\end{gathered}
\end{equation} 
Rewriting the integral $\I_2$ in the new coordinates, we get:
\begin{align}
    \I_2=&\int \prod_{i=1}^{d} \d u_{(i)}\d q_{(i)}\;
    \frac{1}{d}\;\delta\left(u_{(1)}\right)\delta\left(q_{(1)}\right)
    \nonumber
    \\&\qquad\times 
    \exp\left[{-\frac{\beta}{\rho}\sum_{i=1}^{d} \left(\frac{\mu}{2}u_{(i)}^2 +D u_{(i)}q_{(i)}  + \frac{\chi}{2}q_{(i)}^2 \right)}\right]
    \nonumber
    \\  
    =& \frac{1}{d} \int \prod_{i=2}^{d} \d u_{(i)}\d q_{(i)}\;
    \nonumber
    \\ &\qquad\times
     \exp\left[{-\frac{\beta}{\rho}\sum_{i=2}^{d} \left(\frac{\mu}{2}u_{(i)}^2 +D u_{(i)}q_{(i)}  + \frac{\chi}{2}q_{(i)}^2 \right)}\right]\nonumber
     \\
     =& \frac{1}{d} \left[\int \d u\,\d q  \;\ee^{-\frac{\beta}{\rho}\left(\frac{\mu}{2}u^2 +D u  q  + \frac{\chi}{2}q^2 \right)}\right]^{d-1}
     \nonumber
     \\
     = &\frac{1}{d} \left[\frac{2\pi\rho}{\beta\sqrt{\mu\chi-D^2}} \right]^{d-1},
     \label{I2-res} 
\end{align}
where in the last line, we used the Gaussian integral \eqref{gaussian-simple}.

By substituting \cref{I1-res,I2-res} into \cref{Zth-factors}, we derive:
\begin{align}
    \Z_\text{th} =\frac{1}{d\, 2^{d(d-1)/2}}
    \left[\frac{2\pi\rho}{\beta\sqrt{\mu\chi-D^2}} \right]^{(d-1)(d+2)/2}.
    \label{Zth-res}
\end{align}

\subsection{Evaluation of higher moments }
In this section, we evaluate the following integrals when $\Omega \equiv 0$ (note that $\beta$ denotes the inverse temperature and $\mu$ is a free energy parameter, not to be confused with tensor indices):
\subeqn{\label{Z-moments}
\begin{align}
    \int u_\ab u_\mn \tfrac{\kb(\u,\q)}{\ku}\,\d \u\d\q =& \tfrac{\rho}{\beta}\tfrac{\chi}{\mu\chi-D^2} \Z_\text{th} \;\Tiso_\abmn,
    \label{Z-moments-1}
    \\
    \int q_\ab q_\mn \tfrac{\kb(\u,\q)}{\ku}\,\d \u\d\q =& \tfrac{\rho}{\beta}\tfrac{\mu}{\mu\chi-D^2} \Z_\text{th} \;\Tiso_\abmn,
    \\
    \int u_\ab q_\mn \tfrac{\kb(\u,\q)}{\ku}\,\d \u\d\q =& -\tfrac{\rho}{\beta}\tfrac{D}{\mu\chi-D^2} \Z_\text{th} \;\Tiso_\abmn.
\end{align}
}

Here, $\Z_\text{th}$ is defined in \cref{Zth-res}. The tensor $\Tiso$ is a rank-4 tensor that is (i) invariant under rotations, (ii) traceless with respect to the first and last pairs of indices, and (iii) symmetric with respect to the first and last pairs of indices. The unique tensor (up to a constant factor) that meets these criteria is:
\begin{align}\label{Tiso-in-eval}
    \Tiso_\abmn = \frac{1}{2}\left(
    \delta_{\alpha\mu}\delta_{\beta\nu}+\delta_{\alpha\nu}\delta_{\beta\mu}
    \right)
    -\frac{1}{d} \delta_\ab \delta_{\mu\nu}.
\end{align}

Since all integrals in \cref{Z-moments} are proportional to $\Tiso$ due to their common symmetries, evaluating one component of the tensor suffices to determine the proportionality constant. For example, consider the proportionality constant $A = \frac{\rho}{\beta} \frac{\chi}{\mu \chi - D^2} \Z_\text{th}$ in \cref{Z-moments-1}:
\begin{multline}
    A(\Tiso_{1212}+\Tiso_{2121}) = \int (u_{12}^2+ u_{21}^2) \frac{\kb(\u,\q)}{\ku}\,\d\u\, \d\q 
    \\
    = \Z_\text{th} \frac{1}{\I_1} \left( -\frac{2\rho}{\beta}\frac{\d}{\d \mu }\I_1 \right)
    = \Z_\text{th} \frac{\rho}{\beta} \frac{\chi}{\mu \chi - D^2}.
\end{multline}
Here, note that the term $(u_{12}^2 + u_{21}^2)$ appears only in the integral $\I_1$ in \cref{Zth-factors}; see the definition of $\I_1$ in \cref{I1-def}. Given that $\Tiso_{1212} = \Tiso_{2121} = \frac{1}{2}$, the proportionality constant $A = \Z_\text{th} \frac{\rho}{\beta} \frac{\chi}{\mu \chi - D^2}$, as stated in \cref{Z-moments-1}. The remaining integrals in \cref{Z-moments} are evaluated similarly.

\section{Derivation of the constitutive relations}
\label{app-oriola-derivation}

In this appendix, we derive the constitutive relation \cref{sigma-average} as presented in Ref.~\cite{oriola2017}. To obtain an equation for the average stress, we multiply \cref{n-eqn} by $\pd{f(\u,\q)}{u_\mn}$ and integrate over all bound states:

\begin{multline}
    \pd{}{t}\int \pd{f}{u_\mn} n\,\d\u\d\q  +v_\ab \int \pd{n}{u_\ab} \pd{f}{u_\mn} \,\d\u\d\q
    \\
    + \dot{Q}_\ab \int \pd{n}{q_\ab} \pd{f}{u_\mn} \,\d\u\d\q = 
    \\
    =(1-\phi)\int \kb \pd{f}{u_\mn} \,\d\u\d\q - \ku \int n \pd{f}{u_\mn}\,\d\u\d\q.
\end{multline}
Note that $\ku$ is a constant, and $\kb(\u,\q)$ depends on the binding site. Replacing the leftmost and rightmost terms by the average stress tensor defined in \eqref{average-stress-def-0}, and applying integration by parts to the second and third terms on the left-hand side, we obtain:
\begin{multline}
    \pd{}{t}\sigma_\mn - v_\ab \int n  \frac{\partial^2 f}{\partial u_\mn \partial u_\ab} \,\d\u\d\q
    \\
    - \dot{Q}_\ab \int n  \frac{\partial^2 f}{\partial u_\mn \partial q_\ab} \,\d\u\d\q = 
    \\
    =(1-\phi)\int \kb \pd{f}{u_\mn} \,\d\u\d\q - \ku \sigma_\mn.
    \label{agrav}
\end{multline}
Here, the boundary terms vanish because density $n$ is zero at $u_\ab \to \infty$ and $q_\ab \to \infty$.
Given that the free energy is a quadratic function of $\u$ and $\q$ (c.f.~\cref{free-energy}), we have:
\begin{equation}
\begin{aligned}
\frac{\partial f(\u,\q)}{\partial u_\mn} &= \mu u_\mn + D q_\mn,
\\
\frac{\partial^2 f(\u,\q)}{\partial u_\mn \partial u_\ab} &= \mu \delta_{\alpha\mu}\delta_{\beta\nu},
\qquad
\frac{\partial^2 f(\u,\q)}{\partial u_\mn \partial q_\ab} = D \delta_{\alpha\mu}\delta_{\beta\nu}.
\end{aligned}
\end{equation}
Thus, the first two integrals in \cref{agrav} can be evaluated directly. For the third integral, we have:
\begin{nodisplaybreaks}
\begin{multline}
    \int \kb(\u,\q) (\mu u_\mn + D q_\mn) \d\u\d\q = 
    \\
    =\ku \mu \int \ee^{-\beta f(\u,\q)/\rho} u_\mn \d\u\d\q
    \\
    ~~+
    \ku D \int \ee^{-\beta f(\u,\q)/\rho} q_\mn \d\u\d\q
    \\
    +
    \int \ku \Omega(\u,\q) (\mu u_\mn + D q_\mn) \d\u\d\q.
\end{multline}
\end{nodisplaybreaks}
The first two integrals on the right-hand side vanish, because the integrands are odd functions of $\u$ and $\q$ respectively. For the third integral, using the definitions in \cref{omega-uq-def}, we get:
\begin{align}
    \int \kb(\mu u_\mn + D q_\mn) \d\u\d\q = \ku \mu \Omega_u Q_\mn + \ku D \Omega_q Q_\mn.
    \nonumber
\end{align}
Finally, \cref{agrav} becomes:
\begin{multline}
    \pd{}{t}\sigma_\mn - v_\mn \phi \mu 
    - \dot{Q}_\ab \phi D= 
    \\
    =(1-\phi)\ku\left(\mu \Omega_u + D \Omega_q \right) Q_\mn- \ku \sigma_\mn,
\end{multline}
which is exactly \cref{sigma-average}. 

In Ref.~\cite{oriola2017}, a constitutive relation for the nematic field strength is derived using a similar approach. By multiplying \cref{n-eqn} by $\pd{f(\u,\q)}{q_\mn}$ and integrating over all bound states, we obtain:

\begin{align}
        \left(1+\tau \frac{d}{d t}\right) H_\ab=& \gamma \,\dot{Q}_\ab-\nu \,v_\ab
        +\omega Q_\ab,
    \end{align}
where $\gamma = \chi \phi / \ku$, $\omega = (1-\phi)(D\Omega_u + \chi\Omega_q)$, and the remaining coefficients are defined in \cref{coupling}.

\section{Langevin equation with colored noise}
\label{app-colored-noise}
The condition \cref{Y-important}, which states that the conditional average $\Exp{\hatY_t \mid \hatY_0}$ satisfies the linear constitutive relation \cref{Y-average}, is crucial for the white noise spectrum. In this appendix, we show that this condition does not hold when the system is subject to colored noise:
\begin{align}\label{coloredlang}
    \frac{\d}{\d t}\hat Y &= -\hat Y + \hat\eta(t),
\end{align}
where $\hat\eta(t)$ is a colored, i.e. $\la \hat\eta(t) \hat\eta(t')\ra \ne 0$.

Averaging over the noise results in a linear response:
\begin{align}
    \frac{\d}{\d t}\langle \hat Y \rangle  &= -\langle \hat Y \rangle   .
\end{align}
However, this linear relationship does not extend to conditional averages. Consider the formal solution of \cref{coloredlang}:
\begin{align}
    \hat Y(t) = \ee^{-t} Y_0 + \int_{0}^t \ee^{-(t-t')}\hat\eta(t') \d t'.
\end{align}
Taking the conditional average, we obtain:
\begin{align}
    \Exp{ \hat Y(t) \,|\, Y_0} = \ee^{-t}  Y_0 + 
    \int_{0}^t \ee^{-(t-t')}\Exp{\hat\eta(t') \,|\,  Y_0}.
\end{align}
For colored noise, $\Exp{\hat\eta(t') \,|\, Y_0}$ is non-zero, and therefore the simple form $Y_t = \ee^{-t} Y_0$ does not hold.

In \cref{fig-colored-noise}, we illustrate the concept of the conditional average and show that \cref{Y-important} does not apply to \cref{coloredlang}.

\begin{figure}
\centering
    \includegraphics[width=0.5\linewidth]{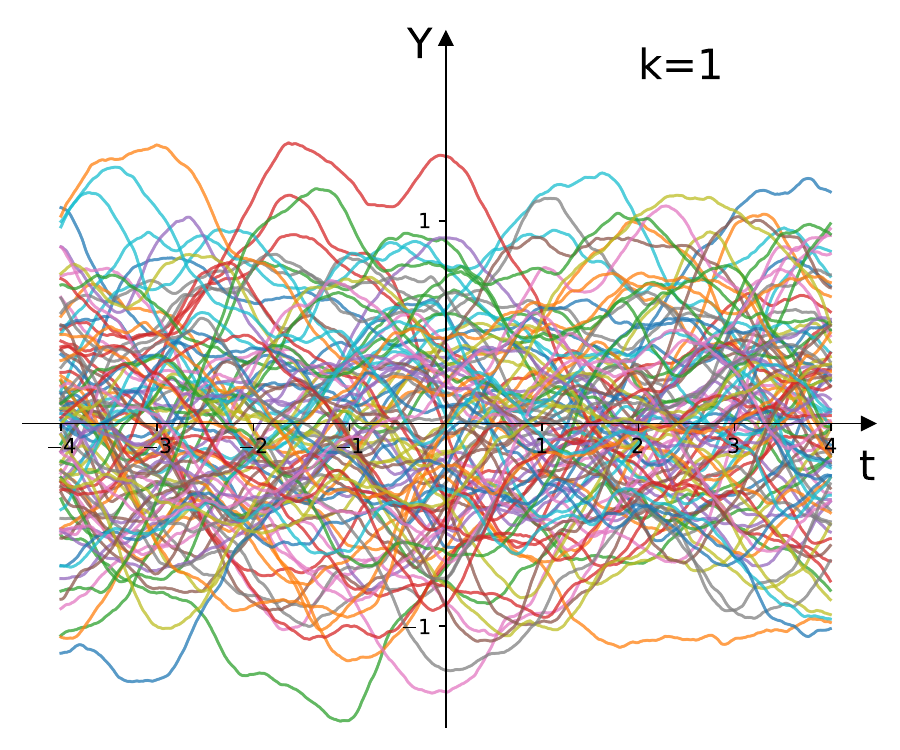}%
    \includegraphics[width=0.5\linewidth]{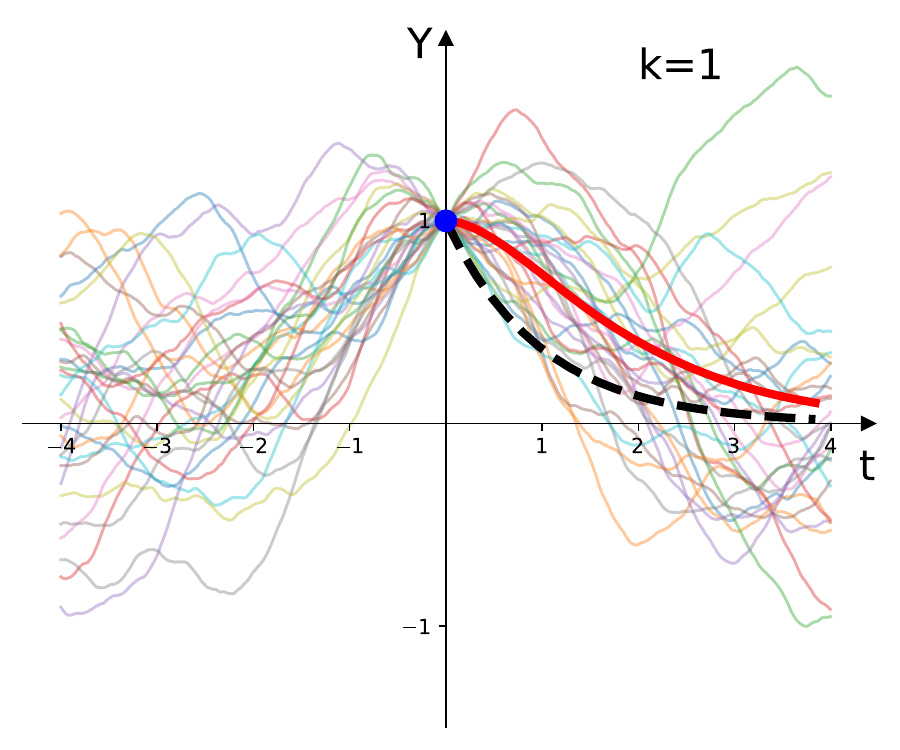}
    \caption{This figure depicts the Ornstein-Uhlenbeck colored noise characterized by $\la \hat\eta(t) \hat\eta(t')\ra = \frac{k}{2} \ee^{-k |t-t'|}$. The left panel shows steady-state trajectories of $Y(t)$, while the right panel presents trajectories conditioned on $Y_0=1$. The red solid line in the right panel represents the conditional average of these trajectories at later times, and the black dashed line indicates the trajectory of the average: $Y_t = \ee^{-t}$.}
    \label{fig-colored-noise}
\end{figure}

\section{Steady-state distribution}
\label{app-ss}
\subsection{Derivation: method of integrating factors}
In this appendix, we solve the equation derived from \cref{n-eqn} by setting $\pd{n}{t}=0$ and $\phi = \frac{\Z}{1+\Z}$:
\begin{align}
        \left(\ku 
    +v_\ab \frac{\partial}{\partial u_\ab}
    +\dot{Q}_\ab \frac{\partial }{\partial q_\ab}
    \right) n_\ss(\u,\q)
    =\frac{\kb(\u,\q)}{1+\Z}.
    \nonumber
\end{align}
To simplify the notation, we introduce generalized variables $\ix = (\u, \q)$ and $\J = (\v, \dot{\Q})$ (1d vectors containing the components of the tensors), transforming the equation into:
\begin{align}
    (\ku + \J \cdot \nabla)n_\ss(\ix) = \frac{\kb(\ix)}{1+\Z}.
\end{align}
Multiplying both sides by the integrating factor $\mu(\ix)$, we obtain:
\begin{align}\label{ruto}
    \ku \mu(\ix) n(\ix) + (\mu(\ix)\J)\cdot\nabla n(\ix) = \frac{\kb(\ix)}{1+\Z} \mu(\ix).
\end{align}
If $\mu(\ix)$ satisfies
\begin{align}\label{ptri}
    \J \cdot \nabla \mu(\ix) = \ku \mu(\ix),
\end{align}
\cref{ruto} simplifies to:
\begin{align} 
\J \cdot\nabla \left( \mu(\ix) n(\ix)\right) = \frac{\kb(\ix)\mu(\ix)}{1+\Z} .
\end{align}
Defining $f(\ix) = \mu(\ix) n(\ix)$ and $g(\ix) = \frac{\kb(\ix) \mu(\ix)}{1+\Z}$, we have:
\begin{align}
    \J \cdot \nabla f(\ix) = g(\ix).
\end{align}
This is a directional derivative. Writing $\ix = \ix_0 + \J \tau$ for some $\ix_0$ and $\tau$, we get:
\begin{align}
    \J \cdot \nabla f(\ix) = \frac{\d}{\d\tau} f(\ix_0 + \J \tau) = g(\ix_0 + \J \tau).
\end{align}
Integrating both sides, we obtain:
\begin{align}
    f(\ix_0) - f(\ix_0-\J \tau_{\rm m})  = \int_0^{\tau_{\rm m}}g(\ix_0 - \J \tau) \d\tau .
\end{align}
Setting $\tau_{\rm m} \rightarrow \infty$ and noting that $f(\infty) = 0$ (since ${n(\ix\to\infty) = 0}$), we have:
\begin{align}
    f(\ix) = \int_0^\infty g(\ix-\J \tau) \d\tau.
\end{align}
In terms of the original functions, this becomes:
\begin{align}
    n(\ix) = \int_0^\infty 
    \frac{\kb(\ix-\J \tau )}{1+\Z} ~
    \frac{\mu(\ix-\J \tau)}{\mu(\ix)} 
   ~ \d\tau.
\end{align}
Choosing $\mu(\ix) = \exp\left( \frac{\ku}{|\J|^2} (\J \cdot \ix) \right)$, which clearly satisfies \cref{ptri}, simplifies the ratio $\frac{\mu(\ix - \J \tau)}{\mu(\ix)}$ to $\ee^{-\ku \tau}$. Thus, we have:
\begin{align}
    \label{n-ss-x}
    n_\ss(\ix) =  \frac{1}{1+\Z}\int_0^\infty\kb(\ix-\J \tau) \;\ee^{-\ku \tau} \;\d\tau.
\end{align}
This corresponds to \cref{ss-10} in the original variables.

\subsection{Probabilistic interpretation of steady-State distribution}
\label{sm-ss-prob}

In \cref{sec-ss}, we argued that the steady-state distribution can be expressed as:
\begin{align}
    n_\ss(\u, \q) = \phi_\ss \la \frac{\kb(\u-\v \tau,\; \q-\dot\Q \tau)}{\ku \Z} \ra_\tau,
    \label{ss-10-intuitive}
\end{align}
where the average is taken over the observed lifetime $\tau$ of the linker. The observed lifetime $\tau$ is the duration for which the linker has been bound, if one just observed that it is currently bound.

We now argue that the observed lifetime $\tau$ follows an exponential distribution with rate $\ku$. When observing a linker, there is a bias towards those that remain bound for longer periods, while linkers that detach quickly are less likely to be observed in the bound state. To derive the distribution of the total lifetime $s$ of an observed linker ($\tau$ plus the time time that the linker remains bound after the observation), denoted as $p(s \,|\, \text{bound})$, consider that if a linker has been bound multiple times with lifetimes $s_1, s_2, s_3, \dots$, the probability that a randomly chosen bound state had lifetime $s_i$ is $({s_i p(s_i)})/({\sum_k s_k p(s_k)})$.
For the continuous case with an exponential lifetime distribution $p(s) = \ku \ee^{-\ku s}$, this becomes:
\begin{align}
    p(s \,|\, \text{bound}) = \frac{s \ee^{-\ku s}}{\int s' \ee^{-\ku s'} \d s'} = \ku^2 \,s\, \ee^{-\ku s}.
\end{align}
The distribution of the observed lifetime $\tau$ is then:
\begin{align}
    p(\tau \,|\, \text{bound}) = \int_0^\infty p(\tau \,|\, s, \text{bound}) \, p(s \,|\, \text{bound}) \d s,
\end{align}
where $p(\tau \,|\, s, \;\text{bound})$ is the conditional distribution of the observed lifetime $\tau$ given that the linker has a total lifetime $s$. This conditional distribution is:
\begin{align}
    p(\tau \,|\, s, \;\text{bound}) = \frac{1}{s} \quad \text{if} \quad \tau < s, \quad \text{otherwise} \quad 0.
\end{align}
Thus, the distribution of the observed lifetime $\tau$ is:
\begin{align}
    p(\tau \,|\, \;\text{bound}) = \int_\tau^\infty \ku^2 \,\ee^{-\ku s} \,\d s = \ku \, \ee^{-\ku \tau}.
\end{align}
Performing the average in \cref{ss-10-intuitive} with respect to this distribution yields the steady-state distribution \eqref{ss-10}.

\subsection{Evaluation of moments}
\label{app-sigma-evaluation}
\subsubsection{Covariance matrix}
In this section, we evaluate the covariance matrix of $\hatY$, denoted as $\Sigma$ in \cref{Ycov}, which includes the variances of the stress $\hat\bsigma$ and $\hat\phi$, as well as their covariance:
\begin{align}
    \label{Sigma-def-sm}
    \Sigma = 
    \begin{pNiceArray}{c|c}
        \la \hat\sigma_\ab \hat\sigma_\mn\ra - \la \hat\sigma_\ab \ra \la  \hat\sigma_\mn\ra  
        & 
        \la \hat\sigma_\ab\smashhat\phi\ra - 
        \la \hat\sigma_\ab \ra\la\smashhat\phi \ra
        \\[1ex]  \hline
        \\[-2ex]
        \la \hat\sigma_\ab\smashhat\phi\ra - 
        \la \hat\sigma_\ab \ra\la\smashhat\phi \ra
        &
        \la \smashhat\phi^2\ra - \la\smashhat\phi \ra ^2
    \end{pNiceArray}.
\end{align}
The components of $\Sigma$ are given by:
\begin{subequations}
\label{sigma-components}
\begin{multline}
    \label{sigma-components-1}
    \la \hat\sigma_\ab \hat\sigma_\mn\ra - \la \hat\sigma_\ab \ra \la  \hat\sigma_\mn\ra =
    \\
    =\frac{1}{N}\left(
    \la \pd{f}{u_\ab}\pd{f}{u_\mn}\ra -
    \la \pd{f}{u_\ab}\ra
    \la \pd{f}{u_\mn}\ra  
    \right),
\end{multline}
\begin{align}
    \label{sigma-components-2}
    \la \hat\sigma_\ab\smashhat\phi\ra - 
        \la \hat\sigma_\ab \ra\la\smashhat\phi \ra
    =& \frac{1}{N}\la \pd{f}{u_\ab}\ra  (1-\bar\phi),
    \\
    \label{sigma-components-3}
    \la \smashhat\phi^2\ra - \la\smashhat\phi \ra ^2 
    =& \frac{1}{N} \bar\phi (1-\bar\phi),
\end{align}
\end{subequations}
where the averages on the right-hand side are taken over the steady-state distribution $n_\ss$ in \cref{ss-10}. Here, $f$ is a function of the random variables $\hat\u$ and $\hat\q$, and we omit its arguments for brevity: $\pd{f}{u_\ab} \equiv \pd{f(\hat\u,\hat\q)}{u_\ab}$. The specific forms of the averages are:
\begin{subequations}
\label{rubo}
\begin{align}
    \label{rubo-1}
    \la\pd{f}{u_\ab}  \ra =& \mu \la u_\ab \ra + D \la q_\ab \ra ,
    \\
     \la\pd{f}{u_\ab} \pd{f}{u_\mn} \ra
    =&
    \mu^2 \la u_\ab u_\mn \ra +
    \nonumber
    \\
    &\hspace{-2cm}+
    D^2 \la q_\ab q_\mn \ra 
    + D \mu \left(
    \la u_\ab q_\mn \ra 
    +\la q_\ab u_\mn \ra \right).
\end{align}
\end{subequations}
We next compute the first and second moments of the distribution $n_\ss(\u,\q)$ in \cref{ss-10}:
\begin{subequations}
    \label{gago}
\begin{align}
    &\la u_\ab \ra = \nonumber
    \\
    &\quad=\textstyle\int \d\u\, \d\q\;u_\ab \frac{1}{1+\Z}\int_0^\infty\d\tau \kb(\u-\v \tau,\; \q-\dot\Q \tau) \;\ee^{-\ku \tau} 
    \nonumber
    \\
    &\quad= \textstyle \frac{1}{1+\Z}\int\limits_0^\infty \d\tau\ee^{-\ku\tau} \int \d\u\, \d\q (u_\ab + v_\ab \tau) \kb(\u,\q)
    \nonumber
    \\
    &\quad=\frac{1}{1+\Z}\left(\Omega^u_\ab + v_\ab \frac{\Z}{\ku}\right),
    \\
    &\la q_\ab \ra = \frac{1}{1+\Z}\left(\Omega^q_\ab + \dot Q_\ab \frac{\Z}{\ku}\right),
\end{align}
\end{subequations}
where $\Omega^u_\ab = \int u_\ab \Omega \,\d\u\, \d\q \equiv \Omega_u Q_\ab$ and $\Omega^q_\ab = \int q_\ab \Omega \,\d\u\, \d\q \equiv \Omega_q Q_\ab$, as defined in \cref{omega-uq-def}. The average stress in the system is then:
\begin{align*}
	\sigma_\ab =& \textstyle\frac{1}{1+\Z}\left(\mu \Omega^u_\ab 
    + D \Omega^q_\ab\right) 
    + \frac{\Z}{1+\Z}\frac{1}{\ku}\left(\mu v_\ab + D \dot Q_\ab\right)
    \\
	\equiv& \zeta Q_\ab + \frac{\phi}{\ku}\left(\mu v_\ab + D \dot Q_\ab\right),
\end{align*}
which is consistent with \cref{sigma-average} in steady-state (with $\frac{\d\bsigma}{\d t}=0$).

The second moments are:

\begin{widetext} 
\begin{subequations}
    \label{rafo}
\begin{align}
    \la u_\ab u_\mn\ra = &
    \int \d\u\, \d\q\;u_\ab u_\mn\; \frac{1}{1+\Z}\int_0^\infty\d\tau \;\kb(\u-\v \tau,\; \q-\dot\Q \tau) \;\ee^{-\ku \tau} 
    \\
    =&\frac{1}{1+\Z}\int \d\u\, \d\q \;\kb(\u,\q) \int_0^\infty \d\tau\;\ee^{-\ku\tau} 
    (u_\ab + v_\ab \tau) (u_\mn + v_\mn \tau) 
    \\
    =&\frac{1}{1+\Z}\int \d\u\, \d\q \;\kb(\u,\q) 
    \frac{1}{\ku}\left(
    u_\ab u_\mn +\frac{1}{\ku}(u_\ab v_\mn + v_\ab u_\mn) + \frac{2}{\ku^2} v_\ab v_\mn 
    \right)
    \\
    =&\frac{1}{1+\Z} \left( 
    \rho\kBT \frac{\chi}{\mu\chi-D^2}(\Z - \Omega_0)\Tiso_\abmn + \Omega^{uu}_\abmn
    +
    \frac{1}{\ku}(\Omega^{u}_\ab v_\mn + v_\ab \Omega^{u}_\mn) + \frac{2\Z}{\ku^2} v_\ab v_\mn 
    \right),
    \\
\la u_\ab q_\mn\ra =&
    \frac{1}{1+\Z} \left( \rho\kBT \frac{-D}{\mu\chi-D^2}(\Z - \Omega_0)\Tiso_\abmn + \Omega^{uq}_\abmn
    +
    \frac{1}{\ku}(\Omega^{u}_\ab \dot Q_\mn + v_\ab \Omega^{q}_\mn) + \frac{2\Z}{\ku^2} v_\ab \dot{Q}_\mn 
    \right),
    \\
\la q_\ab q_\mn\ra=&\frac{1}{1+\Z} \left( 
    \rho\kBT \frac{\mu}{\mu\chi-D^2}(\Z - \Omega_0)\Tiso_\abmn + \Omega^{qq}_\abmn
    +
    \frac{1}{\ku}(\Omega^{q}_\ab \dot{Q}_\mn + \dot{Q}_\ab \Omega^{q}_\mn) + \frac{2\Z}{\ku^2} \dot{Q}_\ab \dot{Q}_\mn 
    \right),
\end{align}
\end{subequations}
where we have used the results \cref{Z-moments} with $\Z_\text{th} = \Z - \Omega_0$. Recall that $\Omega_0 = \int \Omega \,\d\u\, \d\q$ and
\begin{align}
    \begin{pmatrix}
        \Omega^{uu}_{\abmn} & \Omega^{uq}_{\abmn}
        \\
        \Omega^{qu}_{\abmn} & \Omega^{qq}_{\abmn}
    \end{pmatrix}
    =
    \int \d\u\, \d\q \,\Omega \begin{pmatrix}
        u_\ab u_\mn & u_\ab q_\mn
        \\
        q_\ab u_\mn & q_\ab q_\mn
    \end{pmatrix}.
\end{align}

Using these results, we can now compute the components of $\Sigma$ in \cref{Sigma-def-sm}. The expression for block \eqref{sigma-components-3} is already known. By substituting \cref{gago} into \cref{rubo-1} and then into \cref{sigma-components-2}, we obtain:
\begin{align}
    \la \hat\sigma_\ab\smashhat\phi\ra - 
        \la \hat\sigma_\ab \ra\la\smashhat\phi \ra
    ~=&~ 
    \frac{1}{N} \frac{1}{1+\Z} \left( \mu \la u_\ab\ra + D \la q_\ab\ra \right)
    ~=~\frac{1}{N} \frac{1}{(1+\Z)^2}\left(
    (\mu \Omega^u_\ab + D \Omega^q_\ab)
    +
    \frac{\Z}{\ku}(\mu v_\ab + D \dot{Q}_\ab)
    \right) .
\end{align}
Next, substituting \cref{rubo} into \cref{sigma-components-1}, we get:
\begin{multline}
    \la \hat\sigma_\ab \hat\sigma_\mn\ra - 
     \la \hat\sigma_\ab \ra
      \la\hat\sigma_\mn\ra
    =\frac{1}{N} \Big(
    \mu^2 \left(\la u_\ab u_\mn \ra - \la u_\ab  \ra\la u_\mn \ra\right)
    +
    D^2 \left(\la q_\ab q_\mn \ra - \la q_\ab  \ra\la q_\mn \ra\right)
    \\
    \quad+ D \mu \left(
    \la u_\ab q_\mn \ra - \la u_\ab  \ra\la q_\mn \ra
    +\la q_\ab u_\mn \ra - \la q_\ab  \ra\la u_\mn \ra
    \right)
    \Big).
\end{multline}
Further substituting \cref{rafo} and simplifying, we obtain:
\begin{multline}
     \la \hat\sigma_\ab \hat\sigma_\mn\ra - 
     \la \hat\sigma_\ab \ra
      \la\hat\sigma_\mn\ra =
      \frac{\kBT \mu \rho (\Z-\Omega_0)}{N(1+\Z)} \Tiso_\abmn + \frac{1}{N(1+\Z)}\left(\mu^2 \Omega^{uu}_\abmn + \mu D(\Omega^{uq}_\abmn+\Omega^{qu}_\abmn)+ D^2 \Omega^{qq}_\abmn\right)
    \\
    - \frac{1}{N (1+\Z)^2}
    (\mu \Omega^u_\ab + D\Omega^q_\ab)(\mu \Omega^u_\mn + D\Omega^q_\mn)+
    \frac{1}{N \ku^2}\left(1-\frac{1}{(1+\Z)^2}\right)
    (\mu v_\ab +D \dot Q_\ab)
    (\mu v_\mn +D \dot Q_\mn) 
    \\
    +
    \frac{1}{N \ku (1+\Z)} 
    \left[(\mu v_\ab +D \dot Q_\ab)(\mu \Omega^u_\mn + D\Omega^q_\mn)
    +(\mu \Omega^u_\ab + D\Omega^q_\ab)(\mu v_\mn +D \dot Q_\mn)
    \right].
    \label{sigma-var-result-sm}
\end{multline}

\subsubsection{Evaluation of white noise amplitude}
We focus on the amplitude of the noise in the stress equation \cref{sigma-fluct}, which corresponds to the upper left block of \cref{white-noise}:
 
\begin{align}
    \begin{pNiceArray}{c|c}
        \frac{1}{V}\Lambda^{\sigma}_\abmn  
        & 
        \phantom{A}{\cdot}\phantom{A}
        \\[1ex]  \hline
        \cdot & \cdot
    \end{pNiceArray}
    = \frac{1}{\ku^2}\left(\Sigma \,\tr{\M} + \M\,\Sigma\right),
\end{align}
where $\M$ and $\Sigma$ are defined in \cref{M-def,Sigma-def-sm} respectively:
\begin{align}
    \M=\begin{pNiceArray}{c|c}
    \ku &  -\mu v_\ab -D \dot Q_\ab +\ku(\mu\Omega^u_\ab+D\Omega^q_\ab)
    \\[1ex]  \hline
    0 &\ku(1+\Z)
\end{pNiceArray},
\qquad 
    \Sigma = 
    \begin{pNiceArray}{c|c}
        \la \hat\sigma_\ab \hat\sigma_\mn\ra - \la \hat\sigma_\ab \ra \la  \hat\sigma_\mn\ra  
        & 
        \la \hat\sigma_\ab\smashhat\phi\ra - 
        \la \hat\sigma_\ab \ra\la\smashhat\phi \ra
        \\[1ex]  \hline
        \\[-2ex]
        \la \hat\sigma_\ab\smashhat\phi\ra - 
        \la \hat\sigma_\ab \ra\la\smashhat\phi \ra
        &
        \la \smashhat\phi^2\ra - \la\smashhat\phi \ra ^2
    \end{pNiceArray}.
\end{align}
Thus,
\begin{align}
    \frac{1}{V}\Lambda^{\sigma}_\abmn =& 2\ku \left(\la \hat\sigma_\ab \hat\sigma_\mn\ra - 
    \la \hat\sigma_\ab \ra
    \la\hat\sigma_\mn\ra\right)
    +
    \left(-\mu v_\ab -D \dot Q_\ab +\ku(\mu\Omega^u_\ab+D\Omega^q_\ab)\right)
    \left(
    \la \hat\sigma_\mn\smashhat\phi\ra - 
    \la \hat\sigma_\mn \ra\la\smashhat\phi \ra \right)
    \\
    &+
    \left(\la \hat\sigma_\ab\smashhat\phi\ra - 
    \la \hat\sigma_\ab \ra\la\smashhat\phi \ra\right)
    \left(-\mu v_\mn -D \dot Q_\mn +\ku(\mu\Omega^u_\mn+D\Omega^q_\mn)\right).
\end{align}
Substituting \cref{sigma-components}, then \cref{rubo}, then \cref{gago,rafo}, we obtain:
\begin{alignat}{3}
    \Lambda^{\sigma}_\abmn =& \frac{2 \kBT \mu }{\ku} \frac{\Z}{1+\Z}\Tiso_\abmn
    && \text{[thermal]}
    \\
    &+\frac{2}{\rho \ku}\frac{1}{1+\Z}\left(\mu^2 \Omega^{uu}_\abmn + \mu D (\Omega^{uq}_\abmn+\Omega^{qu}_\abmn)+ D^2 \Omega^{qq}_\abmn ~-~ \Omega_0\,\kBT \mu \rho \Tiso_\abmn\right)
    \quad && \text{[active]}
    \\
    &+\frac{2}{\rho \ku^3}\frac{\Z}{1+\Z}(\mu v_\ab +D \dot Q_\ab)
    (\mu v_\mn +D \dot Q_\mn)
    &&\text{[driven]}
    \\
    &+ \frac{1}{\rho\ku^2}\frac{1}{1+\Z}
    \left(
    (\mu v_\ab +D \dot Q_\ab)(\mu \Omega^u_\mn + D\Omega^q_\mn)+
    (\mu \Omega^u_\ab + D\Omega^q_\ab)(\mu v_\mn +D \dot Q_\mn)
    \right)
    \quad &&\text{[cross]}.
\end{alignat}
A simplified version of this expression is used in the main text.

\subsubsection{Stress autocorrelation function and spectral density}
Next, we evaluate the stress correlation function \cref{C-sigma}:
\begin{align}
    {\rm C}^{\sigma}_\abmn(t-t') = \la \hat\sigma_\ab(t) \hat\sigma_\mn(t')\ra - \la \hat\sigma_\ab \ra \la  \hat\sigma_\mn\ra,
\end{align}
which is the upper left block of \cref{CY}:
\begin{align}
    \begin{pNiceArray}{c|c}
        {\rm C}^{\sigma}_\abmn(t-t')
        & 
        \phantom{A}{\cdot}\phantom{A}
        \\[1ex]  \hline
        \cdot & \cdot
    \end{pNiceArray}
    = {\rm C}_\Y(t) = \ee^{-\M t} \Sigma \qquad \text{for} ~~t\ge0.
\end{align} 
Using the identity for block triangular matrices: $\exp\begin{pmatrix}
    A_{11}& A_{12}\\0& A_{22}
\end{pmatrix} = \begin{pmatrix}
    \ee^{A_{11}} & \int_0^1 \ee^{A_{11}(1-s)}\, A_{12} \,\ee^{A_{22} s} \d s
    \\[1ex]
    0 & \ee^{A_{22}}
\end{pmatrix}$, we obtain:
\begin{align}
    \ee^{-\M t} =\begin{pNiceArray}{c|c}
        \ee^{-\ku t} & 
        \left(\mu v_\ab +D \dot Q_\ab -\ku(\mu\Omega^u_\ab+D\Omega^q_\ab)\right)\frac{1}{\ku \Z}\left(\ee^{-\ku t} - \ee^{-(1+\Z)\ku t} \right)
        \\[1ex]  \hline
        \\[-2ex]
        0 & \ee^{-(1+\Z)\ku  t}
    \end{pNiceArray}.
\end{align}
Thus, the correlation function is given by:
\begin{multline}
    {\rm C}^{\sigma}_\abmn(t) = \ee^{-\ku t} ~\left(\la \hat\sigma_\ab \hat\sigma_\mn\ra - \la \hat\sigma_\ab \ra \la  \hat\sigma_\mn\ra \right)\\
    +\left(\mu v_\ab +D \dot Q_\ab -\ku(\mu\Omega^u_\ab+D\Omega^q_\ab)\right)\frac{1}{\ku \Z}\left(\ee^{-\ku t} - \ee^{-(1+\Z)\ku t} \right) 
    ~
    \left(\la \hat\sigma_\mn\smashhat\phi\ra - \la \hat\sigma_\mn \ra\la\smashhat\phi \ra \right).
\end{multline}
Substituting \cref{sigma-components}, then \cref{rubo}, then \cref{gago,rafo}, we obtain:
\begin{subequations}
    \label{C-sigma-result-sm}
\begin{align}
    {\rm C}^{\sigma}_\abmn(t) =&\frac{\kBT \mu \rho (\Z-\Omega_0)}{N(1+\Z)} \Tiso_\abmn \ee^{-\ku t} + \frac{1}{N(1+\Z)}\left(\mu^2 \Omega^{uu}_\abmn + \mu D(\Omega^{uq}_\abmn+\Omega^{qu}_\abmn)+ D^2 \Omega^{qq}_\abmn\right)
     \ee^{-\ku t}
    \\
    &- \frac{1}{N \Z(1+\Z)}
    \left(
    \ee^{-\ku t}-\frac{\ee^{(1+\Z)\ku t}}{1+\Z}
    \right)
    (\mu \Omega^u_\ab + D\Omega^q_\ab)(\mu \Omega^u_\mn + D\Omega^q_\mn)
    \\
    &+
    \frac{1}{N \ku^2}\left(\ee^{-\ku t}-\frac{\ee^{-(1+\Z)\ku t}}{(1+\Z)^2}\right)
    (\mu v_\ab +D \dot Q_\ab)
    (\mu v_\mn +D \dot Q_\mn)
    \\
    &+
    \frac{1}{N \ku \Z(1+\Z)} \left(
    \ee^{-\ku t}-\frac{\ee^{(1+\Z)\ku t}}{1+\Z}
    \right)
    (\mu v_\ab +D \dot Q_\ab)(\mu \Omega^u_\mn + D\Omega^q_\mn)
    \\
    &+
    \frac{\ee^{(1+\Z)\ku t}}{N \ku (1+\Z)^2}
    (\mu \Omega^u_\ab + D\Omega^q_\ab)(\mu v_\mn +D \dot Q_\mn).
\end{align}
\end{subequations}
Using \cref{CY-time-translation}, we extend this to the negative times: ${\rm C}^{\sigma}_\abmn(t) = {\rm C}^{\sigma}_{\mn\ab}(-t)$.

Similarly, using the upper left block of \cref{SY}, we obtain the spectral density of the stress:
\begin{align}
    \begin{pNiceArray}{c|c}
        \rS^{\sigma}_\abmn(\omega)
        & 
        \phantom{A}{\cdot}\phantom{A}
        \\[1ex]  \hline
        \cdot & \cdot
    \end{pNiceArray}
    = \rS_\Y(\omega) =
    (\M + \ii\omega \rI)^{-1}\Sigma + \Sigma\,(\tr{\M} - \ii\omega \rI)^{-1}.
\end{align} 
The inverse matrix is:
\begin{align}
    (\M + \ii\omega \rI)^{-1} = 
    \begin{pNiceArray}{c|c}
        (\ku+\ii\omega)^{-1}
        & (\ku+\ii\omega)^{-1}\left((1+\Z)\ku+\ii\omega\right)^{-1}
        \left(\mu v_\ab D \dot Q_\ab -\ku(\mu\Omega^u_\ab+D\Omega^q_\ab)\right)
        \\[1ex]  \hline
        \\[-2ex]
        0 &         \left((1+\Z)\ku+\ii\omega\right)^{-1}
    \end{pNiceArray}.
\end{align}
Thus, the stress spectral density is:
\begin{multline}
    \rS^{\sigma}_\abmn(\omega) =
    \left((\ku+\ii\omega)^{-1}+(\ku-\ii\omega)^{-1}\right)
    \left(\la \hat\sigma_\ab \hat\sigma_\mn\ra - \la \hat\sigma_\ab \ra \la  \hat\sigma_\mn\ra \right)
    \\
    +(\ku+\ii\omega)^{-1}\left((1+\Z)\ku+\ii\omega\right)^{-1} 
    \left(\mu v_\ab +D \dot Q_\ab -\ku(\mu\Omega^u_\ab+D\Omega^q_\ab)\right)
    \left(
    \la \hat\sigma_\mn\smashhat\phi\ra - 
    \la \hat\sigma_\mn \ra\la\smashhat\phi \ra \right)
    \\
    +
    (\ku-\ii\omega)^{-1}\left((1+\Z)\ku-\ii\omega\right)^{-1} \left(\la \hat\sigma_\ab\smashhat\phi\ra - 
    \la \hat\sigma_\ab \ra\la\smashhat\phi \ra\right)
    \left(\mu v_\mn +D \dot Q_\mn -\ku(\mu\Omega^u_\mn+D\Omega^q_\mn)\right).
\end{multline}
The final expression for the spectral density is:
\begin{equation}
    \label{S4-def}
\begin{aligned}
    \rS^{\sigma}_\abmn(\omega) =& 
    \frac{\kBT \mu \rho  (\Z-\Omega_0)}{N(1+\Z)} \frac{2\ku}{\ku^2+\omega^2}\Tiso_\abmn 
    + \frac{1}{N(1+\Z)}\frac{2\ku}{\ku^2+\omega^2}\left(\mu^2 \Omega^{uu}_\abmn + \mu D(\Omega^{uq}_\abmn+\Omega^{qu}_\abmn)+ D^2 \Omega^{qq}_\abmn\right)
    \\
    &- \frac{2\ku}{N \Z(1+\Z)}
    \left(
    \frac{1}{\ku^2+\omega^2}-\frac{1}{(1+\Z)^2\ku^2+\omega^2}
    \right)
    (\mu \Omega^u_\ab + D\Omega^q_\ab)(\mu \Omega^u_\mn + D\Omega^q_\mn)
    \\
    &+
    \frac{2}{N \ku (1+\Z)}\left(
    \frac{1+\Z}{\ku^2+\omega^2}-\frac{1}{(1+\Z)^2\ku^2+\omega^2}
    \right)
    (\mu v_\ab +D \dot Q_\ab)
    (\mu v_\mn +D \dot Q_\mn)
    \\
    &+
    \frac{1}{N \ku \Z(1+\Z)} \left(
    \frac{\ku-\ii\omega}{\ku^2+\omega^2}+\frac{(\Z-1)\ku + \ii\omega}{(1+\Z)^2\ku^2+\omega^2}
    \right)
    (\mu v_\ab +D \dot Q_\ab)(\mu \Omega^u_\mn + D\Omega^q_\mn)
    \\
    &+
    \frac{1}{N \ku \Z(1+\Z)} \left(
    \frac{\ku+\ii\omega}{\ku^2+\omega^2}+\frac{(\Z-1)\ku - \ii\omega}{(1+\Z)^2\ku^2+\omega^2}
    \right)
    (\mu \Omega^u_\ab + D\Omega^q_\ab)(\mu v_\mn +D \dot Q_\mn).
\end{aligned}
\end{equation}
Indeed, the above expression is the Fourier transform of the correlation function \cref{C-sigma-result-sm} (c.f.~\cref{ft-convention}).
\end{widetext}

\section{Wiener-Khinchin theorem}
\label{app-wiener}

Consider stationary processes $A(t)$ and $B(t)$. Their correlation functions are defined as:
\begin{align*}
    {\rm C}_{AB}(t'-t)=&\la A(t')B(t)\ra
    \\
    =& {\rm C}_{BA}(t-t'),
    \\
    {\rm C}_{AB}(t)=&{\rm C}_{BA}(-t).
\end{align*}
Let $\td{A}(\omega)$, $\td{B}(\omega)$, and $\rS_{AB}(\omega)$ denote the Fourier transforms of $A(t)$, $B(t)$, and ${\rm C}_{AB}(t)$, respectively. The function $\rS$ is referred to as the spectral density. According to the \WK{} theorem:
\begin{align*}
    \la \td{A}(\omega')\td{B}(\omega)\ra =& 2\pi\,\delta(\omega'+\omega)\,\rS_{AB}(\omega')
    \\
    =& 2\pi\,\delta(\omega'+\omega)\,\rS_{AB}(-\omega)
    \\
    =& 2\pi\,\delta(\omega'+\omega)\,\rS_{BA}(\omega).
\end{align*}
The proof is straightforward\cite{risken,mazur}:
\begin{align*}
    &\la \td{A}(\omega')\td{B}(\omega)\ra = \int \int \la A(t')B(t) \ra \ee^{-\ii\omega' t'}\ee^{-\ii\omega t}\d t\d t'
    \\
    &= \int \left(\int {\rm C}_{AB}(t'-t) \ee^{-\ii\omega'(t'-t)} \d t' \right)
    \ee^{-\ii (\omega+\omega')t} \d t
    \\
    &=  \rS_{AB}(\omega') \int \ee^{-\ii (\omega+\omega')t} \d t
    \\
    &= \rS_{AB}(\omega') 2\pi \,\delta(\omega'+\omega).
\end{align*}

\section{Verifying fluctuation-dissipation theorem for passive system}
\label{app-FDT-passive}
In \cref{Sigma-decomposed}, the noise in the stress equation is decomposed into various contributions. For a passive system in equilibrium (i.e., $\Omega=0$, $\v=0$, $\dot\Q = 0$), only the thermal contribution remains:
\begin{align}
    \label{rutio}
        \langle \hat\xi^\sigma_\ab (t)\hat\xi^\sigma_\mn (t')\rangle = 4\eta \kBT ~\Tiso_\abmn \frac{1}{V}\delta(t-t') .
\end{align}
This appendix verifies that this expression is consistent with the Fluctuation-Dissipation Theorem (FDT), a fundamental relation in statistical mechanics. The FDT relates the equilibrium fluctuation spectrum of a thermodynamic variable to its response function \cite{mazur}:
\begin{align}
    \label{FDT-general}
    S^{(x)}_{i j}(\omega)=-\frac{2 k_{\mathrm{B}} T}{\omega} \chi_{i j}^{\prime \prime}(\omega)
    \quad \text{and}\quad
    \tilde{x}_i = \chi_{i j}(\omega) \tilde{f}_j,
\end{align}
where $\chi''_{i j}$ is the imaginary part of the susceptibility $\chi_{i j}$.

We determine the conjugate thermodynamic variables and forces of our system from the entropy production. For the coarse-graining volume $V$, the entropy production is given by:
\begin{align}
    T \dot{S} &= \int  \sigma_\ab \,v_\ab \d V + f_i \dot{x}_i
    \\
    &= \int \sigma_\ab \,\partial_t U_\ab \d V +  f_i \dot{x}_i,
\end{align}
where $f_i$ and $x_i$ are other pairs of thermodynamic forces and variables not of interest. Here, $U_\ab$ is the macroscopic strain which is the proper thermodynamic variable. The fluctuation-dissipation theorem relates the equilibrium fluctuation spectrum of the thermodynamic variable. In our case, we focus on stress fluctuations. We can rewrite the entropy production as: $T \dot{S} = f \dot{x} = \frac{\d}{\d t}(f x) + (-x)\dot f $, where the total time derivative represents the reversible part, and now the thermodynamic force and variable are $-x$ and $f$, respectively. For our specific case (discarding complete differentials):
\begin{align}
    T \dot{S} &= \int -\partial_t\sigma_\ab \,U_\ab \d V +  f_i \dot{x}_i
    \\
    &= (-V U_\ab) \;\partial_t\sigma_\ab  +  f_i \;V\dot{x}_i.
\end{align}
Here, the thermodynamic variable is $-V U_\ab$ and its conjugate force is $\partial_t\sigma_\ab$. Thus, the linear response relationship is written as:
\begin{equation}
    \tilde{\sigma}_\ab(\omega) = \chi_{\ab \mu \nu}(\omega) (-V \tilde{U}_{\mu\nu}(\omega)) + \chi_{\ab i}(\omega) \tilde{f}_i(\omega).
\end{equation}
Comparing this with the Fourier transform of \cref{sigma-average}:
\begin{align}
    \left(1+\ii \omega /\ku \right) \tilde{\sigma}_\ab(\omega)&=2 \eta \tilde{v}_\ab(\omega)-\nu \tilde{\dot{Q}}_\ab(\omega)-\zeta \tilde{Q}_\ab(\omega),
    \nonumber
\end{align}
we find that $\chi_{\ab \mn} = -\frac{2\eta \ii \omega}{1+\ii\omega /\ku} \frac{1}{V}\, \Tiso_{\alpha\beta\mu\nu}$. 
Using the Fluctuation-Dissipation Theorem (FDT) \eqref{FDT-general}, the spectral density of the thermodynamic force, that is the stress tensor, is:
\begin{align}
    S^{(\sigma)}_\abmn(\omega)=&-\frac{2 \kBT }{\omega} \chi''_\abmn(\omega) 
    =
      \frac{4\eta \kBT}{1+(\omega/\ku)^2 } \frac{1}{V}\Tiso_\abmn.
      \nonumber
\end{align}
If we model the stress fluctuations using the Langevin equation \eqref{sigma-fluct}, we obtain
\begin{equation}
\begin{aligned}
    \la \hat\xi_\ab(\omega)\hat\xi_\mn(\omega') \ra =\hspace{-2cm}&
    \\
    =&(1+\ii\omega/\ku)(1+\ii\omega'/\ku)
    \la \hat\sigma_\ab(\omega) \hat\sigma_\mn(\omega') \ra
    \\
    =&(1+\ii\omega/\ku)(1+\ii\omega'/\ku)  2\pi \delta(\omega+\omega') S^{(\sigma)}_\abmn(\omega)
    \\
    =& 2\pi \delta(\omega+\omega') \frac{1}{V} 4\eta \kBT \Tiso_\abmn,
\end{aligned}
\end{equation}
where in the second line, we used the \WK{} theorem \cite{risken,mazur}. This result is consistent with \cref{rutio}.

\section{Visualisation of the steady state distribution of the stress tensor}
\label{app-plots}
In \cref{sec-specific}, we investigated a simple form of activity and, in \cref{fig-sigmadist-1d}, illustrated the resulting steady-state distribution of the stress tensor. This appendix details the numerical procedure used for these visualizations and extends the analysis to present the complete two-dimensional distribution.

In a two-dimensional system, the activity \eqref{activity-specific} is expressed as a function of the independent components of the tensors $\u$ and $\q$:
\begin{align}
    \Omega(u_{11},u_{12},q_{11},q_{12})\!=\!\Omega(u_{11},u_{12})\delta(q_{11}\!-\!Q_{11})\delta(q_{12}\!-\!Q_{12}) .\nonumber
\end{align}
We consider the following form of $\Omega(\u)$:
\begin{align}
\Omega(u_{11},u_{12}) = \frac{\Omega_0}{A\pi} \ee^{-\frac{1}{A}\, (u_{11}^2+u_{12}^2)}.
\end{align}
This form is consistent with \cref{Omega-uq-specific,Omega-uuqq-specific}. For numerical convenience, the delta functions are approximated by sharp Gaussian functions with a small width $C \ll 1$:
\begin{multline}
    \Omega(u_{11},u_{12},q_{11},q_{12})\approx
    \\
    \approx \tfrac{\Omega_0}{A\pi} \ee^{-\frac{1}{A}\, (u_{11}^2+u_{12}^2)}
    \tfrac{1}{C\pi}\ee^{-\frac{1}{C}\left((q_{11}-Q_{11})^2+(q_{12}-Q_{12})^2\right)}  .
\end{multline}
Consequently, the activity depends on the parameters $\Omega_0$, $A$, and $C$.

The ratio of the binding rates is a mixture of Gaussian distributions:
\begin{multline}
    \frac{\kb}{\ku} = 
    \ee^{-\frac{\beta}{\rho} \left(
        \mu(u_{11}^2 + u_{12}^2) 
        +2D(u_{11}q_{11} + u_{12}q_{12}) +
        \chi(q_{11}^2 + q_{12}^2)
    \right)}
    \\
    +\frac{\Omega_0}{AC\pi^2} 
    \ee^{-\frac{1}{A}\, (u_{11}^2+u_{12}^2)-\frac{1}{C}(q_{11}-Q_{11})^2-\frac{1}{C}(q_{12}-Q_{12})^2}.
\end{multline} 
The duty ratio is given by:
\begin{align}
    \Z = \frac{\pi^2\rho^2}{\beta^2(\mu\chi-D^2)} + \Omega_0.
\end{align}
From \cref{ss-10}, the steady-state distribution is:
\begin{widetext}
\begin{align}
    p_\ss(u_{11},u_{12},q_{11},q_{12}) = \int_0^\infty p_\text{active}(u_{11}-v_{11}t,u_{12}-v_{12}t,q_{11}-\dot{Q}_{11}t,q_{12}-\dot{Q}_{12}t) \;\ku\ee^{-\ku t} \d t,
    \\
    \text{where} \quad p_\text{active}(u_{11},u_{12},q_{11},q_{12}) = \frac{1}{1+\Z} \frac{ka(u_{11},u_{12},q_{11},q_{12})}{\ku} .
    \nonumber
\end{align}
This represents the steady-state distribution in the absence of external driving, consisting of active and thermal contributions. The function $ p_\text{active}$ is a mixture of Gaussians; like $n_\ss$ in \cref{ss-10}, this distribution is not normalized. Instead, the integral of the probability density functions (PDFs) yields the fraction of bound linkers.

The steady-state distribution is over the binding states of the linker, but our interest lies in the distribution of stress and the influence of activity and external driving on it. Using the relation \eqref{sigma-1-def}, we have:
\begin{align}
    p_\ss(\sigma_{11},\sigma_{12}) = \int 
    \delta\left(\sigma_{11} - (\mu u_{11} + D q_{11})\right)
    \delta\left(\sigma_{12} - (\mu u_{12} + D q_{12})\right)
    p_\ss(u_{11},u_{12},q_{11},q_{12}) \d u_{11} \d u_{12} \d q_{11} \d q_{12},
    \\
    p_\ss(\sigma_{11}) = \int p_\ss(\sigma_{11},\sigma_{12})\d\sigma_{12} = 
    \int \delta\left(\sigma_{11} - (\mu u_{11} + D q_{11})\right)
    p_\ss(u_{11},u_{12},q_{11},q_{12}) \d u_{11} \d u_{12} \d q_{11} \d q_{12}.
\end{align}
By exchanging the order of integration, we integrate over the state variables $u_{11}, u_{12}, q_{11}, q_{12}$ first, followed by the time variable $t$:
\begin{align}
    &p_\text{active}(\sigma_{11},\sigma_{12}) = \int 
    \delta\left(\sigma_{11} - (\mu u_{11} + D q_{11})\right)
    \delta\left(\sigma_{12} - (\mu u_{12} + D q_{12})\right)
    p_\text{active}(u_{11},u_{12},q_{11},q_{12}) \d u_{11} \d u_{12} \d q_{11} \d q_{12},
    \\
    &p_\text{active}(\sigma_{11}) = \int p_\text{active}(\sigma_{11},\sigma_{12})\d\sigma_{12} = 
    \int \delta\left(\sigma_{11} - (\mu u_{11} + D q_{11})\right)
    p_\text{active}(u_{11},u_{12},q_{11},q_{12}) \d u_{11} \d u_{12} \d q_{11} \d q_{12},
    \\
    &p_\ss(\sigma_{11},\sigma_{12}) = \int_0^\infty p_\text{active}(\sigma_{11}-(\mu v_{11} + D \dot{Q}_{11})t,\sigma_{12}-(\mu v_{12} + D \dot{Q}_{12})t) \;\ku\ee^{-\ku t} \d t,
    \\
    &p_\ss(\sigma_{11}) = \int_0^\infty p_\text{active}(\sigma_{11}-(\mu v_{11} + D \dot{Q}_{11})t) \;\ku\ee^{-\ku t} \d t.
\end{align}
\end{widetext}
Given that \(p_\text{active}(u_{11},u_{12},q_{11},q_{12})\) is a mixture of Gaussians, the evaluation of \(p_\text{active}(\sigma_{11},\sigma_{12})\) is straightforward. The steady-state distributions \(p_\ss(\sigma_{11},\sigma_{12})\) and \(p_\ss(\sigma_{11})\) can be expressed in terms of error functions:
\begin{align}
    \text{erf}(x) = \frac{2}{\sqrt{\pi}} \int_0^x e^{-t^2} \d t.
\end{align}
For instance, we obtain:
\begin{multline}
    \label{pactive}
    \textstyle
    p_\text{active}(\sigma_{11}) = \frac{1}{1+\Z}\left(
    (\Z-\Omega_0)\sqrt{\frac{\beta}{\pi\mu\rho}}\;\ee^{\beta \sigma_{11}^2/(\mu \rho)}\right.
    \\
    \textstyle
    \left.+
    \frac{\Omega_0}{\sqrt{\pi(A\mu^2)}} \ee^{-(\sigma_{11}-D Q_{11})^2/(A\mu^2)}
    \right) .
\end{multline}
where the \(C \to 0\) limit has been taken. For simplicity, we only provide the expression for \(p_\ss(\sigma_{11})\) at \(\Omega_0 = 0\) and \(\dot{Q}_{11} = 0\):
\begin{multline}
\label{pactivedriven}
\textstyle
p_\ss(\sigma_{11}) = \frac{\ku}{2\mu v_{11}}
\left[1+\text{erf}\left(\frac{2v_{11}\sigma_{11}\beta-\rho\ku}{2v_{11}\sqrt{\rho\beta\mu}}\right)\right]
\\
\textstyle
\times\exp\left(-\frac{\sigma_{11}\ku }{v_{11} \mu} + \frac{\ku^2\rho}{4 v_{11}^2 \beta\mu}\right).
\end{multline}

In Figure \ref{fig-sigmadist-1d}, panels (b) and (c) correspond to \cref{pactive} and \cref{pactivedriven}, respectively, including the active contribution. The parameters used are:
\begin{equation}
    \label{params-fig1d}
\begin{aligned}
    &\text{Model:}\quad \rho=0.05,~ \epsilon_0=0,~ \ku=1,~\beta=1,
    \\
    &\hspace{1.35cm} \mu=\chi=1,~ D=0.8,
    \\[1ex]
    &\text{Activity:}\quad
\Omega_0=0.2,\;A=0.01,\; C=0.001,
\\[1ex]
&\text{Strain rate and:}\quad
v_{11}=0.4, ~Q_{11}=1,~ \dot{Q}_{11}=0,
\\[-1ex]
&\text{Orientation} &
\end{aligned}
\end{equation}

\comment{
{\color{blue}
\begin{align}  
\beta^{-1} \sim&\text{units of energy}
\\
\mu\sim& \text{units of energy density = pressure}
\\
\Rightarrow 1/(\beta\mu) \sim& \text{units of volume}
\\
\ku^{-1}\sim&\text{units of time}
\end{align}
}}
This distribution is independent of the other components of orientation and strain rate ($Q_{12}$, $\dot{Q}_{12}$ and $v_{12}$).
These parameters are selected to illustrate the characteristics and properties of the steady-state distribution and do not represent realistic systems.

In Figure \ref{fig-sigmadist-2d}, we visualize the distribution of two independent components of the stress. The parameters used are as follows:
\begin{equation}
    \label{params-fig2d}
\begin{aligned}
        &\text{Model:}\quad \rho=0.05,~ \epsilon_0=0,~ \ku=1,~\beta=1,
        \\
        &\hspace{1.35cm} \mu=\chi=1,~ D=0.8,
        \\[1ex]
        &\text{Activity:}\quad
    \Omega_0=0.2,\;A=0.05,\; C=0.001,
    \\[1ex]
    &\text{Strain rate and:}\quad
    v_{11}=0, v_{12}=0.7, 
    \\
    &\text{Orientation}  \quad\quad\quad Q_{11}=1, Q_{12}=0,~ \dot{Q}_{11}=\dot{Q}_{12}=0,
\end{aligned}
\end{equation}

\begin{figure*}
\centering
	\includegraphics[height=5.5cm]{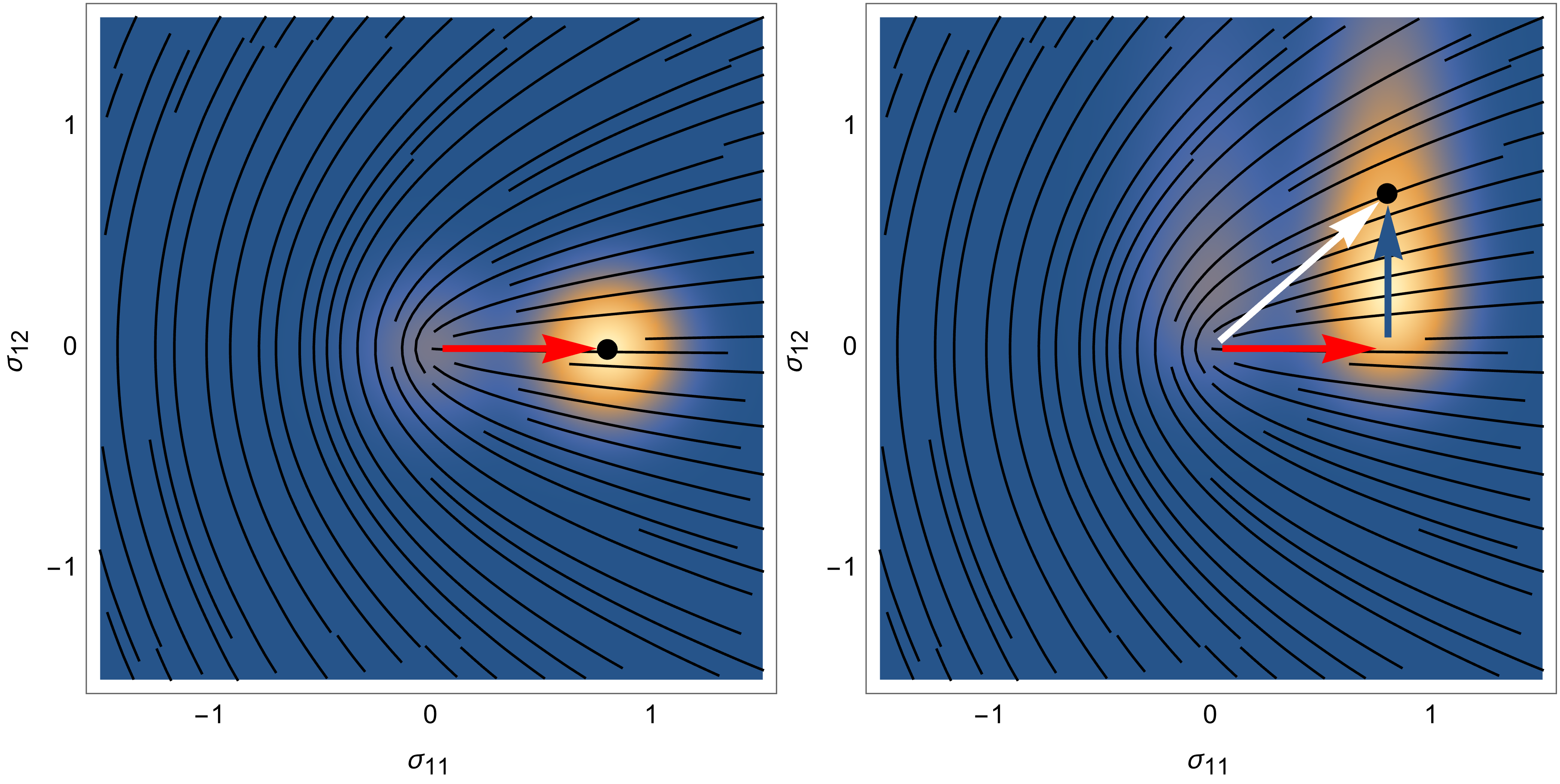}
    \hspace{1cm}
    \raisebox{1.5cm}{\includegraphics[height=3.5cm]{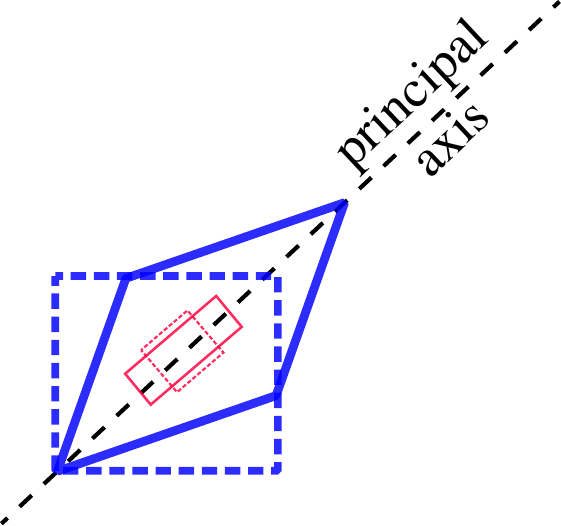}}
	\hspace{-3cm}\raisebox{0.8cm}{
	$\bsigma= \begin{pmatrix}
	0 & 1
	\\
	1 & 0
	\end{pmatrix}$}
	\hspace{1cm}
    \caption{
At steady state, the states of the bound linkers are distributed according to $n_\ss(\u, \q)$ defined in \cref{ss-10}. Each bound linker contributes to the stress tensor $\sigma(\u,\q)$ as defined in \cref{sigma-1-def}.
Here, we visualize the distribution of the stress tensor
in $d=2$ dimensions. The stress tensor has two independent components, $\sigma_{11}$ and $\sigma_{12}$, which are taken as axes. The tangent of the black streamlines represents the principal axis of the corresponding stress tensor(the directions where there is no shear stress). For example, the principal axis of the stress tensor in the right diagram is at 45 degrees, the corresponding point on the left plots is $(\sigma_{11},\sigma_{12}) = (0,1)$, where the tangent of the streamline is also at 45 degrees. Bright spots on the plots indicate regions of high probability density.
Similar to \cref{fig-sigmadist-1d}, we set the activity $\Omega$ to promote binding to states with orientation $\q \approx \Q = \begin{pmatrix}
	1 & 	0
	\\
	0 & 	-1
\end{pmatrix}$.
For the left plot, we set vanishing external drives, $\v=\dot\Q=0$, and we assume that the activity is much stronger than the thermal contribution in \cref{kakd}; latter is responsible for the less bright spot at the center.
As the elastonematic coupling $D=0.8 \ne 0$, there is a high probability density at $\sigma \propto \Q$. The black dot marks the average of the distribution, and the red arrow indicates the probability shift due to activity.
In the right plot, we introduce shear with rate $\v= \begin{pmatrix}
	0 & 1
	\\
	1 & 0
\end{pmatrix}$ in the orthogonal direction in contrast to the \cref{fig-sigmadist-1d}. This distorts the stress distribution. The blue arrow indicates the shift in the average of the distribution induced by the external shear rate $\v$.
Parameter values, chosen for illustration purposes, are given in \cref{params-fig2d}.}
	\label{fig-sigmadist-2d}
\end{figure*}

\section{Isotropic tensors relative to an axis}
\label{app-axis-invariant}

In \cref{sec-reduced-repr}, following the systems's symmetries, we decompose 3-dimensional rank-4 tensors in terms of invariant tensors \eqref{Tm-def}, see \cref{omega-2}.

In this appendix, we show that no additional tensors, linearly independent of those in \eqref{Tm-def}, exist with the specified symmetries.

If we relax the symmetry requirement under the exchange of the first and last pairs of indices, i.e., if we drop the condition $\T_{\alpha\beta\mu\nu} = \T_{\mu\nu\alpha\beta}$, two additional independent tensors become possible:
\begin{equation}
\label{2moretensors}
\begin{aligned}
    \T^{(3)}_{\ab \mu\nu} \!\!=& (n_\alpha n_\mu \epsilon_{\beta\nu\gamma} 
    \!\!+\! n_\alpha n_\nu \epsilon_{\beta\mu\gamma}
    \!\!+\! n_\beta n_\mu \epsilon_{\alpha\nu\gamma}
    \!\!+\! n_\beta n_\nu \epsilon_{\alpha\mu\gamma}
    )n_\gamma,
    \\
    \T^{(4)}_{\ab \mu\nu}\!\!=&
    (\delta_{\alpha\mu} \epsilon_{\beta\nu\gamma} 
    + \delta_{\alpha\nu}\epsilon_{\beta\mu\gamma}
    +\delta_{\beta\mu}\epsilon_{\alpha\nu\gamma}
    +\delta_{\beta\nu}\epsilon_{\alpha\mu\gamma}
    )n_\gamma ,
\end{aligned}
\end{equation}
where $\epsilon_{\alpha\beta\gamma}$ is the totally antisymmetric tensor (Levi-Civita symbol). Tensors in \cref{Tm-def} are symmetric under the exchange of the first and last pairs of indices, while tensors in \cref{2moretensors} are antisymmetric:
\begin{subequations}
\begin{alignat}{4}
    \T^{(i)}_{\alpha\beta\mu\nu}&=\T^{(i)}_{\mu\nu\alpha\beta} \qquad& \text{for }i&=0,1,2,
    \\
    \label{swap-asym}
    \T^{(i)}_{\alpha\beta\mu\nu}&=-\T^{(i)}_{\mu\nu\alpha\beta} \qquad& \text{for }i&=3,4. 
\end{alignat}   
\end{subequations}
To demonstrate that no further linearly independent tensors exist, we formulate the constraints as a system of linear equations and determine the rank of the coefficient matrix using symbolic computation software.

For an infinitely small angle of rotation $\theta$, the rotation matrix can be approximated as
\begin{align}
R(\theta)_\ab \approx \delta_\ab + \theta g_\ab,
\end{align}
where $g_\ab = \epsilon_{\alpha\beta\gamma} n_\gamma$.
Substituting this into the constraint of the rotational invariance $
        R_{\alpha\alpha'}R_{\beta\beta'}R_{\gamma\gamma'}R_{\nu\nu'}\T_{\alpha'\beta'\mu'\nu'}=\T_\abmn$, we obtain
\begin{align}\label{rank4relation}
    g_{\alpha\alpha'}\T_{\alpha'\beta\mu\nu}
    +\!g_{\beta\beta'}\T_{\alpha\beta'\mu\nu}
    +\!g_{\mu\mu'}\T_{\alpha\beta\mu'\nu}
    +\!g_{\nu\nu'}\T_{\alpha\beta\mu\nu'}\!=0,
\end{align}
keeping only the first-order terms in $\theta$. 
\Cref{rank4relation} represents 81 linear equations for the 81 entries of the tensor $\T_{\ab\mn}$. 
Let $x$ be a vector containing all entries of the tensor $\T$; \cref{rank4relation} can be expressed in matrix form as $Ax = 0$,
where $A$ is an $81 \times 81$ matrix containing the coefficients from \cref{rank4relation}. The null space (kernel) of the matrix $A$ is found to be 19-dimensional. This indicates that there are 19 linearly independent rank-4 tensors invariant under rotations around the given axis.

Now, we introduce additional constraints on the tensor $\T_{\ab\mn}$: symmetry with respect to the first two and last two indices, and zero trace with respect to the first and last pair of indices:
\begin{equation}
\begin{aligned}
    &\T_{\alpha\beta\mu\nu}-\T_{\beta\alpha\mu\nu}=0,
    \\
    &\T_{\alpha\beta\mu\nu}-\T_{\alpha\beta\nu\mu}=0,
    \\
    &\T_{\xi\xi\mu\nu}=0, \quad
    \T_{\alpha\beta\xi\xi}=0.
\end{aligned}
\end{equation}
These constraints add $81 + 81 + 9 + 9 = 180$ additional (not necessarily independent) conditions. Incorporating these linear equations into \cref{rank4relation} results in a system that can be expressed in matrix form as $Ax = 0$, where $A$ is a $261 \times 81$ matrix. The kernel of this matrix is 5-dimensional, indicating that there are only 5 linearly independent rank-4 tensors that are symmetric, traceless, and invariant under rotations around the given axis. 

There are several methods for selecting these five tensors. Our approach imposes additional symmetry with respect to the interchange of the first and last pairs of indices: $\T_{\alpha\beta\mu\nu} = \T_{\mu\nu\alpha\beta}$. This further reduces the dimension of the kernel to 3. One evident solution is the isotropic tensor \cref{Tm-0}, which is independent of the symmetry axis $\n$. Another choice is $\T^{(2)}_{\alpha\beta\mu\nu} = N_\ab N_{\mu\nu}$, which clearly satisfies all the constraints. The third tensor, $\T^{(1)}$ (as defined in \cref{Tm-1}), is uniquely determined (up to a constant factor) through an orthogonalization procedure.

The remaining two tensors must satisfy \cref{swap-asym}. Specifically, $\T^{(4)}$ is chosen to be proportional to $\n$, and $\T^{(5)}$ is proportional to $\n \otimes \n \otimes \n$.

\section{Fluctuating fluid dynamics of the gel}
\label{app-fluid}

In this appendix, we solve the fluctuating fluid dynamical equations \eqref{fd-time} subject to the boundary condition \eqref{bc}.

Although \cref{fd-time} is linear, the time-dependent boundary condition \cref{bc} introduces nonlinearities into the problem. To achieve a strictly linear problem, the time dependence of $\hat S(t)$ in \cref{bc} must be neglected \cite{bedeaux1974,bedeaux1977}. This implies that while the particle's surface has a velocity $\hat U_\alpha(t)$, its spatial movements are neglected. This approximation is valid when the particle's displacement is small relative to its size over the relevant timescale (e.g., relaxation time of the gel). Furthermore, we assume the tracer particle fluctuates around $\bR = 0$. Consequently, the linearized boundary condition \eqref{bc} is:
\begin{align} 
    \hat v_\alpha(\r, t) = \hat U_\alpha(t), \quad  |\r| = a.
\end{align} 

We solve the problem in the Fourier domain using the following Fourier transform convention: $\td f(\omega) = \int_{-\infty}^{\infty} f(t) \ee^{-\ii \omega t} \d t$. For clarity, we will omit the `$\tilde{~~}$' notation for all subsequent fields and functions.

\subsection{Derivation of the response function of the active gel}
\label{app-response-function}
In this section, we derive the drag force and the response function of the active gel by considering non-fluctuating dynamical equations in the Fourier domain. The equations and boundary conditions are given by:
\begin{subequations}
    \label{fd-complex-avg}
    \begin{align}
        &\partial_\beta \sigma^\text{tot}_\ab = 0, 
        \quad
        \sigma^\text{tot}_\ab = \sigma_\ab - \delta_\ab  p, 
        \\
        &\partial_\alpha  v_\alpha = 0, 
        \label{incompress-complex-avg}
        \\
        &  \sigma_\ab = 2 \eta'(\omega)  v_\ab - \zeta'(\omega) Q_\ab,  
        \label{noisy-sigma-avg}
        \\
        &  v_\alpha(\r, \omega) =  U_\alpha(\omega) \qquad |\r|=a,
    \end{align}
\end{subequations}
where we defined frequency dependent coupling coefficients as:
\begin{align}
    \label{eta-zeta-prime}
    \eta'(\omega) = \frac{\eta}{1+\ii \omega /\ku}\,,\quad 
    \zeta'(\omega) = \frac{\zeta}{1+\ii \omega /\ku }.
\end{align}
The drag force is
\begin{align}
    F^\text{gel}_\alpha(\omega) = \oint\limits_{|\r|=a} \sigma^\text{tot}_\ab(\r, \omega)  \d S_\beta,
\end{align}
which equals the external force $F^\text{ext}_\alpha(\omega)$ acting on the tracer particle when inertia is neglected (c.f.~\cref{tracer-balance}).

Suppose the external force $F^\text{ext}_\alpha(\omega)$ induces a motion of the tracer particle with a velocity $U_\alpha(\omega)$. The solution to \cref{fd-complex-avg} is know as Stokes flow \cite{landau-fluid}:
\begin{equation}
\label{stokes-flow}
    p(\r,\omega)\!=\! \frac{3}{2}\eta'(\omega) \frac{a}{r^3}\,U_\gamma(\omega)\, r_\gamma,
    \quad
    v_\alpha(\r, \omega)\!=\! P_{\alpha\gamma}(\r)\;U_\gamma(\omega),
\end{equation}
where $P_{\alpha\gamma}(\r)$ is defined in \cref{P-def}.
Having the flow field, the force exerted on the tracer particle by the fluid, known as the Stokes drag, is:
\begin{align}
    \label{F-avg-def}
F^\text{gel}_\alpha(\omega) = \oint\limits_{|\r|=a} \sigma^\text{tot}_\ab(\r, \omega)  \d S_\beta = -6\pi a \eta'(\omega) U_\alpha(\omega).
\end{align}
In the time domain, this corresponds to \cref{stokes-time}.

Using the force balance $F^\text{ext}_\alpha(\omega) = -F^\text{gel}_\alpha(\omega)$ and the definition of the response function \eqref{chi-def}, we obtain \cref{chi-final}.

\subsection{Position fluctuations of the free tracer particle}
\label{app-position-fluct}
In this section, we calculate the position fluctuation spectrum $\rS^R_\ab(\omega)$ defined in \eqref{S-def}. We extend the derivation of Refs. \cite{bedeaux1974,bedeaux1977} to viscoelastic medium.
The position fluctuation spectrum is related to the position autocorrelation function in Furier space by \WK{} theorem (\cref{app-wiener}):
\begin{align}\label{wk-SR}
 \la \hat R_\alpha(\omega) \hat R_\beta(\omega') \ra = 2\pi \delta(\omega+\omega') S^R_\ab(\omega).
\end{align}
We proceed to calculate the velocity autocorrelation function $\la \hat U_\alpha(\omega) \hat U_\beta(\omega') \ra = \ii\omega \,\ii\omega' \la \hat R_\alpha(\omega) \hat R_\beta(\omega') \ra$.

The linearized system of fluctuating equations is given by:
\begin{subequations}
\begin{align}
    &\partial_\beta \hat\sigma^\text{tot}_\ab = 0,
    \quad
    \hat\sigma^\text{tot}_\ab = \hat\sigma_\ab - \delta_\ab \hat p,
    \label{sigma-def-complex}
    \\
    &\partial_\alpha \hat v_\alpha = 0,
    \label{incompress-complex}
    \\
    & \hat \sigma_\ab = 2 \eta'(\omega) \hat v_\ab - \zeta'(\omega) Q_\ab+ \hat s_\ab,
    \label{noisy-sigma}
    \\
    & \hat v_\alpha(\r, \omega) = \hat U_\alpha(\omega) \qquad |\r|=a, 
    \label{bc-complex}
\end{align}
\end{subequations} 
where $\eta'$, $\zeta'$ are same as in \cref{eta-zeta-prime}, and  $\hat s_\ab$ represents the total noise in the constitutive relation.
Note that in \cref{fd-time-2}, the viscosity and active stress coefficients are fluctuating quantities, as indicated by \cref{transport-fluct}. In contrast, the transport coefficients in \cref{noisy-sigma} are their steady-state averages, and all fluctuating terms are incorporated into $\hat s_\ab$:
\begin{align}
    \hat s_\ab = \tfrac{\hat \xi^\sigma_\ab}{1+\ii\omega/\ku} 
    - \tfrac{\hat \zeta - \zeta}{1+\ii \omega /\ku} Q_\ab
    + 2  \tfrac{\hat\eta - \eta}{1+\ii \omega /\ku} {\hat v_\ab}.
\end{align} 
The last term, which includes fluctuations of the strain rate, is higher order compared to the other two terms, and thus neglected.
From the \WK{} theorem, the spectrum of the noise $\hat s_\ab$ is given by:
\begin{align}\label{wk-s}
    &\la \hat s_\ab(\r, \omega) \hat s_\mn(\r',\omega')\ra 
    =2\pi \delta (\r-\r')\delta(\omega+\omega')\rS^\text{partial}_{\ab\mn}(\omega),
\end{align}
where $\rS^\text{partial}_{\ab\mn}(\omega)$ is the \emph{partial stress spectrum}; the complete stress spectrum, in addition to $\hat{s}_\ab$, includes the fluctuations of the strain rate $\hat v_\ab$ generated by the fluctuating velocity field $\hat v_\alpha$ (c.f.~\cref{noisy-sigma}).
The partial stress spectrum is defined as:
\begin{align}
    \label{partial-stress-spectrum-def}
    \rS^\text{partial}_{\ab\mn}(\omega) = \left.\frac{1}{V} \rS^\sigma_{\ab\mn}(\omega) \right|_{\dot Q_\ab=0,v_\ab=0},
\end{align}
where $\rS^\sigma$ is the stress fluctuation spectrum defined for a coarse-grained volume $V$ in \cref{S4-def}; the prefactor $\frac{1}{V}$ gives the spectrum per unit volume. We set $\dot Q_\ab=0$ due to the consideration of a constant uniform nematic order, and $v_\ab=0$ to account for small velocity fluctuations around zero. 

To derive the fluctuations of the tracer particle's velocity $\hat  U_\alpha(\omega)$, we need a solution to the averaged equations of motion \cref{fd-complex-avg} as a reference.
Suppose the tracer particle is driven externally, such that it has a velocity $\bar U(\omega)$. It generates a flow field $\bar v_\alpha(\r, \omega)$ given by \cref{stokes-flow}. We call this the `driven' dynamics, and the associated variables are denoted with a bar.

The derivation proceeds by ``projecting'' the fluctuating dynamics onto the driven dynamics.
There is no connection between the fluctuating and driven dynamics, except that the tracer particle moves around $\bR = 0$.
The starting point is the expression $\bar F_\alpha(\omega) \hat U_\alpha(\omega)$ (omitting arguments for simplicity):
\begin{align}
    \bar F_\alpha \hat U_\alpha =&\, 
    \bar F_\alpha \hat U_\alpha - \hat F_\alpha \bar U_\alpha.
\end{align}
Here $\hat F_\alpha=0$ because the tracer particle is free.
Expressing the forces in terms of stress tensors and using the boundary condition \eqref{bc-complex}, we get
\begin{align}
  =& \oint\limits_{|\r|=a} 
    \left[
        \bar \sigma^\text{tot}_\ab \hat v_\alpha 
        - \hat \sigma^\text{tot}_\ab \bar v_\alpha
    \right]\d S_\beta &
    \\
    =& - \int\limits_{|\r|>a} \d^3 \r  \left[
        \bar \sigma^\text{tot}_\ab \partial_\beta \hat v_\alpha 
        - \hat \sigma^\text{tot}_\ab \partial_\beta \bar v_\alpha
    \right]
    \nonumber
    \\
    &\qquad\qquad +
    \oint\limits_{|\r|=R_\infty} 
    \left[
        \bar \sigma^\text{tot}_\ab \hat v_\alpha 
        - \hat \sigma^\text{tot}_\ab \bar v_\alpha
    \right]\d S_\beta,
\end{align} 
where in the second line we used the divergence theorem to transform the surface integral into a volume integral. 
Note that the surface element $\d S_\beta$ is outward oriented, pointing inside the volume of integration.
Setting the velocity at the outer boundary to zero, the surface integral at infinity vanishes. Now, substituting $\hat\sigma^\text{tot}_\ab$ from \cref{sigma-def-complex}, we get
\begin{equation}
=\!-\!\!\!\!\int\limits_{|\r|>a}\!\!\! \d^3 \r  \left[
        \bar \sigma_\ab (\partial_\beta \hat v_\alpha)
        \!+\! \bar p (\partial_\alpha \hat v_\alpha )
        \!-\! \hat \sigma_\ab (\partial_\beta \bar v_\alpha)
        \!-\! \hat p (\partial_\alpha \bar v_\alpha )
    \right].
\end{equation}
The pressure terms vanish due to the incompressibility conditions \cref{incompress-complex,incompress-complex-avg}. Using the constitutive relations \cref{noisy-sigma-avg,noisy-sigma}, we obtain:
\begin{multline}
    \label{cancelings}
    = - \int\limits_{|\r|>a} \d^3 \r  \Big[
        2 \eta'(\omega) \bar v_\ab (\partial_\beta \hat v_\alpha)
        -\zeta'(\omega) Q_\ab (\partial_\beta \hat v_\alpha )
        \\
        - 2 \eta'(\omega) \hat v_\ab (\partial_\beta \bar v_\alpha)
        +\zeta'(\omega) Q_\ab (\partial_\beta \bar v_\alpha )
        - \hat s_\ab (\partial_\beta \bar v_\alpha) 
    \Big].
\end{multline} 
Most terms above cancel out, as follows.
Since $\bar v_\ab=\frac{1}{2}\left(\partial_\alpha \bar v_\beta + \partial_\beta \bar v_\alpha\right)$ is symmetric, we can rewrite 
\begin{align}
    \bar v_\ab (\partial_\beta \hat v_\alpha) = \bar v_\ab  \tfrac{1}{2} \left(
        \partial_\beta \hat v_\alpha \!+\! \partial_\alpha \hat v_\beta
    \right) = \bar v_\ab \hat v_\ab = (\partial_\beta \bar v_\alpha) \hat v_\ab.
\end{align}
Thus the terms proportional to viscosity $\eta'$ cancel. Additionally, with uniform nematic order, we have:
\begin{multline}
    \int\limits_{|\r|>a} \d^3 \r \,Q_\ab (\partial_\beta \hat v_\alpha) = Q_\ab \int\limits_{|\r|>a} \d^3 \r (\partial_\beta \hat v_\alpha) =
    \\
    =  -Q_\ab \oint\limits_{|\r|=a} \hat v_\alpha \d S_\beta = -Q_\ab \hat U_\alpha \oint\limits_{|\r|=a} \d S_\beta = 0.
\end{multline}
Similarly, $\int_{|\r|>a} \d^3 \r\, Q_\ab (\partial_\beta \bar v_\alpha) = 0$. Eventually, in \cref{cancelings} only the stress fluctuations $\hat s_\ab$ remain:
\begin{align}
    \bar F_\alpha(\omega) \hat U_\alpha(\omega) = \int\limits_{|\r|>a} \d^3 \r\;
          (\partial_\beta \bar v_\alpha(\r,\omega)) \hat s_\ab(\omega).
\end{align}
Multiplying by itself and taking the average, we get:

\begin{multline}
    \bar F_\theta(\omega)\avg{\hat U_\theta(\omega)\hat U_\phi(\omega') }
    \bar F_\phi(\omega') =
    \\ 
    \int\!\! \d^3\r\,\d^3 \r'\;
    \partial_\beta \bar v_\alpha(\r,\omega)
    \partial_\nu \bar v_\mu(\r'\!,\omega') 
    \Big\langle\!\hat s_\ab(\r,\omega) \hat s_\mn(\r'\!,\omega')\!\Big\rangle.
\end{multline}
Substituting the velocity field \eqref{stokes-flow} and the drag force \eqref{F-avg-def} into the equation above, we obtain:
\begin{multline}
    (6 \pi a)^2\eta'(\omega)\eta'(\omega')\;
    \bar U_\theta(\omega)\bar U_\phi(\omega')
    \;
    \avg{\hat U_\theta(\omega)\hat U_\phi(\omega') }
    \\
    =
    \bar U_\theta(\omega)\bar U_\phi(\omega') 
    \\
     \times \!\!\!\!\!\int\limits_{|\r|,|\r'|>a}\!\!\!\!\!\!
    \d^3 \r\,\d^3 \r'\;
    P_{\alpha \theta,\beta}(\r)P_{\mu \phi,\nu}(\r')
    \Big\langle\hat s_\ab(\r,\omega) \hat s_\mn(\r',\omega')\!\Big\rangle ,
\end{multline}
where we defined
\begin{align}
    P_{\alpha\gamma,\beta}(\r) = \frac{\partial}{\partial r_\beta }P_{\alpha\gamma}(\r).
\end{align}
Since this holds for any driven flow $\bar U_\alpha$, we conclude:
\begin{align}
    &(6 \pi a)^2\eta'(\omega)\eta'(\omega')
    \;
    \avg{\hat U_\theta(\omega)\hat U_\phi(\omega') }=
    \nonumber
    \\
    &= \!\!\!\int\limits_{|\r|,|\r'|>a}\!\!\!\!\!
     \d^3 \r\,\d^3 \r'\;
    P_{\alpha \theta,\beta}(\r)P_{\mu \phi,\nu}(\r')
    \,
    \Big\langle\hat s_\ab(\r,\omega) \hat s_\mn(\r',\omega')\Big\rangle 
    \nonumber
    \\
    &=\int\limits_{|\r|,|\r'|>a}
     \d^3 \r\,\d^3 \r'
    P_{\alpha \theta,\beta}(\r)P_{\mu \phi,\nu}(\r')
    \nonumber
    \\[-2ex]
    &\hspace{18ex}\times
    2\pi \delta (\r\!-\!\r')\delta(\omega\!+\!\omega')\rS^\text{partial}_{\ab\mn}(\omega)
    \nonumber
    \\
    &=2\pi \delta(\omega+\omega') \int\limits_{|\r|>a}
     \d^3 \r\;
    P_{\alpha \theta,\beta}(\r)P_{\mu \phi,\nu}(\r)
    \;
    \rS^\text{partial}_{\ab\mn}(\omega),
\end{align}
where we used \cref{wk-s}. 
Applying \cref{wk-SR}, we get:
\begin{align} 
    \la \hat U_\theta(\omega) \hat U_\phi(\omega') \ra =&
    \ii\omega\, \ii\omega' \la R_\theta(\omega) R_\phi(\omega') \ra 
    \nonumber
    \\
    =& \omega^2 \, 2\pi \delta(\omega+\omega') \rS^R_{\theta\phi}(\omega),
\end{align}
where $\omega' = -\omega$ due to the delta function.
Finally, for the position fluctuation spectrum, we obtain \cref{spectrum-reduction}:
\begin{align}
    \label{spectrum-reduction-sm}
    \rS^R_{\theta\phi}(\omega) = \frac{1+\omega^2/\ku^2}{(6\pi a \eta \omega)^2} \!\!\int\limits_{|\r|>a}\!\!\!
     \d^3 \r\;
    P_{\alpha \theta,\beta}(\r)P_{\mu \phi,\nu}(\r)
    \;\rS^\text{partial}_{\ab\mn}(\omega).
\end{align}
This equation provides a method to reduce the rank-4 tensor of stress fluctuations to the rank-2 tensor (matrix) of position fluctuations of the tracer particle. 
Below, we will provide the recipe for evaluating the integral in \cref{spectrum-reduction-sm}.

Using the results from \cref{sec-reduced-repr} and \cref{app-axis-invariant}, the partial stress tensor is expressed as \cref{decompose}, or more generally:
\begin{align}
    \label{s-decomp}
    \rS^\text{partial}_{\ab\mn}(\omega) = \sum_{i=0}^5 S^{(i)}(\omega) \T^{(i)}_{\ab\mn},
\end{align}
where the tensors $\T^{(i)}_\abmn$ are defined in \cref{Tm-def,2moretensors}. 
The exact coefficients $S^{(i)}(\omega)$ depend on the model, which is discussed in the subsequent section.

The decomposition \eqref{s-decomp} enables us to evaluate the integral separately:
\begin{equation}
\begin{aligned}
    \T^{(0)}_{\ab\mn}
     \int_{|\r|>a}\d^3 \r\, 
    P_{\alpha\theta,\beta}(\r)
    P_{\mu\phi,\nu}(\r) =& 3a\pi \;\delta_{\theta\phi},
    \\
    \T^{(1)}_{\ab\mn}
     \int_{|\r|>a}\d^3 \r\, 
    P_{\alpha\theta,\beta}(\r)
    P_{\mu\phi,\nu}(\r) =& \frac{3}{4} a\pi \;N_{\theta\phi},
    \\
    \T^{(2)}_{\ab\mn}
     \int_{|\r|>a}\d^3 \r\, 
    P_{\alpha\theta,\beta}(\r)
    P_{\mu\phi,\nu}(\r) =& \frac{2}{5} a\pi\;\delta_{\theta\phi} +  \frac{3}{7} a\pi\;N_{\theta\phi},
    \\
    \T^{(3)}_{\ab\mn}
     \int_{|\r|>a}\d^3 \r\, 
    P_{\alpha\theta,\beta}(\r)
    P_{\mu\phi,\nu}(\r) =& \frac{3}{5}a\pi \;\epsilon_{\theta\phi\kappa}n_\kappa,
    \\
    \T^{(4)}_{\ab\mn}
     \int_{|\r|>a}\d^3 \r\, 
    P_{\alpha\theta,\beta}(\r)
    P_{\mu\phi,\nu}(\r) =& 3a\pi \;\epsilon_{\theta\phi\kappa}n_\kappa.    
\end{aligned}
\end{equation}
Consequently, the spectrum of the position fluctuations is given by:
\begin{align}
     \rS^{R}_{\theta\phi}(\omega) =& \frac{1+\omega^2/\ku^2}{(6\pi a \eta \,\omega)^2} 
     \pi a
     \Big[
     \left(
     3 S^{(0)}(\omega)+\tfrac{2}{5}S^{(2)}(\omega) \right)\delta_{\theta\phi}
     \nonumber 
     \\
     &+\left(
     \tfrac{3}{4} S^{(1)}(\omega)+\tfrac{3}{7}S^{(2)}(\omega) \right) N_{\theta\phi}
     \nonumber
     \\
     &+
     \left(\tfrac{3}{5} S^{(3)}(\omega)+3 S^{(4)}(\omega)\right)\epsilon_{\theta\phi\kappa}n_\kappa
     \Big].
     \label{F-spectrum}
\end{align}

In the following section, we consider the partial stress spectrum $\rS^\text{partial}_{\ab\mn}(\omega)$ associated with the active gel model and explicitly calculate the functions $S^{(i)}(\omega)$ in \cref{s-decomp}. Moreover, we show Fluctuation-Dissipation relation \eqref{FDT} for the tracer particle, when the detailed balance is maintained in the gel.

\subsection{Orientation-selective activity}
\label{app-tracer-special}

In \cref{sec-specific}, we consider a simple form of activity given in \cref{activity-specific}. Here, we compute the integral \eqref{spectrum-reduction-sm} for this specific activity.

We find the partial stress spectrum $\rS^\text{partial}_{\ab\mn}(\omega)$ from \cref{partial-stress-spectrum-def} by additionally setting
\begin{equation}
\begin{aligned}
    &\Omega_0 = \Omega_0,\quad \Omega^{q}_\ab = \Omega_0 Q_\ab,\quad
    \\
    &\Omega^{uu}_\abmn = A\Omega_0 \Tiso_\abmn, \quad
    \Omega^{qq}_\abmn = \Omega_0 Q_\ab Q_\mn,
\end{aligned}
\end{equation}
consistent with \cref{Omega-uq-specific,Omega-uuqq-specific}.
We identify the coefficients in \cref{s-decomp}:
\begin{align*}
    &  S^{(0)}(\omega) \!\!=\!\!  \big[
    \tfrac{\kBT \mu (\Z-\Omega_0)}{\ku(1+\Z)}
    \!+\!
    \tfrac{A\mu^2\Omega_0}{\rho\ku(1+\Z)}
    \big] \tfrac{2}{1+\omega^2/\ku^2},
    \vphantom{\Bigg[}
    \\
    &  S^{(2)}(\omega) \!\!=\!\! \big[\tfrac{\Z-\Omega_0}{\rho\ku\Z(1+\Z)}
    \!+\!
    \tfrac{\Omega_0}{\rho\ku\Z(1+\Z)}\tfrac{\ku^2+\omega^2}{(1+\Z)^2\ku^2+\omega^2} 
    \big]
    \tfrac{2D^2 \Omega_0 |\Q|^2}{1+\omega^2/\ku^2},
\end{align*}
where $|\Q|$ denotes the strength of the nematic order.
The rest of the coefficients vanish. Using \cref{F-spectrum}, we obtain the spectrum of the position fluctuations:
\begin{multline}
\rS^{R}_{\theta\phi}(\omega) = \frac{1+\omega^2/\ku^2}{(6\pi a \eta \,\omega)^2} 
     3\pi a
     \Big[
     S^{(0)}(\omega)\delta_{\theta\phi} 
     \\
     +S^{(2)}(\omega)\left(\tfrac{2}{15}\delta_{\theta\phi} + \tfrac{1}{7} N_{\theta\phi}\right)
     \Big],
\end{multline}
which simplifies to \cref{SR-final}.
}

\bibliography{literature.bib}

@book{kim2013microhydrodynamics,
  title={Microhydrodynamics: principles and selected applications},
  author={Kim, Sangtae and Karrila, Seppo J},
  year={2013},
  publisher={Butterworth-Heinemann}
}

@article{basu2008thermal,
  title={Thermal and non-thermal fluctuations in active polar gels},
  author={Basu, A and Joanny, JF and J{\"u}licher, F and Prost, J},
  journal={The European Physical Journal E},
  volume={27},
  number={2},
  pages={149--160},
  year={2008},
  publisher={Springer},
  doi={10.1140/epje/i2008-10364-9},
}

@article{narinder2026,
  title = {Time Irreversibility, Entropy Production, and Effective Temperature Are Independently Regulated in the Actin Cortex of Living Cells},
  author = {Narinder, N and Fischer-Friedrich, Elisabeth},
  journal = {Phys. Rev. X},
  volume = {16},
  issue = {1},
  pages = {011007},
  numpages = {12},
  year = {2026},
  month = {Jan},
  publisher = {American Physical Society},
  doi = {10.1103/5zyn-kgs3},
  url = {https://link.aps.org/doi/10.1103/5zyn-kgs3}
}

@article{mulla2022,
  title={Weak catch bonds make strong networks},
  author={Mulla, Yuval and Avellaneda, Mario J and Roland, Antoine and Baldauf, Lucia and Jung, Wonyeong and Kim, Taeyoon and Tans, Sander J and Koenderink, Gijsje H},
  journal={Nature materials},
  volume={21},
  number={9},
  pages={1019--1023},
  year={2022},
  publisher={Nature Publishing Group UK London},
  doi={10.1038/s41563-022-01288-0},
}

@article{han2021,
  title={Fluctuating hydrodynamics of chiral active fluids},
  author={Han, Ming and Fruchart, Michel and Scheibner, Colin and Vaikuntanathan, Suriyanarayanan and De Pablo, Juan J and Vitelli, Vincenzo},
  journal={Nature Physics},
  volume={17},
  number={11},
  pages={1260--1269},
  year={2021},
  publisher={Nature Publishing Group UK London},
  doi={10.1038/s41567-021-01360-7},
}

@article{solon2022,
doi = {10.1088/1751-8121/ac5d82},
year = {2022},
month = {apr},
publisher = {IOP Publishing},
volume = {55},
number = {18},
pages = {184002},
author = {Solon, Alexandre and Horowitz, Jordan M},
title = {On the {Einstein} relation between mobility and diffusion coefficient in an active bath},
journal = {Journal of Physics A: Mathematical and Theoretical}
}

@article{gnesotto2018,
  title={Broken detailed balance and non-equilibrium dynamics in living systems: a review},
  author={Gnesotto, Federico S and Mura, Federica and Gladrow, Jannes and Broedersz, Chase P},
  journal={Reports on Progress in Physics},
  volume={81},
  number={6},
  pages={066601},
  year={2018},
  publisher={IOP Publishing},
  doi={10.1088/1361-6633/aab3ed},
}

@article{ben-isaac2011,
  title={Effective temperature of red-blood-cell membrane fluctuations},
  author={Ben-Isaac, Eyal and Park, YongKeun and Popescu, Gabriel and Brown, Frank LH and Gov, Nir S and Shokef, Yair},
  journal={Physical review letters},
  volume={106},
  number={23},
  pages={238103},
  year={2011},
  publisher={APS},
  doi={10.1103/physrevlett.106.238103},
}

@article{levine2009,
  title={The mechanics and fluctuation spectrum of active gels},
  author={Levine, Alex J and MacKintosh, FC},
  journal={The Journal of Physical Chemistry B},
  volume={113},
  number={12},
  pages={3820--3830},
  year={2009},
  publisher={ACS Publications},
  doi={10.1021/jp808192w},
}

@article{mura2019,
  title={Mesoscopic non-equilibrium measures can reveal intrinsic features of the active driving},
  author={Mura, Federica and Gradziuk, Grzegorz and Broedersz, Chase P},
  journal={Soft Matter},
  volume={15},
  number={40},
  pages={8067--8076},
  year={2019},
  publisher={Royal Society of Chemistry},
  doi={10.1039/c9sm01169b},
}

@article{umeda2023,
  title={Activity-dependent glassy cell mechanics {II}: Nonthermal fluctuations under metabolic activity},
  author={Umeda, Katsuhiro and Nishizawa, Kenji and Nagao, Wataru and Inokuchi, Shono and Sugino, Yujiro and Ebata, Hiroyuki and Mizuno, Daisuke},
  journal={Biophysical journal},
  volume={122},
  number={22},
  pages={4395--4413},
  year={2023},
  publisher={Elsevier},
  doi={10.1016/j.bpj.2023.10.018},
}

@article{hurst2021,
  title={Intracellular softening and increased viscoelastic fluidity during division},
  author={Hurst, Sebastian and Vos, Bart E and Brandt, Matthias and Betz, Timo},
  journal={Nature Physics},
  volume={17},
  number={11},
  pages={1270--1276},
  year={2021},
  publisher={Nature Publishing Group UK London},
  doi={10.1038/s41567-021-01368-z},
}

@article{toyota2011,
  title={Non-{Gaussian} athermal fluctuations in active gels},
  author={Toyota, Toshihiro and Head, David A and Schmidt, Christoph F and Mizuno, Daisuke},
  journal={Soft Matter},
  volume={7},
  number={7},
  pages={3234--3239},
  year={2011},
  publisher={Royal Society of Chemistry},
  doi={10.1039/c0sm00925c},
}

@article{bertrand2012,
  title={Active, motor-driven mechanics in a {DNA} gel},
  author={Bertrand, Olivier JN and Fygenson, Deborah Kuchnir and Saleh, Omar A},
  journal={Proceedings of the National Academy of Sciences},
  volume={109},
  number={43},
  pages={17342--17347},
  year={2012},
  publisher={National Academy of Sciences},
  doi={10.1073/pnas.1208732109},
}

@article{weber2012,
  title={Nonthermal {ATP}-dependent fluctuations contribute to the in vivo motion of chromosomal loci},
  author={Weber, Stephanie C and Spakowitz, Andrew J and Theriot, Julie A},
  journal={Proceedings of the National Academy of Sciences},
  volume={109},
  number={19},
  pages={7338--7343},
  year={2012},
  publisher={National Academy of Sciences},
  doi={10.1073/pnas.1119505109},
}

@article{chu2017,
  title={On the origin of shape fluctuations of the cell nucleus},
  author={Chu, Fang-Yi and Haley, Shannon C and Zidovska, Alexandra},
  journal={Proceedings of the National Academy of Sciences},
  volume={114},
  number={39},
  pages={10338--10343},
  year={2017},
  publisher={National Academy of Sciences},
  doi={10.1073/pnas.1702226114},
}

@article{plotnikov2012,
  title={Force fluctuations within focal adhesions mediate {ECM}-rigidity sensing to guide directed cell migration},
  author={Plotnikov, Sergey V and Pasapera, Ana M and Sabass, Benedikt and Waterman, Clare M},
  journal={Cell},
  volume={151},
  number={7},
  pages={1513--1527},
  year={2012},
  publisher={Elsevier},
  doi={10.1016/j.cell.2012.11.034},
}

@article{ebata2023,
  title={Activity-dependent glassy cell mechanics {I}: Mechanical properties measured with active microrheology},
  author={Ebata, Hiroyuki and Umeda, Katsuhiro and Nishizawa, Kenji and Nagao, Wataru and Inokuchi, Shono and Sugino, Yujiro and Miyamoto, Takafumi and Mizuno, Daisuke},
  journal={Biophysical journal},
  volume={122},
  number={10},
  pages={1781--1793},
  year={2023},
  publisher={Elsevier},
  doi={10.1016/j.bpj.2023.04.011},
}

@article{smith2015,
  title={Nonthermal fluctuations of the mitotic spindle},
  author={Smith, Kevin and Griffin, Brian and Byrd, Henry and MacKintosh, FC and Kilfoil, Maria L},
  journal={Soft Matter},
  volume={11},
  number={22},
  pages={4396--4401},
  year={2015},
  publisher={Royal Society of Chemistry},
  doi={10.1039/c5sm00149h},
}

@article{das2024,
  title={Chromatin remodeling due to transient-link-and-pass activity enhances subnuclear dynamics},
  author={Das, Rakesh and Sakaue, Takahiro and Shivashankar, GV and Prost, Jacques and Hiraiwa, Tetsuya},
  journal={Physical Review Letters},
  volume={132},
  number={5},
  pages={058401},
  year={2024},
  publisher={APS},
  doi={10.1103/physrevlett.132.058401},
}

@article{janes2022,
  title={First-principle coarse-graining framework for scale-free bell-like association and dissociation rates in thermal and active systems},
  author={Janes, Josip Augustin and Monzel, Cornelia and Schmidt, Daniel and Merkel, Rudolf and Seifert, Udo and Sengupta, Kheya and Smith, Ana-Suncana},
  journal={Physical Review X},
  volume={12},
  number={3},
  pages={031030},
  year={2022},
  publisher={APS},
  doi={10.1103/physrevx.12.031030},
}

@article{turlier2016,
  title={Equilibrium physics breakdown reveals the active nature of red blood cell flickering},
  author={Turlier, Herv{\'e} and Fedosov, Dmitry A and Audoly, Basile and Auth, Thorsten and Gov, Nir S and Sykes, C{\'e}cile and Joanny, J-F and Gompper, Gerhard and Betz, Timo},
  journal={Nature physics},
  volume={12},
  number={5},
  pages={513--519},
  year={2016},
  publisher={Nature Publishing Group UK London},
  doi={10.1038/nphys3621},
}

@article{alert2016,
  title={Bleb nucleation through membrane peeling},
  author={Alert, Ricard and Casademunt, Jaume},
  journal={Physical review letters},
  volume={116},
  number={6},
  pages={068101},
  year={2016},
  publisher={APS},
  doi={10.1103/physrevlett.116.068101},
}

@article{shu2024,
  title={Mesoscale molecular assembly is favored by the active, crowded cytoplasm},
  author={Shu, Tong and Mitra, Gaurav and Alberts, Jonathan and Viana, Matheus Palhares and Levy, Emmanuel and Hocky, Glen M and Holt, Liam J},
  journal={Biophysical Journal},
  volume={123},
  number={3},
  pages={493a},
  year={2024},
  publisher={Elsevier},
  doi={10.1016/j.bpj.2023.11.2988},
}

@article{moriel2023,
title = {Characteristic energy scales of active fluctuations in adherent cells},
journal = {Biophysical Reports},
volume = {3},
number = {1},
pages = {100099},
year = {2023},
issn = {2667-0747},
doi = {https://doi.org/10.1016/j.bpr.2022.100099},
author = {Avraham Moriel and Haguy Wolfenson and Eran Bouchbinder},
}

@article{maes2020,
  title={Fluctuating motion in an active environment},
  author={Maes, Christian},
  journal={Physical Review Letters},
  volume={125},
  number={20},
  pages={208001},
  year={2020},
  publisher={APS},
  doi={10.1103/physrevlett.125.208001},
}

@article{seyforth2022,
  title={Nonequilibrium fluctuations and nonlinear response of an active bath},
  author={Seyforth, Hunter and Gomez, Mauricio and Rogers, W Benjamin and Ross, Jennifer L and Ahmed, Wylie W},
  journal={Physical review research},
  volume={4},
  number={2},
  pages={023043},
  year={2022},
  publisher={APS},
  doi={10.1103/physrevresearch.4.023043},
}

@article{massana-cid2024,
  title={Multiple temperatures and melting of a colloidal active crystal},
  author={Massana-Cid, Helena and Maggi, Claudio and Gnan, Nicoletta and Frangipane, Giacomo and Di Leonardo, Roberto},
  journal={Nature Communications},
  volume={15},
  number={1},
  pages={6574},
  year={2024},
  publisher={Nature Publishing Group UK London},
  doi={10.1038/s41467-024-50937-2},
}

@article{majhi2025,
  title={Decoding active force fluctuations from spatial trajectories of active systems},
  author={Majhi, Anisha and Das, Biswajit and Gupta, Subhadeep and Ranjan, Anand Dev and Mallick, Amirul Islam and Paul, Shuvojit and Banerjee, Ayan},
  journal={Physical Review E},
  volume={111},
  number={6},
  pages={065411},
  year={2025},
  publisher={APS},
  doi={10.1103/pqrl-splf},
}

@article{vilallobos-concha2025,
  title={Active bacterial baths in droplets},
  author={Villalobos-Concha, Cristian and Liu, Zhengyang and Ramos, Gabriel and Goral, Martyna and Lindner, Anke and L{\'o}pez-Le{\'o}n, Teresa and Cl{\'e}ment, Eric and Soto, Rodrigo and Cordero, Mar{\'\i}a Luisa},
  journal={Proceedings of the National Academy of Sciences},
  volume={122},
  number={31},
  pages={e2426096122},
  year={2025},
  publisher={National Academy of Sciences},
  doi={10.1073/pnas.2426096122},
}

@article{frechette2025,
title = {Active-noise-induced dynamic clustering of passive colloidal particles},
journal = {Newton},
volume = {1},
number = {7},
pages = {100167},
year = {2025},
issn = {2950-6360},
doi = {https://doi.org/10.1016/j.newton.2025.100167},
author = {Layne B. Frechette and Aparna Baskaran and Michael F. Hagan},
}

@article{seara2021,
  title={Irreversibility in dynamical phases and transitions},
  author={Seara, Daniel S and Machta, Benjamin B and Murrell, Michael P},
  journal={Nature communications},
  volume={12},
  number={1},
  pages={392},
  year={2021},
  publisher={Nature Publishing Group UK London},
  doi={10.1038/s41467-020-20281-2},
}

@article{sorkin2024,
  title={Second law of thermodynamics without {Einstein} relation},
  author={Sorkin, Benjamin and Diamant, Haim and Ariel, Gil and Markovich, Tomer},
  journal={Physical Review Letters},
  volume={133},
  number={26},
  pages={267101},
  year={2024},
  publisher={APS},
  doi={10.1103/physrevlett.133.267101},
}

@article{johnsrud2025,
  title={Fluctuation dissipation relations for active field theories},
  author={Johnsrud, Martin Kj{\o}llesdal and Golestanian, Ramin},
  journal={Physical Review Research},
  volume={7},
  number={3},
  pages={L032053},
  year={2025},
  publisher={APS},
  doi={10.1103/xx4z-lj5c},
}

@article{kirkpatrick2025,
  title={Fluctuation-response relation in nonequilibrium systems and active matter},
  author={Kirkpatrick, TR and Belitz, D},
  journal={Physical Review E},
  volume={111},
  number={1},
  pages={014102},
  year={2025},
  publisher={APS},
  doi={10.1103/physreve.111.014102},
}

@article{basu2008,
  title={Thermal and non-thermal fluctuations in active polar gels},
  author={Basu, A and Joanny, JF and J{\"u}licher, F and Prost, J},
  journal={The European Physical Journal E},
  volume={27},
  number={2},
  pages={149--160},
  year={2008},
  publisher={Springer},
  doi={10.1140/epje/i2008-10364-9},
}

@article{Basu2012,
   author = {Abhik Basu and Jean-François Joanny and Frank Jülicher and Jacques Prost},
   doi = {10.1088/1367-2630/14/11/115001},
   issn = {1367-2630},
   issue = {11},
   journal = {New Journal of Physics},
   month = {11},
   pages = {115001},
   publisher = {IOP Publishing},
   title = {Anomalous behavior of the diffusion coefficient in thin active films},
   volume = {14},
   url = {http://iopscience.iop.org/1367-2630/14/11/115001/article/},
   year = {2012}
}

@article{caprini2021,
  title={Fluctuation--dissipation relations in active matter systems},
  author={Caprini, Lorenzo and Puglisi, Andrea and Sarracino, Alessandro},
  journal={Symmetry},
  volume={13},
  number={1},
  pages={81},
  year={2021},
  publisher={MDPI},
  doi={10.3390/sym13010081},
}

@article{fodor2016how,
  title={How far from equilibrium is active matter?},
  author={Fodor, {\'E}tienne and Nardini, Cesare and Cates, Michael E and Tailleur, Julien and Visco, Paolo and Van Wijland, Fr{\'e}d{\'e}ric},
  journal={Physical review letters},
  volume={117},
  number={3},
  pages={038103},
  year={2016},
  publisher={APS},
  doi={10.1103/physrevlett.117.038103},
}

@article{kanazawa2020,
  title={Loopy L{\'e}vy flights enhance tracer diffusion in active suspensions},
  author={Kanazawa, Kiyoshi and Sano, Tomohiko G and Cairoli, Andrea and Baule, Adrian},
  journal={Nature},
  volume={579},
  number={7799},
  pages={364--367},
  year={2020},
  publisher={Nature Publishing Group UK London},
  doi={https://doi.org/10.1038/s41586-020-2086-2},
}

@article{mackintosh2008,
  title={Nonequilibrium mechanics and dynamics of motor-activated gels},
  author={MacKintosh, Fred C and Levine, Alex J},
  journal={Physical review letters},
  volume={100},
  number={1},
  pages={018104},
  year={2008},
  publisher={APS},
  doi={10.1103/physrevlett.100.018104},
}

@article{mizuno2007,
  title={Nonequilibrium mechanics of active cytoskeletal networks},
  author={Mizuno, Daisuke and Tardin, Catherine and Schmidt, Christoph F and MacKintosh, Frederik C},
  journal={Science},
  volume={315},
  number={5810},
  pages={370--373},
  year={2007},
  publisher={American Association for the Advancement of Science},
  doi={10.1126/science.1134404},
}

@article{rupprecht2018,
  title={Maximal fluctuations of confined actomyosin gels: dynamics of the cell nucleus},
  author={Rupprecht, J-F and Singh Vishen, A and Shivashankar, GV and Rao, Madan and Prost, Jacques},
  journal={Physical Review Letters},
  volume={120},
  number={9},
  pages={098001},
  year={2018},
  publisher={APS},
  doi={10.1103/physrevlett.120.098001},
}

@article{ariga2021,
  title={Noise-induced acceleration of single molecule kinesin-1},
  author={Ariga, Takayuki and Tateishi, Keito and Tomishige, Michio and Mizuno, Daisuke},
  journal={Physical review letters},
  volume={127},
  number={17},
  pages={178101},
  year={2021},
  publisher={APS},
  doi={10.1103/physrevlett.127.178101},
}

@article{wolgemuth2020,
  title={Active random forces can drive differential cellular positioning and enhance motor-driven transport},
  author={Wolgemuth, Charles W and Sun, Sean X},
  journal={Molecular Biology of the Cell},
  volume={31},
  number={20},
  pages={2283--2288},
  year={2020},
  publisher={The American Society for Cell Biology},
  doi={10.1091/mbc.e19-11-0629},
}

@article{ezber2020,
  title={Dynein harnesses active fluctuations of microtubules for faster movement},
  author={Ezber, Yasin and Belyy, Vladislav and Can, Sinan and Yildiz, Ahmet},
  journal={Nature physics},
  volume={16},
  number={3},
  pages={312--316},
  year={2020},
  publisher={Nature Publishing Group UK London},
  doi={10.1038/s41567-019-0757-4},
}

@article{fakhri2014,
  author    = {Fakhri, Nikta and Wessel, Andreas D. and Willms, Chris and Pasquali, Matteo and Klopfenstein, Dietmar R. and MacKintosh, Fred C. and Schmidt, Christoph F.},
  title     = {High-resolution mapping of intracellular fluctuations using carbon nanotubes},
  journal   = {Science},
  year      = {2014},
  month     = {May},
  volume    = {344},
  number    = {6187},
  pages     = {1031--1035},
  doi       = {10.1126/science.1250170},
  pmid      = {24876498},
  url={https://doi.org/10.1126/science.1250170},
}

@article{guo2014,
  title={Probing the stochastic, motor-driven properties of the cytoplasm using force spectrum microscopy},
  author={Guo, Ming and Ehrlicher, Allen J and Jensen, Mikkel H and Renz, Malte and Moore, Jeffrey R and Goldman, Robert D and Lippincott-Schwartz, Jennifer and Mackintosh, Frederick C and Weitz, David A},
  journal={Cell},
  volume={158},
  number={4},
  pages={822--832},
  year={2014},
  publisher={Elsevier},
  doi={10.1016/j.cell.2014.06.051},
}

@article{brangwynne2009,
  title={Intracellular transport by active diffusion},
  author={Brangwynne, Clifford P and Koenderink, Gijsje H and MacKintosh, Frederick C and Weitz, David A},
  journal={Trends in cell biology},
  volume={19},
  number={9},
  pages={423--427},
  year={2009},
  publisher={Elsevier},
  doi={10.1016/j.tcb.2009.04.004},
}

@article{matevosyan2021nonequilibrium,
  title={Nonequilibrium, weak-field-induced cyclotron motion: A mechanism for magnetobiology},
  author={Matevosyan, Ashot and Allahverdyan, Armen E},
  journal={Physical Review E},
  volume={104},
  number={6},
  pages={064407},
  year={2021},
  publisher={APS},
  doi={10.1103/physreve.104.064407},
}

@article{abbasi2023non,
  title={Non-markovian modeling of nonequilibrium fluctuations and dissipation in active viscoelastic biomatter},
  author={Abbasi, Amir and Netz, Roland R and Naji, Ali},
  journal={Physical Review Letters},
  volume={131},
  number={22},
  pages={228202},
  year={2023},
  publisher={APS},
  doi={10.1103/physrevlett.131.228202},
}

@article{brugues2014physical,
  title={Physical basis of spindle self-organization},
  author={Brugu{\'e}s, Jan and Needleman, Daniel},
  journal={Proceedings of the National Academy of Sciences},
  volume={111},
  number={52},
  pages={18496--18500},
  year={2014},
  publisher={National Academy of Sciences},
  doi={10.1073/pnas.1409404111},
}

@article{ramaswamy2010mechanics,
  title={The mechanics and statistics of active matter},
  author={Ramaswamy, Sriram},
  journal={Annu. Rev. Condens. Matter Phys.},
  volume={1},
  number={1},
  pages={323--345},
  year={2010},
  publisher={Annual Reviews},
  doi={10.1146/annurev-conmatphys-070909-104101},
}

@book{grimmett2020,
  title={Probability and random processes},
  author={Grimmett, Geoffrey and Stirzaker, David},
  year={2020},
  publisher={Oxford university press},
}

@article{lubensky2002,
  title={Symmetries and elasticity of nematic gels},
  author={Lubensky, TC and Mukhopadhyay, Ranjan and Radzihovsky, Leo and Xing, Xiangjun},
  journal={Physical review E},
  volume={66},
  number={1},
  pages={011702},
  year={2002},
  publisher={APS},
  doi={10.1103/physreve.66.011702},
}

@article{walcott2010,
  title={A mechanical model of actin stress fiber formation and substrate elasticity sensing in adherent cells},
  author={Walcott, Sam and Sun, Sean X},
  journal={Biophysical Journal},
  volume={98},
  number={3},
  pages={365a},
  year={2010},
  publisher={Elsevier},
  doi={10.1016/j.bpj.2009.12.1969},
}

@article{parry2014,
  title={The bacterial cytoplasm has glass-like properties and is fluidized by metabolic activity},
  author={Parry, Bradley R and Surovtsev, Ivan V and Cabeen, Matthew T and O-Hern, Corey S and Dufresne, Eric R and Jacobs-Wagner, Christine},
  journal={Cell},
  volume={156},
  number={1},
  pages={183--194},
  year={2014},
  publisher={Elsevier},
  doi={10.1016/j.cell.2013.11.028},
}

@article{marchetti2013,
  title={Hydrodynamics of soft active matter},
  author={Marchetti, M Cristina and Joanny, Jean-Fran{\c{c}}ois and Ramaswamy, Sriram and Liverpool, Tanniemola B and Prost, Jacques and Rao, Madan and Simha, R Aditi},
  journal={Reviews of modern physics},
  volume={85},
  number={3},
  pages={1143},
  year={2013},
  publisher={APS},
  doi={10.1103/revmodphys.85.1143},
}

@article{julicher2018,
  title={Hydrodynamic theory of active matter},
  author={J{\"u}licher, Frank and Grill, Stephan W and Salbreux, Guillaume},
  journal={Reports on Progress in Physics},
  volume={81},
  number={7},
  pages={076601},
  year={2018},
  publisher={IOP Publishing},
  doi={10.1088/1361-6633/aab6bb},
}

@article{prost2015,
  title={Active gel physics},
  author={Prost, Jacques and J{\"u}licher, Frank and Joanny, Jean-Fran{\c{c}}ois},
  journal={Nature physics},
  volume={11},
  number={2},
  pages={111--117},
  year={2015},
  publisher={Nature Publishing Group UK London},
  doi={10.1038/nphys3224},
}

@article{fodor2015,
  title={Activity-driven fluctuations in living cells},
  author={Fodor, {\'E} and Guo, M and Gov, NS and Visco, P and Weitz, DA and Van Wijland, F},
  journal={Europhysics Letters},
  volume={110},
  number={4},
  pages={48005},
  year={2015},
  publisher={IOP Publishing},
  doi={10.1209/0295-5075/110/48005},
}

@article{fodor2016,
  title={Nonequilibrium dissipation in living oocytes},
  author={Fodor, {\'E}tienne and Ahmed, Wylie W and Almonacid, Maria and Bussonnier, Matthias and Gov, Nir S and Verlhac, M-H and Betz, Timo and Visco, Paolo and van Wijland, Fr{\'e}d{\'e}ric},
  journal={Europhysics Letters},
  volume={116},
  number={3},
  pages={30008},
  year={2016},
  publisher={IOP Publishing},
  doi={10.1209/0295-5075/116/30008},
}

@article{ahmed2018,
  title={Active mechanics reveal molecular-scale force kinetics in living oocytes},
  author={Ahmed, Wylie W and Fodor, {\'E}tienne and Almonacid, Maria and Bussonnier, Matthias and Verlhac, Marie-H{\'e}l{\`e}ne and Gov, Nir and Visco, Paolo and van Wijland, Fr{\'e}d{\'e}ric and Betz, Timo},
  journal={Biophysical journal},
  volume={114},
  number={7},
  pages={1667--1679},
  year={2018},
  publisher={Elsevier},
  doi={10.1016/j.bpj.2018.02.009},
}

@article{lau2003,
  title={Microrheology, stress fluctuations, and active behavior of living cells},
  author={Lau, Andy WC and Hoffman, Brenton D and Davies, A and Crocker, John C and Lubensky, Thomas C},
  journal={Physical review letters},
  volume={91},
  number={19},
  pages={198101},
  year={2003},
  publisher={APS},
  doi={10.1103/physrevlett.91.198101},
}

@article{bacanu2023,
  title={Inferring scale-dependent non-equilibrium activity using carbon nanotubes},
  author={Bacanu, Alexandru and Pelletier, James F and Jung, Yoon and Fakhri, Nikta},
  journal={Nature nanotechnology},
  volume={18},
  number={8},
  pages={905--911},
  year={2023},
  publisher={Nature Publishing Group UK London},
  doi={10.1038/s41565-023-01395-2},
}

@article{muenker2024eLife,
  author    = {Muenker, Till M. and Vos, Bart E. and Betz, Timo},
  title     = {Intracellular mechanical fingerprint reveals cell type specific mechanical tuning},
  journal   = {eLife},
  year      = {2024},
  month     = {May},
  publisher = {eLife Sciences Publications, Ltd},
  doi       = {10.7554/eLife.97416.1},
}

@article{muenker2024natmat,
  title={Accessing activity and viscoelastic properties of artificial and living systems from passive measurement},
  author={Muenker, Till M and Knotz, Gabriel and Kr{\"u}ger, Matthias and Betz, Timo},
  journal={Nature Materials},
  volume={23},
  number={9},
  pages={1283--1291},
  year={2024},
  publisher={Nature Publishing Group UK London},
  doi={10.1038/s41563-024-01957-2},
}

@article{bernheim2018,
  title={Living matter: mesoscopic active materials},
  author={Bernheim-Groswasser, Anne and Gov, Nir S and Safran, Samuel A and Tzlil, Shelly},
  journal={Advanced Materials},
  volume={30},
  number={41},
  pages={1707028},
  year={2018},
  publisher={Wiley Online Library},
  doi={10.1002/adma.201707028},
}

@article{julicher1997,
  title={Modeling molecular motors},
  author={J{\"u}licher, Frank and Ajdari, Armand and Prost, Jacques},
  journal={Reviews of Modern Physics},
  volume={69},
  number={4},
  pages={1269},
  year={1997},
  publisher={APS},
  doi={10.1103/revmodphys.69.1269},
}

@article{chen2020,
  title={Motor-free contractility in active gels},
  author={Chen, Sihan and Markovich, Tomer and MacKintosh, Fred C},
  journal={Physical Review Letters},
  volume={125},
  number={20},
  pages={208101},
  year={2020},
  publisher={APS},
  doi={10.1103/physrevlett.125.208101},
}

@article{hennig2020,
  title={Stick-slip dynamics of cell adhesion triggers spontaneous symmetry breaking and directional migration of mesenchymal cells on one-dimensional lines},
  author={Hennig, K and Wang, I and Moreau, P and Valon, L and DeBeco, S and Coppey, M and Miroshnikova, YA and Albiges-Rizo, C and Favard, Cyril and Voituriez, R and others},
  journal={Science Advances},
  volume={6},
  number={1},
  pages={eaau5670},
  year={2020},
  publisher={American Association for the Advancement of Science},
  doi={10.1126/sciadv.aau5670},
}

@article{dalcengio2021,
  title={Fluctuation--dissipation relations in the absence of detailed balance: formalism and applications to active matter},
  author={Dal Cengio, Sara and Levis, Demian and Pagonabarraga, Ignacio},
  journal={Journal of Statistical Mechanics: Theory and Experiment},
  volume={2021},
  number={4},
  pages={043201},
  year={2021},
  publisher={IOP Publishing},
  doi={10.1088/1742-5468/abee22},
}

@article{dalcengio2019,
  title={Linear response theory and {Green-Kubo} relations for active matter},
  author={Dal Cengio, Sara and Levis, Demian and Pagonabarraga, Ignacio},
  journal={Physical review letters},
  volume={123},
  number={23},
  pages={238003},
  year={2019},
  publisher={APS},
  doi={10.1103/physrevlett.123.238003},
}

@article{bedeaux1974,
  title={Brownian motion and fluctuating hydrodynamics},
  author={Bedeaux, D and Mazur, P},
  journal={Physica},
  volume={76},
  number={2},
  pages={247--258},
  year={1974},
  publisher={Elsevier},
  doi={10.1016/0031-8914(74)90198-0},
}

@article{bedeaux1977,
  title={Brownian motion and fluctuating hydrodynamics {II}; A fluctuation-dissipation theorem for the slip coefficient},
  author={Bedeaux, D and Albano, AM and Mazur, P},
  journal={Physica A: Statistical Mechanics and its Applications},
  volume={88},
  number={3},
  pages={574--582},
  year={1977},
  publisher={Elsevier},
  doi={10.1016/0378-4371(77)90128-5},
}

@book{landau-fluid,
  author    = {Landau, Lev D. and Lifshitz, Evgenii M.},
  title     = {Fluid Mechanics},
  series    = {Course of Theoretical Physics},
  volume    = {6},
  publisher = {Pergamon Press},
  year      = {1987},
  url={https://phys.au.dk/~srf/hydro/Landau+Lifschitz.pdf}
}

@book{landau-stat,
  author    = {Landau, Lev D. and Lifshitz, Evgenii M.},
  title     = {Statistical Physics, Part 1},
  series    = {Course of Theoretical Physics},
  volume    = {5},
  publisher = {Pergamon Press},
  year      = {1980},
  url={https://perso.crans.org/sylvainrey/Biblio%20Physique/Physique/Physique%20statistique/%5BLandau%2C%20Lifshitz%5D%2005%20Statistical%20Physics%20Part.1.pdf}
}

@article{paoluzzi2024,
  title={Noise-Induced Phase Separation and Time Reversal Symmetry Breaking in Active Field Theories Driven by Persistent Noise},
  author={Paoluzzi, Matteo and Levis, Demian and Crisanti, Andrea and Pagonabarraga, Ignacio},
  journal={Physical Review Letters},
  volume={133},
  number={11},
  pages={118301},
  year={2024},
  publisher={APS},
  doi={10.1103/physrevlett.133.118301},
}

@article{mabillard2023,
  title={Heat fluctuations in chemically active systems},
  author={Mabillard, Jo{\"e}l and Weber, Christoph A and J{\"u}licher, Frank},
  journal={Physical Review E},
  volume={107},
  number={1},
  pages={014118},
  year={2023},
  publisher={APS},
  doi={10.1103/physreve.107.014118},
}

@article{oriola2017,
  title={Fluidization and active thinning by molecular kinetics in active gels},
  author={Oriola, David and Alert, Ricard and Casademunt, Jaume},
  journal={Physical Review Letters},
  volume={118},
  number={8},
  pages={088002},
  year={2017},
  publisher={APS},
  doi={10.1103/physrevlett.118.088002},
}

@article{kruse2005generic,
  title={Generic theory of active polar gels: a paradigm for cytoskeletal dynamics},
  author={Kruse, Karsten and Joanny, Jean-Francois and J{\"u}licher, Frank and Prost, Jacques and Sekimoto, Ken},
  journal={The European Physical Journal E},
  volume={16},
  pages={5--16},
  year={2005},
  publisher={Springer},
  doi={10.1140/epje/e2005-00002-5},
}

@book{mazur,
  title={Non-equilibrium thermodynamics},
  author={De Groot, Sybren Ruurds and Mazur, Peter},
  year={2013},
  publisher={Dover Publications},
  url={https://pierre.ag.gerard.web.ulb.be/textbooks/books/Nonequilibrium_thermodynamics.pdf}
}

@incollection{risken,
  title={{Fokker-Planck} equation},
  author={Risken, Hannes},
  booktitle={The Fokker-Planck Equation},
  pages={63--95},
  year={1996},
  publisher={Springer},
}

@article{turlier2019,
    title = {{Unveiling the Active Nature of Living-Membrane Fluctuations and Mechanics}},
    year = {2019},
    journal = {Annual Review of Condensed Matter Physics},
    author = {Turlier, Hervé and Betz, Timo},
    number = {1},
    month = {3},
    pages = {213--232},
    volume = {10},
    publisher = {Annual Reviews},
    url = {https://www.annualreviews.org/doi/10.1146/annurev-conmatphys-031218-013757},
    doi = {10.1146/annurev-conmatphys-031218-013757},
    issn = {1947-5454}
}

@article{almonacid2015nucleus-positioning,
  title={Active diffusion positions the nucleus in mouse oocytes},
  author={Almonacid, Maria and Ahmed, Wylie W and Bussonnier, Matthias and Mailly, Philippe and Betz, Timo and Voituriez, Rapha{\"e}l and Gov, Nir S and Verlhac, Marie-H{\'e}l{\`e}ne},
  journal={Nature cell biology},
  volume={17},
  number={4},
  pages={470--479},
  year={2015},
  publisher={Nature Publishing Group UK London},
  doi={10.1038/ncb3131},
}

@article{brangwynne2008,
  title={Nonequilibrium microtubule fluctuations in a model cytoskeleton},
  author={Brangwynne, Clifford P and Koenderink, Gijsje H and MacKintosh, Frederick C and Weitz, David A},
  journal={Physical review letters},
  volume={100},
  number={11},
  pages={118104},
  year={2008},
  publisher={APS},
  doi={10.1103/physrevlett.100.118104},
}

@article{bursac2005,
  title={Cytoskeletal remodelling and slow dynamics in the living cell},
  author={Bursac, Predrag and Lenormand, Guillaume and Fabry, Ben and Oliver, Madavi and Weitz, David A and Viasnoff, Virgile and Butler, James P and Fredberg, Jeffrey J},
  journal={Nature materials},
  volume={4},
  number={7},
  pages={557--561},
  year={2005},
  publisher={Nature Publishing Group UK London},
  doi={10.1038/nmat1404},
}

@article{wilhelm2008,
  title={Out-of-equilibrium microrheology inside living cells},
  author={Wilhelm, Claire},
  journal={Physical review letters},
  volume={101},
  number={2},
  pages={028101},
  year={2008},
  publisher={APS},
  doi={10.1103/physrevlett.101.028101},
}

@article{nishizawa2017,
  title={Feedback-tracking microrheology in living cells},
  author={Nishizawa, Kenji and Bremerich, Marcel and Ayade, Heev and Schmidt, Christoph F and Ariga, Takayuki and Mizuno, Daisuke},
  journal={Science advances},
  volume={3},
  number={9},
  pages={e1700318},
  year={2017},
  publisher={American Association for the Advancement of Science},
  doi={10.1126/sciadv.1700318},
}

@article{nardini2017,
  title={Entropy production in field theories without time-reversal symmetry: quantifying the non-equilibrium character of active matter},
  author={Nardini, Cesare and Fodor, {\'E}tienne and Tjhung, Elsen and Van Wijland, Fr{\'e}d{\'e}ric and Tailleur, Julien and Cates, Michael E},
  journal={Physical Review X},
  volume={7},
  number={2},
  pages={021007},
  year={2017},
  publisher={APS},
  doi={10.1103/physrevx.7.021007},
}

@article{horowitz2020,
  title={Thermodynamic uncertainty relations constrain non-equilibrium fluctuations},
  author={Horowitz, Jordan M and Gingrich, Todd R},
  journal={Nature Physics},
  volume={16},
  number={1},
  pages={15--20},
  year={2020},
  publisher={Nature Publishing Group UK London},
  doi={10.1038/s41567-019-0702-6},
}

@article{mayer2010,
  title={Anisotropies in cortical tension reveal the physical basis of polarizing cortical flows},
  author={Mayer, Mirjam and Depken, Martin and Bois, Justin S and J{\"u}licher, Frank and Grill, Stephan W},
  journal={Nature},
  volume={467},
  number={7315},
  pages={617--621},
  year={2010},
  publisher={Nature Publishing Group UK London},
  doi={10.1038/nature09376},
}

\clearpage

\clearpage

\makeatletter
\global\let\l@section\@gobbletwo
\global\let\l@subsection\@gobbletwo
\global\let\l@subsubsection\@gobbletwo
\makeatother

\pagenumbering{arabic}

\widetext

\setcounter{section}{0}

\def\thesection{\S \arabic{section}}
\def\thesubsection{\S \arabic{section}.\arabic{subsection}}
\def\thesubsubsection{\S \arabic{section}.\arabic{subsection}.\arabic{subsubsection}}

\renewcommand{\theequation}{S\arabic{equation}}
\setcounter{equation}{0}
\numberwithin{equation}{section}
\renewcommand{\theequation}{S\arabic{section}.\arabic{equation}}

\section*{ Supplementary material:
\\
Solution of mesoscopic dynamics}
In the Supplementary Material, we present more direct approach to the problem.
We explicitly model the stochastic dynamics of each linker and solve it at the mesoscopic scale. We then utilize these solutions to determine the fluctuations at the hydrodynamic scale, which aligns with the results presented in the main text. 

This approach is useful because one can define any hydrodynamic variable out of mesoscopic variables and calculate the statistics. Additionally, this approach is more flexible to modifications of the model. However, this comes with a cost of more complex algebra and technical details, which hides the physical picture.

In \ref{sm-general}, we introduce the generalized notation for the mesoscopic variables, which allows us to express the model in a more compact form.

To facilitate this, we first define our system in terms of generalized thermodynamic variables and fluxes, thereby simplifying the algebra.

\section{The model in generalized notation}
\label{sm-general}
The crosslink is modeled as a mesoscopic thermodynamic system characterized by the thermodynamic variables $\u$ and $\q$. Let $\ix$ denote the vector that contains all the entries of the $d$-dimensional rank-2 tensors $\u$ and $\q$:
\begin{align}
    \label{x-def}
    \ix = \left(
    u_{11}\dots,u_{1d},
    u_{21}\dots,u_{2d},
    \dots, u_{dd},\;q_{11},\dots,q_{dd}\right)\equiv (\u,\q).
\end{align}
Thus, $\ix$ is a $2d^2$-dimensional vector. Therefore, the binding/unbinding dynamics \cref{kakd-new} is rewritten as:
\begin{align}
    \label{kakd-ix}
    {\kb(\ix)} = \ku \left( \ee^{-\beta f(\ix)/\rho} + \Omega(\ix) \right)\,g(\ix).
\end{align}
Since $\ix$ contains components of symmetric and traceless tensors, it is confined to a lower-dimensional manifold. Accordingly, we introduced a \emph{density of states} $g(\ix)$, which is nonzero only for values permitted by symmetry. Further technical details are provided in \cref{app-integral}, where integrals over $\ix$ are encountered; this aspect was omitted from the main text for simplicity.

The free energy density $f(\ix)$ is \cref{free-energy} expressed in terms of $\ix$ as:
\begin{align}
        f(\ix) = \frac{1}{2}\ix\tr{} \A  \ix , \label{f-def-ix}
\end{align}
where $\A$ is a $2d^2\times 2d^2$ block matrix:
\begin{align}
    \label{A-def-old}
    \A = \begin{pmatrix}
        \mu \mathbf{I} & D \mathbf{I}
        \\
        D \mathbf{I} & \chi \mathbf{I}
    \end{pmatrix},
    \qquad
    \A^{-1} = \frac{1}{\mu\chi-D^2}\begin{pmatrix}
        \chi \mathbf{I} & -D \mathbf{I}
        \\
        -D \mathbf{I} & \mu \mathbf{I}
    \end{pmatrix}.
\end{align}
Here, $\mathbf{I}$ is the $d^2 \times d^2$ identity matrix. However, using the symmetry properties of $\u$ and $\q$, it is more convenient to express $\A$ in the symmetrized form:
\begin{align}
    \label{A-def}
    \A = \begin{pmatrix}
        \mu \;\Tiso & D \;\Tiso
        \\
        D \;\Tiso & \chi \;\Tiso
    \end{pmatrix},
    \qquad
    \A^{-1} = \frac{1}{\mu\chi-D^2}\begin{pmatrix}
        \chi \;\Tiso & -D \;\Tiso
        \\
        -D \;\Tiso & \mu \;\Tiso
    \end{pmatrix},
\end{align}
where  $\Tiso$ is isotropic traceless rank-4 tensor in $d$-dimentions:
\begin{align}
    \Tiso_\abmn = \frac{1}{2}\left(
    \delta_{\alpha\mu}\delta_{\beta\nu}+\delta_{\alpha\nu}\delta_{\beta\mu}
    \right)
    -\frac{1}{d} \delta_\ab \delta_{\mu\nu}.
\end{align}
For further discussion on tensors, see \cref{app-axis-invariant}. Note that the block matrix in \cref{A-def} has rank-4 tensors as its blocks, and the vector $\ix$ consists of two blocks of rank-2 tensors (matrices). Each block in the matrix acts on a block of the vector through tensor contraction. For example:
\begin{align}
    \begin{pmatrix}
        \T^{(a)}_\abmn & \T^{(b)}_\abmn
        \\
        \T^{(c)}_\abmn & \T^{(d)}_\abmn
    \end{pmatrix}\ix
    =
    \big(\, \T^{(a)}_\abmn u_\mn + \T^{(b)}_\abmn q_\mn\,,\,
    \T^{(c)}_\abmn u_\mn + \T^{(d)}_\abmn q_\mn \,\big),
\end{align}
specifically,
\begin{align}
    \A\ix =& \Big(\mu \,\Tiso_\abmn u_\mn +D \,\Tiso_\abmn q_\mn\;,~D \,\Tiso_\abmn u_\mn +\chi \,\Tiso_\abmn q_\mn \Big)
    \\
    =&\Big(\mu \,\u +D \,\q\;,~D \,\u +\chi \,\q \Big).
\end{align}
Notice with these definitions, \cref{free-energy} and \cref{f-def-ix} are equivalent.

The conjugate thermodynamic flux corresponding to $\ix$ is defined as $\F^{(1)}(\ix) = \partial f(\ix) / \partial \ix\equiv \nabla f(\ix)$. This can be expressed as:
\begin{align}\label{F1-def}
    \F^{(1)} = \A \ix = \left({\bs\sigma}^{(1)},\; \mathbf{H}^{(1)}\right),
    \qquad
    \sigma^{(1)}_\ab(\ix) = \frac{1}{N}\frac{\partial}{\partial u_\ab} f(\ix),
    \qquad
    H^{(1)}_\ab(\ix) = \frac{1}{N} \frac{\partial}{\partial q_\ab} f(\ix),
\end{align}
where $\sigma^{(1)}_\ab(\ix)$ and $H^{(1)}_\ab(\ix)$ represent (respectively) the stress and nematic field contributions from a single linker, analogous to $\sigma^\text{linker}_\ab$ in  \cref{sigma-1-def} in the main text.

We further combine the macroscopic shear rate $v_\ab$ and the rate of change of orientation $\dot{Q}_\ab$ into a $2d^2$-dimensional generalized flux   vector $\J$:
\begin{align}
    \J = \left(\v,\; \dot{\Q} \right).
\end{align}

\subsection{Derivation of constitutive relation}
\label{sm-subsec-derivation}
Rewriting the fraction of the bound linkers in generalized variables: $n(\u,\q,t)=n(\ix,t)$, \cref{n-eqn} reads
\begin{align}\label{n-eqn-gen}
    \pd{}{t}n(\ix, t) +\J\cdot\nabla n(\ix,t) = (1-\phi(t))\kb(\ix) - n(\ix,t)\ku
\end{align}
Noting that the total fraction of bound linkers is $\phi(t) = \int n(\ix,t) \d\ix$, integrating the equation above, we end up with \cref{phi-average} in the main text.

Instead of average stress defined in \cref{average-stress-def-0}, define average thermodynamic force, which include both average stress tensor and nematic field force:
\begin{align}
    \F(t) = \la \hat\F(t) \ra =& \int n(\ix,t)\nabla f(\ix) \d\ix
\end{align}
Taking the time derivative, we obtain:
\begin{align}
    \frac{\d}{\d t} \F(t) =& \int \d \ix\,\left(\nabla f(\ix)\right)
    \pd{}{t}n(\ix,t)
    \\
    =&
    \int 
    \d \ix\,\left(\nabla f(\ix)\right)
    \left(- \J \cdot \nabla n(\ix,t) 
    -\ku\,n(\ix,t ) + (1-\phi(t))\,\kb(\ix) 
    \right)
\end{align}
where in the second line, we substituted \cref{n-eqn-gen}. Integrating by parts, we obtain:
\begin{align}
    \frac{\d}{\d t} \F(t) =& -\ku \F(t) + \int \left(\nabla \left(\nabla f(\ix)\right)\J\right) n(\ix,t) \,\d \ix
    + (1-\phi(t)) \int \kb(\ix) \nabla f(\ix)\d\ix
\end{align}
Recalling the definition of the quadratic free energy \eqref{f-def-ix}: $ \nabla f(\ix)=\A \ix$ and $\nabla (\nabla f(\ix)) = \A$ and independent of $\ix$. We obtain
\begin{align}
    \frac{\d}{\d t} \F(t) =& -\ku \F(t) +  \A \J \,\phi(t)
    + (1-\phi(t)) \A\int \ix\kb(\ix) \d\ix
\end{align}
Rearranging terms, we obtain:
\begin{align}
\label{constitutive-general-0}
    \left(1 + \frac{1}{\ku} \frac{\d}{\d t}\right) \F = (1-\phi) \A \bs{\Omega}_{\ix} + \frac{\phi}{\ku} \A\J
\end{align}
where $\F = ({\bs\sigma}, {\bs H})$ and $\bs{\Omega}_{\ix} = \left(\,\Omega_u \Q,\; \Omega_q \Q \,\right)$ is the first moment of activity  (c.f.~\cref{omega-uq-def})
\begin{align}
    \label{Omega-ix-def}
     \bs{\Omega}_{\ix} = \int \frac{\kb(\ix)}{\ku}\, \ix\;\d\ix =  \int \ix\, \Omega(\ix)\,g(\ix)\;\d\ix 
\end{align}
where free energy contribution vanishes due to the rotational symmetry of $f(\ix)$ (c.f.~\cref{f-def-ix}).  

In \cref{constitutive-general-0}, the right- and left-hand sides are vectors. The first component reduces to the constitutive relation for stress (c.f.~\cref{sigma-average}), and the second component gives the constitutive relation for the nematic field strength:
\begin{align}
\label{rubik}
    \left(1+\tau \frac{d}{d t}\right) \sigma_\ab=& 2 \eta \,v_\ab-\nu \,\dot{Q}_\ab
    +\zeta Q_\ab
    \\
    \left(1+\tau \frac{d}{d t}\right) H_\ab=& \gamma \,\dot{Q}_\ab-\nu \,v_\ab
    +\omega Q_\ab
\end{align}
where we defined (also in \cref{coupling})
\begin{gather}
    \eta= \mu \phi / (2 \ku) 
    \qquad 
    \tau=1/\ku
    \qquad
    \nu=-D\phi/\ku \qquad \gamma = \chi \phi / \ku
    \\
    \zeta=(1-\phi)\left(\mu\Omega_u+D\Omega_q\right)
    \qquad
    \omega=(1-\phi)\left(D\Omega_u+\chi\Omega_q\right)
\end{gather}

\section{Stochastic dynamics of the linkers}
\label{sm-stochastic}
We assume that the dynamics of individual linkers are independent. Therefore, we focus on the stochastic dynamics of a single linker, which can exist in either the unbound state or one of the bound states within a continuous state space. The dynamics of the linker are modeled as a continuous-time Markov chain in this state space, denoted by the state variable $M_t$. 
\Cref{fig-alternative} visualizes the state space of the Markov chain and the state variable $M_t$.

At any time $t$, $M_t$ is a function of $\ix$. For an unbound linker at time  $t$,  $M_t$ is identically zero; $M_t(\ix)\equiv 0$. For linker bound to state $\ix_0$ at time $t$, $M_t(\ix)$ is delta-peaked around $\ix_0$; $M_t(\ix) = \delta(\ix - \ix_0)$. Here, $\delta$ is the Dirac delta function defined such that $\int \varphi(\ix) \delta(\ix - \ix_0) \, \d\ix = \varphi(\ix_0)$ for any function $\varphi$.

The linker binds to state $\ix$ with rate $\kb(\ix)$. 
From the bound state, the linker can only unbind. Therefore, in order to bind to another state, it must first unbind from the current bound state. We assume unbinding rate $\ku$ is constant, i.e., independent of the state variable $\ix$. This forms a continuous-time Markov chain for the state space including of all possible bound states and unbound state.

Let's see how the hydrodynamic variables are defined in terms of $M_t$. We consider a volume $V$ containing $N$ linkers. Assuming the linkers are independent form each other, we assign an independent and identical copy of the process $M_t^{(i)}$ to each linker.
 Consequently, the fraction of bound linkers and the total force in the system are given by:
\subeqn{
    \label{obs-in-em}
\begin{align}
    \label{phi-def-sm}
    \hat\phi(t)=&\frac{1}{N} \sum_{i=1}^N \int M_t^{(i)}(\ix) \,\d \ix
    \\
    \label{F-def-sm}
    \hat\F(t) =& \sum_{i=1}^N \int \F^{(1)}(\ix)  M_t^{(i)}(\ix) \,\d \ix
    \quad= \frac{1}{N} \sum_{i=1}^N \int \frac{\partial f(\ix)}{\partial \ix} M_t^{(i)}(\ix) \,\d \ix
    \quad = \frac{1}{N} \sum_{i=1}^N \A \int \ix \, M_t^{(i)}(\ix) \,\d \ix   
\end{align}
}
In \cref{phi-def-sm}, the integral evaluates to 0 or 1, corresponding to the linker's unbound or bound states, respectively. 
In \cref{F-def-sm}, when linker $i$ is bound to state $\ix_0$, the integral evaluates to the force $\F^{(1)}(\ix_0)$ because of delta function property. Thus, the total force $\hat{\F}$ is the sum of the forces from all bound linkers.

$M_t^{(i)}$ is a stochastic process, for fixed time it is a random function. 
Therefore, $\hat\phi(t)$ and $\hat\F(t)$ defined above are also random variables, fluctuating quantities.

In the main text we modeled the state of the linker with an indicator function $\ind_i(t)$; see definition \eqref{def-ind}. It is related to the stochastic process $M_t^{(i)}$ as follows:
\begin{align}
    \ind_i(t) = \int M_t^{(i)}(\ix)\d\ix
\end{align}
$M_t^{(i)}(\ix)$ has the complete information about the state of the linker $i$ at time $t$,  whereas $\ind_i(t)$ only indicates whether it is bound or unbound. In the main text, $\ind_i(t)$ always appears alongside the state variable $(\hat\u, \hat\q) \equiv \ix$.
Notice the consistence between Eqs. (\ref{phi-1},\ref{sigma-1}) and Eqs.~(\ref{phi-def-sm},\ref{F-def-sm}) respectively.

Now, we model the dynamics of a single linker. First we define the probability densities describing the stochastic process $M_t$.
Let $p_t(\ix)$ denote the probability density of the linker being in the bound state $\ix$ at time $t$.
Equivalently, $p_t(\ix)$ is the probability density that $M_t$ is the function $\ix' \mapsto \delta(\ix'-\ix)$. Let the probability of the linker being in the unbound state at time $t$ be denoted as $p_t(\unbound)$, or equivalently, the probability that $M_t$ is identically zero. 
The total probability must equal to one:
\begin{align}
    p_t(\unbound) + \int p_t(\ix) \,\d \ix= 1
    \label{def-unbound}
\end{align} 
In the following sections, we will encounter the average of macroscopic variables over the stochastic process $M_t$. For example, the average of \eqref{F-def-sm} contains terms like:
\begin{align}
\label{first-moment}
 \la \varphi(\ix) M_t(\ix) \ra = \int p_t(\ix') \varphi(\ix)\delta(\ix-\ix')\,\d\ix' \quad= p_t(\ix) \varphi(\ix)
\end{align}
where $\la ~\ra$ denotes the average over the distribution $p_t(\ix)$. Here, with probability $p_t(\ix')$, $M_t(\ix)$ takes the value $\delta(\ix - \ix')$.

Now we write a  Kolmogorov forward equation for evolution of $p_t(\ix)$ \cite{grimmett2020}.
The state of the linker changes by two processes: binding/unbinding and external fluxes:
\begin{align} \label{kolmogor}
    \frac{\partial  p_t(\ix)}{\partial t}=&
    - \J \cdot \frac{\partial p_t(\ix)}{\partial \ix} 
    -\ku\,p_t(\ix ) + p_t(\unbound)\,\kb(\ix) 
\end{align}
Here, the first term on the right-hand side accounts for the effect of external fluxes, while the other two terms model the binding/unbinding dynamics with $\kb(\ix) = \kb(\u, \q)$.
Integrating this over all states and using \cref{def-unbound}, we obtain a closed equation for $p_t(\unbound)$:
\begin{align}
    \partialt p_t(\unbound)= \ku - (1+\Z) \ku \,p_t(\unbound)
    \label{punbound-deq}
\end{align}
where $\Z$ is defined in \cref{Z-def}.

\Cref{kolmogor}  bears a resemblance to \cref{n-eqn}. 
Taking an initial condition $p_{t=0}(\ix) = \delta(\ix - \ix_0)$ (thus $p_{t=0}(\unbound) = 0$), the solution of \cref{kolmogor} is the propagator $p_t(\ix \, | \, \ix_0)$, representing the distribution over bound states at time $t$ given that the linker was bound at $\ix_0$ at $t = 0$.
This propagator appears in the second moment of the stochastic process $M_t$. Similar to \cref{first-moment}, we have
\begin{align}
    \la \varphi(\ix,\ix_0) M_t(\ix)M_0(\ix_0)\ra =&
    \int\!\! \d\ix'\,\d\ix'_0\; 
     p_0(\ix'_0) p_t(\ix'\,|\,\ix'_0)
      \;\varphi(\ix,\ix_0)\,\delta(\ix - \ix')\,\delta(\ix_0 - \ix'_0)\;
    \quad
    \\
    =& \; \varphi(\ix,\ix_0) p_0(\ix_0) p_t(\ix\,|\,\ix_0) \label{M-p}
\end{align}

In the following section, we solve the Kolmogorov equation \eqref{kolmogor} to obtain the propagator $p_t(\ix \, | \, \ix_0)$.

\begin{figure}
    \includegraphics[width=0.5\linewidth]{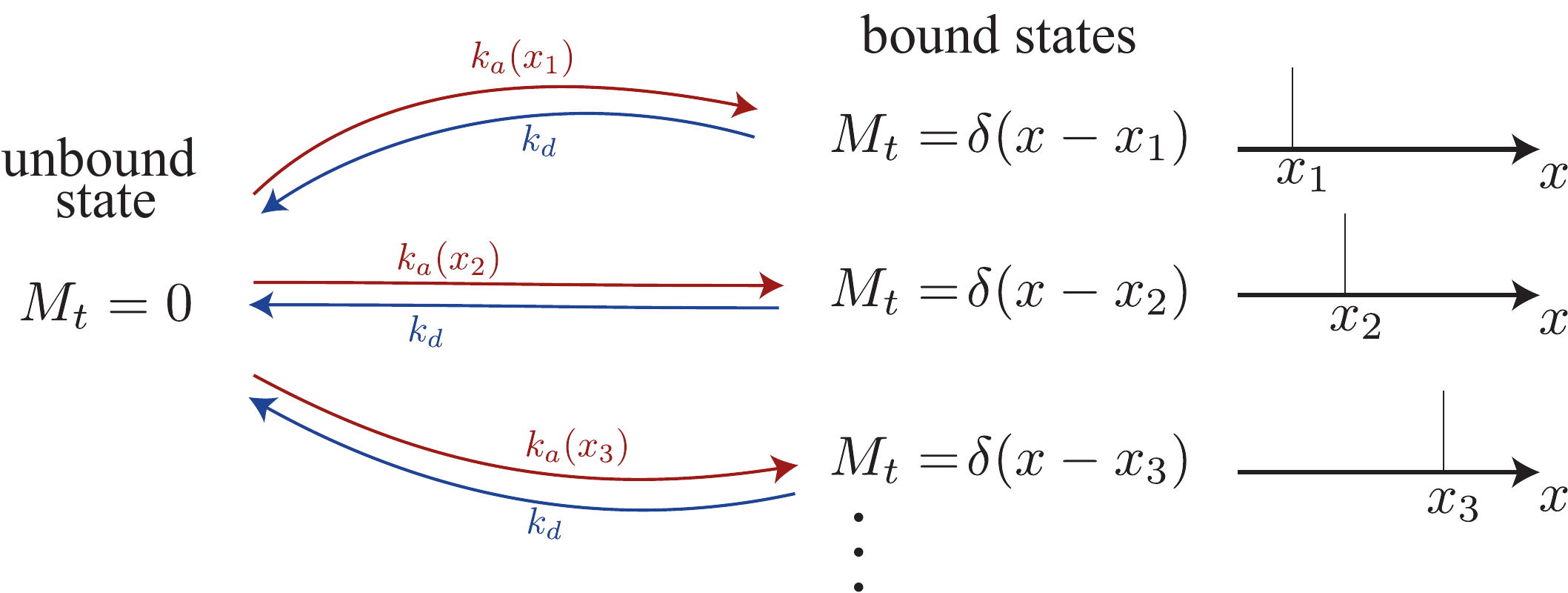}
    \caption{State space of the single-linker continuous-time Markov chain and the state variable $M_t$. The unbound state corresponds to $M_t(\ix)\equiv 0$, while a linker bound at $\ix_0$ has $M_t(\ix)=\delta(\ix-\ix_0)$. Transitions occur with binding rate $\kb(\ix)$ and unbinding rate $\ku$.}
    \label{fig-alternative}
\end{figure}

\section{First moments: equations for average fields}
In this section, we will derive hydrodynamics equation \eqref{constitutive-general-0} starting from mesoscopic definitions of observables in \eqref{obs-in-em}. Then, we define the fluctuating hydrodynamic equations by introducing noise term in the hydrodynamics equation for averages. Finally, we will establish that the first moments of the noise vanish.
The higher order moments (autocorrelation functions) of observables and noise are considered in \ref{sm-af}

\subsection{Hydrodynamic equations for averages}
In this section, we derive \cref{constitutive-general-0}. Take the average of \cref{phi-def-sm} over the distribution $p_t(\ix)$:
\begin{align}
    \label{zutro}
    \phi(t) = \la \hat\phi(t) \ra =& \frac{1}{N} \sum_{i=1}^N  \int \la M_t(\ix) \ra \,\d \ix 
    = \frac{1}{N} \sum_{i=1}^N \int p_t(\ix) \,\d \ix 
    =\int p_t(\ix) \,\d \ix =1 - p_t(\unbound)
\end{align}
where we used \cref{first-moment} for the third equality. Additionally, note that the summands are identical in the sum.
Similarly, for \cref{F-def-sm}, we get
\begin{align}\label{futro}
    \F(t) = \la \hat\F(t) \ra =& \frac{1}{N} \sum_{i=1}^N \int \la \F^{(1)}(\ix) M_t(\ix) \ra \,\d \ix 
    = \int \F^{(1)}(\ix) p_t(\ix) \,\d \ix 
\end{align}
Taking the time derivative, we obtain:
\begin{align}
    \frac{\d}{\d t} \F(t) =& \int \d \ix\,\F^{(1)}(\ix)
    \frac{\partial  p_t(\ix)}{\partial t} 
    \\
    =&
    \int \F^{(1)}(\ix) 
    \left(- \J \cdot \frac{\partial p_t(\ix)}{\partial \ix} 
    -\ku\,p_t(\ix ) + p_t(\unbound)\,\kb(\ix) 
    \right)
\end{align}
where in the second line, we substituted Kolmogorov equation \eqref{kolmogor}. Integrating by parts, and substituting $p_t(\unbound)$ from \cref{zutro}, we obtain:
\begin{align}
    \frac{\d}{\d t} \F(t) =& -\ku \F(t) + \int \left(\nabla \F^{(1)}(\ix)\J\right) p_t(\ix) \,\d \ix
    + (1-\phi(t)) \int F^{(1)}(\ix)\kb(\ix) \d\ix
\end{align}
Recalling the definition \eqref{F1-def} of $\F^{(1)}$ and the quadratic free energy \eqref{f-def-ix}: $ \F^{(1)}(\ix)=\A \ix$ and $\nabla \F^{(1)}(\ix) = \A$ and independent of $\ix$. We obtain
\begin{align}
    \frac{\d}{\d t} \F(t) =& -\ku \F(t) +  \A \J \,\phi(t)
    + (1-\phi(t)) \A\int \ix\kb(\ix) \d\ix
\end{align}
From \cref{kakd-ix,Omega-ix-def}, we have $\int \ix \kb(\ix) \d\ix = \ku \Omega_\ix$, because the contribution from the first summand of \cref{kakd-ix} vanished due to rotational symmetry of free energy \eqref{f-def-ix}. Therefore, the above equation coincides with \cref{constitutive-general-0}.

In a similar way, we can derive the equation for $\phi(t)$. Taking the time derivative of \cref{zutro} and substituting \cref{kolmogor}, we have:
\begin{align}
    \frac{\d}{\d t} \phi(t) =& \int \frac{\partial p_t(\ix)}{\partial t} \,\d \ix  = 
    \int \left(- \J \cdot \frac{\partial p_t(\ix)}{\partial \ix} 
    -\ku\,p_t(\ix ) + p_t(\unbound)\,\kb(\ix) 
    \right) \d\ix
    \\
    \frac{1}{\ku}\frac{\d}{\d t} \phi(t) =& -(1+\Z) \phi(t) + \Z
    \label{phi-eq-sm}
\end{align}
where $\Z$ is defined in \cref{Z-def}.

\subsection{Fluctuating hydrodynamic equations}
\label{sm-with-noise}
Our goal is to write a stochastic differential equation (SDE) that describes the dynamics of the fluctuationg observables defined in \cref{obs-in-em}.
To do this, in \cref{constitutive-general-0} we replace the average fields with their fluctuating counterparts, and introduce a stochastic noise term:
\begin{align}
\label{constitutive-general-noisy}
    \left(1 + \frac{1}{\ku} \frac{\d}{\d t}\right) \hat\F = (1-\hat\phi) \A \bs{\Omega}_{\ix} + \frac{\hat\phi}{\ku} \A\J + \hat\bxi(t)
\end{align}
As $\hat\phi(t)$ also fluctuates, we write a SDE for it. Starting from \cref{phi-eq-sm}, we get:
\begin{align}
    \label{phi-eq-noisy-sm}
    \frac{1}{\ku}\frac{\d}{\d t} \hat\phi(t) =& -(1+\Z) \hat\phi(t) +  \Z + \hat\xi^\phi(t)
\end{align}
where $\hat\xi^\phi(t)$ is a stochastic noise term.

We do not impose any properties on the noise terms $\hat\bxi(t)$ and $\hat\xi^\phi(t)$
Essentially, \cref{constitutive-general-noisy,phi-eq-noisy-sm} define the noise terms; substituting \cref{obs-in-em}, we obtain:
\subeqn{
    \label{meso-noise-def}
\begin{align}
    \hat\xi^\phi(t) =& \frac{1}{N}\sum_{i=1}^N\left[
        (1+\Z)\int M_t^{(i)}(\ix) \,\d \ix
        + \frac{1}{\ku}\int \left(\tfrac{\d}{\d t} M_t^{(i)}(\ix)\right) \,\d \ix 
        - \Z
    \right]
    \\
     \hat\bxi(t) =& \frac{1}{N}\sum_{i=1}^N\A \left[
        \int\ix M_t^{(i)}(\ix) \,\d \ix
        + \frac{1}{\ku}\int \ix \left(\tfrac{\d}{\d t} M_t^{(i)}(\ix)\right) \,\d \ix
        +(\Omega_\ix - \J/\ku)\int M_t^{(i)}(\ix) \,\d \ix
        - \Omega_\ix
     \right]
    \label{meso-F-noise-def}
\end{align}
}
\Cref{meso-noise-def} defines the noise terms in terms of mesoscopic state. The necessary condition for the noise terms zero average. Taking the average over the non-stationary distribution $p_t(\ix)$, we have:
\begin{alignat}{3}
    \la \hat\xi^\phi(t) \ra =&  \frac{1}{N}\sum_{i=1}^N\left[
        (1+\Z)\int \la M_t^{(i)}(\ix) \ra \,\d \ix
        + \frac{1}{\ku}\int \left(\frac{\d}{\d t} \la M_t^{(i)}(\ix) \ra \right) \,\d \ix 
        - \Z
    \right]
    & 
    \\
    =&\frac{1}{N}\sum_{i=1}^N \left[
        (1+\Z)\int p_t(\ix)\d\ix + \frac{1}{\ku}\frac{\d}{\d t} \int p_t(\ix) \d\ix - \Z
        \right]
        & \left[\text{using \cref{first-moment}}\right]
    \\
    =&\frac{1}{N}\sum_{i=1}^N \left[
        (1+\Z)(1-p_t(\unbound)) - \frac{1}{\ku}\frac{\d}{\d t} p_t(\unbound)- \Z
        \right]
        &  \left[\text{using \cref{def-unbound}}\right]
    \\
    =&\, 0 & \left[\text{using \cref{punbound-deq}} \right]
\end{alignat}
where in the first line we assumed integral, time derivative and average are interchangeable.
Similarly, the average of \cref{meso-F-noise-def} with respect to $p_t$ reads:
\begin{alignat}{3}
    \la \hat\bxi(t) \ra  =&\frac{1}{N}\sum_{i=1}^N\A \left[
        \int\ix M_t^{(i)}(\ix) \,\d \ix
        + \frac{1}{\ku}\int \ix \left(\tfrac{\d}{\d t} M_t^{(i)}(\ix)\right) \,\d \ix
        +(\Omega_\ix - \J/\ku)\int M_t^{(i)}(\ix) \,\d \ix
        - \Omega_\ix
     \right]
     \\
    =& \frac{1}{N}\sum_{i=1}^N\A \left[
        \int  \la\ix M_t^{(i)}(\ix) \ra \,\d \ix
        + \frac{1}{\ku}\int  \left(\frac{\d}{\d t} \la\ix M_t^{(i)}(\ix) \ra\right) \,\d \ix
        +(\Omega_\ix - \J/\ku)\int \la M_t^{(i)}(\ix) \ra\,\d\ix
        - \Omega_\ix
     \right]
     \\
     =&\frac{1}{N}\sum_{i=1}^N\A \left[
        \int  \ix p_t(\ix)\d \ix
        + \frac{1}{\ku}\int  \tfrac{\d}{\d t}\left( \ix p_t(\ix) \right) \,\d \ix
        +(\Omega_\ix - \J/\ku)\int p_t(\ix) \,\d\ix
        - \Omega_\ix
     \right]
\end{alignat}
Substituting \cref{kolmogor}, $\ix p_t(\ix)$ terms directly cancel each other, and we obtain:
\begin{align}
    \la \hat\bxi(t) \ra  =\frac{1}{N}\sum_{i=1}^N\A\Bigg\{&
    -\frac{1}{\ku}\int \J p_t(\ix)\d\ix - \frac{1}{\ku}
    \int \ix \left(\J\cdot\frac{\partial p_t(\ix)}{\partial \ix}\right) \d\ix
    \\
    &\Omega_\ix\int p_t(\ix)\d\ix +p_t(\unbound)\int \frac{\kb(\ix)}{\ku}\ix \d\ix - \Omega_\ix \Bigg\}
\end{align}
The terms in the first line cancel each other after integrating by parts. The second line vanished due to the definition of $\Omega_\ix$ in \cref{Omega-ix-def}. Therefore, we show that $\la \hat\bxi(t) \ra = 0$.

An important remark is that the average of the noise terms vanish even in non-stationary state.

\section{Solution of kolmogorov equation}
\label{sm-solution}
In this section, we solve the partial differential equation \eqref{kolmogor} to obtain the propagator $p_t(\ix|\ix_0)$. The later is used to evaluate autocorrelation functions, see \cref{M-p}. 

We employ the Fourier transform to eliminate the partial derivatives associated with the probability fluxes. Notably, there is no need to perform the inverse Fourier transform, as the autocorrelations of observables can be expressed in terms of the Fourier transform of the propagator.

We solve \cref{kolmogor} with the initial conditions:
\begin{align}
    \label{ic-sm}
    p_{t=0}(\unbound) = 0 \quad \text{and} \quad p_{t=0}(\ix) = \delta(\ix - \ix_0).
\end{align}
which means the linker is initially bound to state $\ix_0$. Recall the closed equation for the unbound state \cref{punbound-deq}. With the initial conditions above, it has the following solution:
\begin{align}\label{punbound-sol}
    p_t(\unbound\,|\,\ix_0) =&\frac{1}{1+\Z}\left(1-\,\ee^{-(\Z+1)\ku t}\right)
\end{align}
To solve the remaining partial differential equation (PDE) \cref{kolmogor}, we apply the Fourier transform with respect to $\ix$:
\begin{align}\label{ptil-def}
    \ptil_t(\tix) = \int p_t(\ix) \ee^{- \ii \tix\cdot\ix} \d\ix 
\end{align}
where $\tix$ is the conjugate variable of $\ix$. The transformed PDE and initial conditions become:
\begin{align}
    \partialt {\ptil}_t(\tix)=& -\left(\ku+\ii\;\J\!\cdot\!\tix\right)\,\ptil_t(\tix) + p_t(\unbound)\,\tka(\tix) 
    \\
    =& -\left(\ku+\ii\;\J\!\cdot\!\tix \right)\,\ptil_t(\tix) + \frac{1}{1+\Z}\left(1-\,\ee^{-(\Z+1)\ku t}\right)\,\tka(\tix) 
    \label{pqt-deq}
    \\
    \ptil_{t=0}(\tix) =&\, \ee^{-\ii\tix\cdot \ix_0}
\end{align}
where we used the solution \cref{punbound-sol} for $p_t(\unbound)$.
Here, $\tka(\tix)$ is the Fourier transform of $\kb(\ix)$.
This is now an ordinary differential equation and can be solved using standard methods. The solution is:
\begin{align}\label{pq-sol}
    \ptil_t(\tix\,|\,\ix_0) =&\, 
    \ee^{-\ii\tix\cdot \ix_0} \ee ^ {-(\ku+\ii\,\J\cdot\tix) t} 
    + \frac{1}{1+\Z} \frac{\tka(\tix)}{\ku+\ii\;\J\!\cdot\!\tix} \left(1-\ee^{-(\ku+\ii\,\J\cdot\tix) t}\right)
    +\frac{1}{1+\Z} \frac{\tka(\tix)}{\Z \ku-\ii\;\J\!\cdot\!\tix}\left( \ee^{-\ku(1+Z) t}-\ee^{-(\ku+\ii\,\J\cdot\tix) t}\right)
\end{align}
In particular, when $t\rightarrow\infty$, the system approaches a steady state with the distribution
\begin{align}\label{pss-def}
    \ptil_\ss( \tix) = \ptil_\infty(\tix) = \frac{1}{1+\Z} \frac{\tka(\tix)}{\ku+\ii\;\J\!\cdot\!\tix}
\end{align}
which is independent of the initial condition. This indicates that the system forgets its initial state. The external fluxes $\J$ affect the steady-state distribution, deviating it from a Gibbs distribution. 
In \cref{app-ss} we derive the steady-state distribution \cref{n-ss-x}, which is the inverse Fourier transform of \cref{pss-def}:
\begin{align}\label{p-ss-x}
    p_\ss(\ix) =  \frac{1}{1+\Z}\int_0^\infty\kb(\ix-\J \tau) \;\ee^{-\ku \tau} \;\d\tau
\end{align}

The steady-state probability of the unbound state can also be calculated using \cref{def-unbound}:
\begin{align}
    p_\ss(\unbound) = 1 - \int p_\ss(\ix) \,\d\ix = 1 - \int p_\ss(\ix) \ee^{-\ii \ix \cdot \zero} \,\d\ix
    = 1 - \ptil_\ss(\zero)
    = \frac{1}{1+\Z} 
    \label{punbound-ss}
\end{align}
where we used \cref{pss-def} with $\tka(\zero) =\int\kb(\ix)\d\ix =  \Z \ku$.

\section{Hydrodynamic quantities at steady state}
Having derived the steady-state distribution \cref{pss-def} and the propagator \cref{pq-sol}, we now have the two-time steady-state joint distribution function of the mesoscopic state. This means, we can calculate the autocorrelation functions of the observables defined in \cref{obs-in-em} at steady state.

The inverse Fourier transform consistent with \cref{ptil-def} is
\begin{align}\label{inverse-foourier}
    p_t(\ix) = \frac{1}{(2\pi)^n}\int \ptil_t(\tix)  \ee^{ \ii \tix\cdot\ix} \d\tix 
\end{align}
where $n = 2d^2$ is the dimension of vectors $\ix$ and $\tix$.

We will need the following expressions derived from \cref{pss-def}:
\begin{align}
    \ptil_\ss(\zero) =&\frac{\Z}{1+\Z}
    \label{pss-zero}
    \\
    \frac{\d}{\d \tix}\ptil_\ss(\zero) =& -\frac{\ii}{1+\Z}\frac{\k+\Z\J}{\ku}
    \label{pss-zero-1}
    \\
    \frac{\d^2}{\d \tix^2}\ptil_\ss(\zero) =&\frac{1}{1+\Z}\left(
    -\frac{\kk}{\ku}-\frac{\k \J\tr{} + \J {\k}\tr{}}{\ku^2}
    -\frac{2\Z \J\J\tr{}}{\ku^2}
    \right)
    \label{pss-zero-2}
\end{align}
where $\zero$ is the null vectorm and the second-order derivative is represented by the Hessian matrix:
\begin{align}
    \label{hessian-def}
    \left(\frac{\d^2}{\d \tix^2}\,f\right)_{ij}\equiv \frac{\d^2}{\d \tix_i \d\tix_j}f
\end{align}

During the derivation, we encounter the derivatives of the Fourier-transformed rate $\tka(\tix)$ evaluated at the null vector $\zero$:
\begin{align}
    \tka(\zero) &= \int \kb(\ix) \d\ix = \ku \Z
    \\
    \frac{\d}{\d \tix}\tka(\zero)=& \int -\ii\,\ix\,\kb(\ix)\,\d\ix = -\ii \k
    \\
    \frac{\d^2}{\d \tix_i \d\tix_j} \tka(\zero) =& \int -\ix_i\ix_j\,\kb(\ix)\,\d\ix = -(\kk)_{ij}
\end{align}
where the vector $\k$ and matrix $\kk$ are the moments of the binding rate $\kb(\ix)$, defined as:
\begin{align}
    \text{vector}\quad \k =& \int \kb(\ix)\,\ix \,\d\ix = \int \ku\Omega(\ix)\,\ix \,\d\ix
    =\ku \bs{\Omega}_{\ix}
    \label{k1-def}
    \\
    \text{matrix}\quad \kk = & \int \kb(\ix)\,\ix\ix\tr{} \,\d\ix
    \label{k2-def}
\end{align}
Note that in \cref{k1-def}, $\k$ is related to the first moment of the molecular activity $\Omega$.

\subsection{Averages at steady state}
Taking $p_t(\unbound) = p_\ss(\unbound)$ in \eqref{zutro}, we obtain
\begin{align}
\label{phi-mean-def-sm}
\bar{\phi} \equiv \la \hat\phi(t) \ra_\ss = 1 - p_\ss(\unbound) = \frac{\Z}{1+\Z}
\end{align}
where we used \cref{punbound-ss}.
For the steady-state value of $\F$, we use the steady-state distribution $p_t (\ix) = p_\ss(\ix)$ in \eqref{futro}:
\begin{align}
    \bar\F = \la \hat\F(t) \ra_\ss 
    =\A\int \ix p_\ss(\ix) \,\d \ix 
\end{align}
With this example, we demonstrate how we use the fourier transform of the distribution, $\ptil_\ss(\tix)$  to evaluate the average of the observable $\hat\F$:
\begin{alignat}{3}
    \la \smash{\hat\F} \ra 
    =& \A \int \d\tix\d\ix \; \frac{1}{(2\pi)^n}\,\ptil_\ss(\tix)\,\ix\,\ee^{\ii \tix\cdot\ix}
    &&\text{[inverse fourier transform, \cref{inverse-foourier}]}
    \\
    =&\A\int \d\tix\d\ix\; \frac{1}{(2\pi)^n}(-\ii)\,  \ptil_\ss(\tix) \frac{\d}{\d \tix}\ee^{\ii \tix\cdot\ix}
        \qquad &&\text{[ $\ix\ee^{\ii \tix\cdot\ix}=-\ii \tfrac{\d}{\d\tix}\ee^{\ii \tix\cdot\ix}$]}
    \\
    =&\A\int \d\tix\left(\int \frac{1}{(2\pi)^n} \ee^{\ii \tix\cdot\ix}\,\d\ix\right) \ii\, \frac{\d}{\d \tix}\ptil_\ss(\tix)
    \qquad &&\text{[Integration by parts]}
    \\
    =&\A\int \d\tix\,\delta(\tix)\, \ii\,\frac{\d}{\d \tix}\ptil_\ss(\tix)    
    &&\text{[using Fourier transform of Dirac Delta $\delta(\ix)$]}
    \\
    \label{gyot-0}
    =&\A\,\ii \left(\tfrac{\d}{\d \tix}\ptil_\ss(\zero) \right)
    &&\text{[contraction with $\delta$]}
    \\
    =&\frac{1}{1+\Z}\frac{\A(\k+\Z\J)}{\ku}
    &&\text{[using \cref{pss-zero-1}]}
    \label{F-mean}
\end{alignat}
The above is just an illustration, because it is much straightforward to use \cref{p-ss-x}. Even simpler method is to just solve \eqref{constitutive-general-0} with time-derivative set to zero.
However, the method above is essential for calculating the autocorrelation functions of the observables, as we will see in the following.

\subsection{Autocorrelation function}
\label{sm-af}
In this section, we evaluate the autocorrelation functions of $\hat{\F}$ and $\hat{\phi}$, as well as their cross-correlation function. 
Here, we must assume that linkers are independent from each other, i.e., the processes $\{ M_t^{(i)} \}_{i=1}^N$ are independent. This is a reasonable assumption for sufficiently large $N$ from the mathematical point of view (Law of Large Numbers). 
The autocorrelation function of the fraction of bound linkers $\hat{\phi}$ is defined as
\begin{align}
    C_\phi(t, t') = \la (\hat\phi(t)-\bar{\phi})(\hat\phi(t')-\bar{\phi}) \ra = \la \smashhat\phi(t)\smashhat\phi(t')\ra  - (\bar{\phi})^2
\end{align}
The average is over steady-state distribution. The steady state distribution is time-translation invariant, therefore, we can write $C_\phi(t, t') = C_\phi(t-t')$ and 
\begin{align}
    C_\phi(t) = \la (\hat\phi(t)-\bar{\phi})(\hat\phi(0)-\bar{\phi}) \ra = \la \hat\phi(t)\hat\phi(0)\ra  - (\bar{\phi})^2
\end{align}

Using the definition in \cref{phi-def-sm}, we have
\begin{align}
    \la \phi(t)\phi(0)\ra =& 
    \la \Bigg(\frac{1}{N} \sum_{i=1}^N \int M_t^{(i)}(\ix) \,\d \ix\Bigg)
    \Bigg(\frac{1}{N} \sum_{j=1}^N \int M_0^{(j)}(\ix_0) \,\d \ix_0\Bigg) \ra 
    \\
    =& \int \frac{1}{N^2}\sum_{{i,j=1}}^{N}\la M_t^{(i)}(\ix)M_0^{(j)}(\ix_0) \ra \;\d\ix\,\d\ix_0
    \\
    =&\int \left[\frac{1}{N}\la M_t(\ix)M_0(\ix_0)\ra + \frac{N-1}{N}\la M_t(\ix)\ra \la M_0(\ix_0)\ra\right]\;\d\ix\,\d\ix_0
\end{align}
Here, we used the independence of $M_t$ processes when $i \ne j$. Therefore,
\begin{align}
    \label{Cphi-M}
    C_\phi(t) =& \frac{1}{N}\int \Big[\la M_t(\ix)M_0(\ix_0)\ra -\la M_t(\ix)\ra \la M_0(\ix_0)\ra\Big]\;\d\ix\,\d\ix_0
    \\
    =& \frac{1}{N}\int \Big[
    p_\ss(\ix_0)p_t(\ix\,|\,\ix_0)
    -p_\ss(\ix)p_\ss(\ix_0)\Big]\;\d\ix\,\d\ix_0
\end{align}
where we used \cref{first-moment,M-p}. Now, we calculate the integrals:
\begin{alignat}{3}
    \int p_\ss(\ix)p_\ss(\ix_0)\;\d\ix\,\d\ix_0 =& \big(1-p_\ss(\unbound)\big)^2 = (\bar{\phi})^2
    \qquad &&\text{[using \cref{def-unbound}]}
    \\
    \int p_\ss(\ix_0)p_t(\ix\,|\,\ix_0)\;\d\ix\,\d\ix_0 =& \int p_\ss(\ix_0) \big(1-p_t(\unbound\,|\,\ix_0)\big) \d\ix_0
    \qquad &&\text{[using \cref{def-unbound}]}
    \\
    =& \;\bar{\phi} - \bar{\phi}(1-\bar{\phi})\left(1-\,\ee^{-(\Z+1)\ku t}\right) &&\text{[using \cref{punbound-sol} and \cref{punbound-ss}]}
\end{alignat}
Thus, the autocorrelation function of $\phi(t)$ is
\begin{align}
    \label{C-phi}
    C_\phi(t) =& \frac{1}{N}\frac{\Z}{(1+\Z)^2}\,\ee^{-\ku(\Z+1) t}
\end{align}

Next, consider the autocorrelation of the process $\hat{\F}(t)$:
\begin{align}
    \C_{\F}(t) = 
    \la \hat\F(t) \,\tr{\smashhat\F(0)} \ra - 
    \bar{\F}\tr{\bar{\F}}
\end{align}
Note that $\mathbf{C}_{\F}(t)$ is a matrix. Similar to \cref{Cphi-M}, using definition \cref{F-def-sm}, we obtain:
\begin{align}
    \label{involved}
    \C_{\F}(t) = \frac{1}{N}\,
    \A\left(\int \la M_t(\ix)M_0(\ix_0)\ra\,\ix\tr{\ix_0}\,\d \ix\,\d \ix_0 - \int \la M_t(\ix)\ra\la M_0(\ix_0)\ra\,\ix\tr{\ix_0} \,\d \ix\,\d \ix_0\right)\tr{\A}
\end{align}
The second integral simplifies to:
\begin{align}
    \A\left(\int \la M_t(\ix)\ra\la M_0(\ix_0)\ra\,\ix\tr{\ix_0} \,\d \ix\,\d \ix_0\right)\tr{\A} = \bar{\F}\tr{\bar{\F}}    
    \quad
    = \frac{1}{(1+\Z)^2}\frac{1}{\ku^2}\;\A\,(\k+\Z\J)\,\tr{(\k+\Z\J)}\tr{\A}
\end{align}
where we used \cref{F-mean}. The first integral in \cref{involved} is more involved:
\begin{alignat}{3}
    \int &\la M_t(\ix)M_0(\ix_0)\ra\,\ix\tr{\ix_0}\,\d \ix\,\d \ix_0 =
    \\
    =& \int \d\ix\d\ix_0\; p_\ss(\ix_0)p_t(\ix\,|\,\ix_0)\,\ix\tr{\ix_0}
    \qquad&&\text{[using \cref{M-p}]}
    \\
    =&\int \d\ix\d\ix_0\;\ix\tr{\ix_0}\left(\int\d\tix_0 \frac{1}{(2\pi)^n}\ptil_\ss(\tix_0)\ee^{\ii\tix_0\cdot\ix_0}\right)
    \left(\int\d\tix \frac{1}{(2\pi)^n}\ptil_t(\tix\,|\,\ix_0)\ee^{\ii\tix\cdot\ix}\right)
    \qquad&&\text{[using \cref{inverse-foourier}]}
    \\
    =&\int \d\ix\d\ix_0\;\d\tix\d\tix_0\;\frac{1}{(2\pi)^{2n}}\ix\tr{\ix_0}
    \;\ptil_\ss(\tix_0)\ptil_t(\tix\,|\,\ix_0)
    \;\ee^{\ii\tix_0\cdot\ix_0}\ee^{\ii\tix\cdot\ix}\label{balu}
\end{alignat}
For clarity, decompose $\ptil_t(\tix\,|\,\ix_0)$ from \cref{pq-sol} into:
\begin{align}\label{p-split}
    \ptil_t(\tix\,|\,\ix_0) =&\,\ee^{-\ii\tix\cdot \ix_0} \;\ptil'_t(\tix) +\ptil''_t(\tix) 
    \\
    \ptil'_t(\tix) =& \,\ee ^ {-(\ku+\ii\,\J\cdot\tix) t} 
    \label{p-prime1}
    \\
    \ptil''_t(\tix) =&\,\frac{1}{1+\Z} \frac{\tka(\tix)}{\ku+\ii\;\J\!\cdot\!\tix} \left(1-\ee^{-(\ku+\ii\,\J\cdot\tix) t}\right)
    +\frac{1}{1+\Z} \frac{\tka(\tix)}{\Z \ku-\ii\;\J\!\cdot\!\tix}\left( \ee^{-\ku(1+Z) t}-\ee^{-(\ku+\ii\,\J\cdot\tix) t}\right)
    \label{p-prime2}
\end{align}
Then we have:
\begin{align}
\int \la M_t(\ix)M_0(\ix_0)\ra\,\ix\tr{\ix_0}\,\d \ix\,\d \ix_0
=&\int \d\ix\d\ix_0\;\d\tix\d\tix_0\;\frac{1}{(2\pi)^{2n}}\ix\tr{\ix_0}
    \;\ptil_\ss(\tix_0)\ptil'_t(\tix)
    \;\ee^{\ii\tix_0\cdot\ix_0}\ee^{\ii\tix\cdot(\ix-\ix_0)}\label{tvar-1}
    \\
    &+\int \d\ix\d\ix_0\;\d\tix\d\tix_0\;\frac{1}{(2\pi)^{2n}}\ix\tr{\ix_0}
    \;\ptil_\ss(\tix_0)\ptil''_t(\tix)
    \;\ee^{\ii\tix_0\cdot\ix_0}\ee^{\ii\tix\cdot\ix}
    \label{tvar-2}
\end{align}
The second integral can be written as:
\begin{alignat}{3}
    \text{line \eqref{tvar-2}}=&
    \int \d\ix\d\ix_0\;\d\tix\d\tix_0\;\frac{-1}{(2\pi)^{2n}}\;\ptil_\ss(\tix_0)\ptil''_t(\tix)
    \left(\tfrac{\d}{\d \tix}\ee^{\ii\tix\cdot\ix}\right)
    \tr{\left(\tfrac{\d}{\d \tix_0}\ee^{\ii\tix_0\cdot\ix_0}\right)}
    && \text{[$\ix \ee^{\ii \tix\cdot\ix} = -\ii \tfrac{\d}{\d \tix}\ee^{\ii\tix\cdot\ix}$, etc.]}
    \\
    =&\int \d\ix\d\ix_0\;\d\tix\d\tix_0\;\frac{-1}{(2\pi)^{2n}}\left(\tfrac{\d}{\d \tix}\ptil''_t(\tix)\right)
    \tr{\left(\tfrac{\d}{\d \tix_0}\ptil_\ss(\tix_0)\right)}
    \ee^{\ii\tix_0\cdot\ix_0}\;\ee^{\ii\tix\cdot\ix}
    \quad &&\text{[integration by parts, twice]}
    \\
    =&-\int \d\tix\d\tix_0 
    \left(\tfrac{\d}{\d \tix}\ptil''_t(\tix)\right)
    \tr{\left(\tfrac{\d}{\d \tix_0}\ptil_\ss(\tix_0)\right)}
    \delta(\tix)\delta(\tix_0)
    &&\text{[integrate over $\ix$ and $\ix_0$]}
    \\
    =&-\left(\tfrac{\d}{\d \tix}\ptil''_t(\zero)\right)
    \tr{\left(\tfrac{\d}{\d \tix_0}\ptil_\ss(\zero)\right)}
    &&\text{[contract $\delta$  functions]}
    \label{gyot-1}
\end{alignat}
For the line \cref{tvar-1}, we rewrite $\ix\tr{\ix_0} = (\ix-\ix_0)\tr{\ix_0} + \ix_0 \tr{\ix_0}$:
\begin{align}
    \text{line \eqref{tvar-1}}=&
    \int \d\ix\d\ix_0\;\d\tix\d\tix_0\;\frac{1}{(2\pi)^{2n}}(\ix-\ix_0)\tr{\ix_0}
    \;\ptil_\ss(\tix_0)\ptil'_t(\tix)
    \;\ee^{\ii\tix_0\cdot\ix_0}\ee^{\ii\tix\cdot(\ix-\ix_0)}
    \\
    &+\int \d\ix\d\ix_0\;\d\tix\d\tix_0\;\frac{1}{(2\pi)^{2n}}\ix_0\tr{\ix_0}
    \;\ptil_\ss(\tix_0)\ptil'_t(\tix)
    \;\ee^{\ii\tix_0\cdot\ix_0}\ee^{\ii\tix\cdot(\ix-\ix_0)}
    \\
    =&-\left(\tfrac{\d}{\d \tix}\ptil'_t(\zero)\right)
    \tr{\left(\tfrac{\d}{\d \tix_0}\ptil_\ss(\zero)\right)}
    \qquad\text{[changing variables, and same calculaiton as \cref{gyot-1}]}
    \label{gyot-2}
    \\
    &+\int \d\ix_0\d\tix_0\frac{1}{(2\pi)^{n}}\ix_0\tr{\ix_0}
    \;\ptil_\ss(\tix_0)\ptil'_t(\zero)
    \;\ee^{\ii\tix_0\cdot\ix_0}\label{ez-1}
    \qquad\text{[integrated over $\ix$, then $\tix$]}
\end{align}
The integral can be further simplified into:
\begin{alignat}{3}
    \int \d\ix_0\d\tix_0&\frac{1}{(2\pi)^{n}}\ix_{0,i}\ix_{0,j} 
    \;\ptil_\ss(\tix_0)\ptil'_t(\zero)
    \;\ee^{\ii\tix_0\cdot\ix_0}=
    \\
    =&-\int \d\ix_0\d\tix_0\frac{1}{(2\pi)^{n}} 
    \;\ptil_\ss(\tix_0)\ptil'_t(\zero)
    \;\frac{\d^2}{\d\tix_{0,i}\d\tix_{0,j}}\ee^{\ii\tix_0\cdot\ix_0}
    \qquad&&\text{[insert $\ix_{0,i}\ix_{0,j}$ into derivative]}
    \\
    =&-\int \d\ix_0\d\tix_0\frac{1}{(2\pi)^{n}} 
    \ee^{\ii\tix_0\cdot\ix_0} 
    \;\ptil'_t(\zero)
    \frac{\d^2}{\d\tix_{0,i}\d\tix_{0,j}}\ptil_\ss(\tix_0)
    \qquad &&\text{[integration by parts, twice]}
    \\
    =&-\int \d\tix_0\;\delta(\tix_0) 
    \;\ptil'_t(\zero)
    \frac{\d^2}{\d\tix_{0,i}\d\tix_{0,j}}\ptil_\ss(\tix_0)
    &&\text{[integrate over $\ix_0$, resulting in $\delta(\tix_0)$]}
    \\
    =&-\ptil'_t(\zero)
    \tfrac{\d^2}{\d\tix_{0,i}\d\tix_{0,j}}\ptil_\ss(\zero)
    &&\text{[contracting $\delta$ function]}
    \\
    =& -\ptil'_t(\zero)
    \left(\tfrac{\d^2}{\d\tix^2}\ptil_\ss(\zero)\right)_{ij} 
    \label{gyot-3}
    &&\text{[see the definition of Hessian \cref{hessian-def}]}
\end{alignat}
Combining Equations (\ref{gyot-1}, \ref{gyot-2},\ref{gyot-3}, \ref{gyot-0}), we finally get
\begin{align}
    \C_{\F}(t) = \frac{1}{N}\A &\left[
    \left(\tfrac{\d}{\d \tix}\ptil_\ss(\zero)-\tfrac{\d}{\d \tix}\ptil'_t(\zero)-\tfrac{\d}{\d \tix}\ptil''_t(\zero)\right)
    \tr{\left(\tfrac{\d}{\d \tix}\ptil_\ss(\zero)\right)} -\ptil'_t(\zero)
    \left(\tfrac{\d^2}{\d\tix^2}\ptil_\ss(\zero)\right)
    \right]\tr{\A}
\end{align}
Recall definitions (\ref{p-prime1}, \ref{p-prime2}), then:
\begin{align}
    \ptil'_t(\zero)=&\,\ee^{-\ku t} 
    \\
    \tfrac{\d}{\d \tix}\ptil'_t(\zero) =& -\ii\, \J t\,\ee^{-\ku t}
    \\
    \tfrac{\d}{\d \tix}\ptil''_t(\zero) =&-\frac{\ii}{1+\Z} \frac{\k+\Z \J}{\ku}\left(1-\ee^{-\ku t}\right)
    \\
    &-\frac{\ii}{1+\Z} \frac{\k- \J}{\Z\ku}\left(\ee^{-\ku(1+\Z) t}-\ee^{-\ku t}\right)
    \\
    &+\ii\,\J t \, \ee^{-\ku t}
\end{align}
Thus, the autocorrelation function reads:
\begin{align}
    \C_{\F}(t) = \frac{1}{N}\A\Bigg[&\kk \frac{\ee^{-\ku t}}{\ku (1+\Z)}
    - \k \tr{\k}\frac{1}{\ku^2(1+\Z)\Z}\left(\ee^{-\ku t}-\frac{\ee^{-\ku(1+\Z)t}}{1+\Z}\right)
    \\
    &+\left(\k \tr{\J} + \J \tr{\k}\right)\frac{\ee^{-\ku(1+\Z)t}}{\ku^2(1+\Z)^2}
    + \J \tr{\k}\;\frac{\ee^{-\ku t}-\ee^{-\ku(1+\Z)t}}{\ku^2\Z(1+\Z)}
    \\
    &+\J\tr{\J}\frac{1}{\ku^2} \left(\ee^{-\ku t}-\frac{\ee^{-\ku(1+\Z)t}}{(1+\Z)^2}\right)
    \Bigg]\tr{\A}
    \label{C-F}
\end{align}

With a similar but somewhat simpler calculation, we compute the cross-correlation function between the processes $\hat\phi(t)$ and $\hat\F(t)$:
\begin{align}
    \C_{\F\phi}(t) = \la \hat\F(t) \hat\phi(0)\ra - \bar{\F} \bar{\phi}
\end{align}
Substituting \cref{F-def-sm,phi-def-sm}, we obtain:
\begin{align}
    C_{\F\phi}(t) =& \frac{1}{N}\left(-\bar{\phi}\bar{\F}+\int \la M_t(\ix)M_0(\ix_0)\ra\,\A\ix\,\d \ix\,\d \ix_0  \right)
    \label{zur}
\end{align}
Evaluating the integral:
\begin{align}
    \int \la M_t(\ix)M_0(\ix_0)\ra  \,\ix\,\d \ix\,\d \ix_0 =&\int\d \ix\,\d \ix_0\,p_\ss(\ix_0)p_t(\ix\,|\,\ix_0) \ix
    \\
    =&\int \d\ix\d\ix_0\;\d\tix\d\tix_0\;\frac{1}{(2\pi)^{2n}}\ix 
    \;\ptil_\ss(\tix_0)\ptil_t(\tix\,|\,\ix_0)
    \;\ee^{\ii\tix_0\cdot\ix_0}\ee^{\ii\tix\cdot\ix}
    \\
    =&\int \d\ix\d\ix_0\;\d\tix\d\tix_0\;\frac{1}{(2\pi)^{2n}}\left((\ix-\ix_0)+\ix_0\right)
    \;\ptil_\ss(\tix_0)\ptil'_t(\tix)
    \;\ee^{\ii\tix_0\cdot\ix_0}\ee^{\ii\tix\cdot(\ix-\ix_0)}\label{tvar-3}
    \\
    &+\int \d\ix\d\ix_0\;\d\tix\d\tix_0\;\frac{1}{(2\pi)^{2n}}\ix 
    \;\ptil_\ss(\tix_0)\ptil''_t(\tix)
    \;\ee^{\ii\tix_0\cdot\ix_0}\ee^{\ii\tix\cdot\ix}
    \label{tvar-4}
\end{align}
where we decomposed $\ptil_t(\tix\,|\,\ix_0)$ as in \cref{p-split}. Using the same methods employed in the calculation of $\C_\F(t)$, we simplify:
\begin{align}
    \text{line \eqref{tvar-3}} =&\,\ii \left(\tfrac{\d}{\d \tix} \ptil'_t(\zero)\right)
    \ptil_\ss(\zero)
    +
    \ii \,\ptil'_t(\zero)\left(\tfrac{\d}{\d \tix}\ptil_\ss(\zero)\right)
    \\
    \text{line \eqref{tvar-4}} =& \,\ii \left(\tfrac{\d}{\d \tix} \ptil''_t(\zero)\right)
    \ptil_\ss(\zero)
\end{align}
Thus, combining these results with \cref{gyot-0,phi-mean-def-sm,pss-zero}, \cref{zur} becomes:
\begin{align}
    \C_{\F\phi}(t) =& \frac{\ii}{N}\A \left[
    \left(
    \tfrac{\d}{\d \tix} \ptil'_t(\zero)
    +\tfrac{\d}{\d \tix} \ptil''_t(\zero)
    -\tfrac{\d}{\d \tix}\ptil_\ss(\zero)
    \right)\ptil_\ss(\zero)
    +\ptil'_t(\zero)\left(\tfrac{\d}{\d \tix}\ptil_\ss(\zero)\right)
    \right]
    \\
    =&\frac{\A}{N}\left[
    \frac{\k}{\ku(1+\Z)^2}\ee^{-\ku(1+\Z)t}
    +
    \frac{\J}{\ku(1+\Z)}\left(
    \ee^{-\ku t}
    -\frac{\ee^{-\ku(1+\Z)t}}{1+\Z}
    \right)
    \right]
    \label{C-F-phi}
\end{align}
where in the last line we used equations (\ref{p-prime1}, \ref{p-prime2}, \ref{pss-zero}).
With symmetry arguments, we obtain the other cross correlation function:
\begin{align}
   \C_{\F\phi}(-t)\equiv C_{\phi\F}(t) =& \la \phi(t)\F(0) \ra - \bar{\F} \bar{\phi}
    \\
    =&\frac{1}{N}\!\left(-\bar{\phi}\la \F \ra+\int \la M_t(\ix)M_0(\ix_0)\ra\A\ix_0\,\d \ix\,\d \ix_0  \right)
    \\
    =&\frac{\A}{N}\,
    \frac{\k+\Z \J}{\ku(1+\Z)^2}\,\ee^{-\ku(1+\Z)t}
    \label{C-phi-F}
\end{align}

\section{Summary of the results}
\label{sec-calc-summary}
We summarize the correlation functions derived above along with their corresponding spectral densities. Additionally, we extend these functions to include negative arguments.

Equations (\ref{C-phi}, \ref{C-F-phi}, \ref{C-phi-F}, \ref{C-F}) are summarized as follows:
\begin{align}
    C_\phi(t) =& \frac{1}{N}\frac{\Z}{(1+\Z)^2}\,\ee^{-\ku(\Z+1) t}
    \\
    C_{\F\phi}(t) =& \frac{\A}{N}\left[
    \frac{\k}{\ku(1+\Z)^2}\ee^{-\ku(1+\Z)t}
    +
    \frac{\J}{\ku(1+\Z)}\left(
    \ee^{-\ku t}
    -\frac{\ee^{-\ku(1+\Z)t}}{1+\Z}
    \right)
    \right]
    \\
    C_{\phi\F}(t) =& \frac{\A}{N}\,
    \frac{\k+\Z \J}{\ku(1+\Z)^2}\,\ee^{-\ku(1+\Z)t}
    \\
    C_{\F}(t) =& \frac{1}{N}\A\Bigg[\kk \frac{\ee^{-\ku t}}{\ku (1+\Z)}
    \\
    &- \k {\k}\tr{}\frac{1}{\ku^2(1+\Z)\Z}\left(\ee^{-\ku t}-\frac{\ee^{-\ku(1+\Z)t}}{1+\Z}\right)
    \\
    &+\J\tr{\J}\frac{1}{\ku^2} \left(\ee^{-\ku t}-\frac{\ee^{-\ku(1+\Z)t}}{(1+\Z)^2}\right)
    \\
    &+\k \tr{\J} \frac{\ee^{-\ku(1+\Z)t}}{\ku^2(1+\Z)^2}
    \\
    & +\J \tr{\k}\left(\frac{\ee^{-\ku t}}{\ku^2\Z(1+\Z)}-\frac{\ee^{-\ku(1+\Z)t}}{\ku^2\Z(1+\Z)^2}
    \right)\Bigg]\tr{\A}
\end{align}
Now, a bit simplified:
\begin{align}
    C_\phi(t) =& \frac{1}{N}\frac{\Z}{(1+\Z)^2}\,\ee^{-\ku(\Z+1) t}
    \\
    C_{\F\phi}(t) =& \frac{\A}{N}\left[
    \frac{\k-\J}{\ku(1+\Z)^2}\ee^{-\ku(1+\Z)t}
    +
    \frac{\J}{\ku(1+\Z)}
    \ee^{-\ku t}
    \right]
    \\
    C_{\phi\F}(t) =& \frac{\A}{N}\left[
    \frac{\k-\J}{\ku(1+\Z)^2}\ee^{-\ku(1+\Z)t}
    +
    \frac{\J}{\ku(1+\Z)}
    \ee^{-\ku (1+\Z)t}
    \right]
    \\
    C_{\F}(t) =& \frac{1}{N}\A\Bigg[\kk \frac{\ee^{-\ku t}}{\ku (1+\Z)}
    \\
    &+(\k-\J)(\k-\J)\tr{} \frac{\ee^{-\ku(1+\Z)t}}{\ku^2\Z(1+\Z)^2}
    \\
    &+(\k-\J)\J\tr{} \frac{\ee^{-\ku(1+\Z)t}}{\ku^2\Z(1+\Z)}
    \\
    &-(\k-\J)\tr{\k} \frac{\ee^{-\ku t}}{\ku^2\Z(1+\Z)}
    \\
    &+\J\J^T\frac{1}{\ku^2} \ee^{-\ku t}
    \Bigg]\tr{\A}
\end{align}
These expressions are initially derived for positive $t$. Using the time-translation invariance of the steady state, we extend them to negative times:
\begin{align}
    C_\phi(-t) =& \la \phi(-t) \phi(0) \ra = \la \phi(0) \phi(t) \ra =C_\phi(t)
    \\
    C_{\F\phi}(-t) =& \la \F(-t) \phi(0) \ra = \la \F(0) \phi(t) \ra = C_{\phi\F}(t)
    \\
    C_{\phi\F}(-t) =&\, C_{\F\phi}(t)
    \\
    C_{\F}(-t) =& \la \F(-t) \F(0)\tr{} \ra =\la \F(0) \F(t)\tr{} \ra = C_{\F}(t)\tr{} 
\end{align}
Thus, the correlation functions are valid for all $t$. The corresponding spectral densities (the Fourier transform of the autocorrelation function) are:
\subeqn{
    \label{spectra-sm}
\begin{align}
    \label{S-phi}
    S_\phi(\omega) =& \frac{1}{N}\frac{\Z}{1+\Z}\frac{2\ku}{ \ku^2(1+\Z)^2+\omega^2  }
    \\
    S_{\F\phi}(\omega) =&\frac{1}{N}\frac{1}{1+Z}\frac{1}{\ku^2(1+\Z)^2+\omega^2}\A\left[
    2\k + \Z\J\left(1+\frac{\ku(2+\Z)}{\ku+\ii\,\omega}\right)\right]
    \\
    S_{\phi\F}(\omega) =&\frac{1}{N}\frac{1}{1+Z}\frac{1}{\ku^2(1+\Z)^2+\omega^2}\A\left[
    2\k + \Z\J\left(1+\frac{\ku(2+\Z)}{\ku-\ii\,\omega}\right)\right]
    \\
    \label{S-F}
    S_{\F}(\omega)=&\A\Bigg\{
    \frac{2 \,\kk}{N(1+\Z)}\frac{1}{\ku^2+\omega^2}
    \\
    &-\frac{2 ~\k{\k}\tr{}}{N\Z(1+\Z)\ku}
    \left[\frac{1}{\ku^2+\omega^2}-\frac{1}{\ku^2(1+\Z)^2+\omega^2}\right]
    \\
    &+\frac{\k \J\tr{} + \J {\k}\tr{}}{N\ku (1+\Z)\Z}\left[
    \frac{1}{\ku^2+\omega^2}
    +
    \frac{\Z-1}{\ku^2(1+\Z)^2+\omega^2}
    \right]
    \\
    &+\frac{\k \J\tr{} - \J {\k}\tr{}}{N\ku^2 (1+\Z)\Z}\left[
    \frac{\ii \omega}{\ku^2+\omega^2}
    -
    \frac{\ii \omega}{\ku^2(1+\Z)^2+\omega^2}
    \right]
    \\
    &+\frac{2\,\J\J\tr{}}{N\ku(1+\Z)}\left[\frac{1+\Z}{\ku^2+\omega^2}
    -\frac{1}{ \ku^2(1+\Z)^2+\omega^2  }
    \right]\Bigg\}\A\tr{}
\end{align}
}
Recall that $\F = (\bsigma, \bs H)$, thus, the spectral density of the stress tensor $\bsigma$ is given by the upper left block of the matrix $S_{\F}(\omega)$:
\begin{align}
S^\sigma_\abmn&(\omega)=4\eta \kBT \frac{\ku^2}{\ku^2+\omega^2} \frac{1}{V}\;\Tiso_\abmn    
\\
&+\frac{2\ku}{\rho V (1+\Z)}\frac{1}{\ku^2+\omega^2}\left[
    \rho\kBT\mu\Omega_0\Tiso_\abmn  \atop
    - \mu^2 \Omega^{uu}_\abmn
    -\mu D \left(\Omega^{uq}_\abmn+\Omega^{qu}_\abmn\right)
    -D^2\Omega^{qq}_\abmn  \right]
\\
&-\frac{2\ku(\mu \Omega_u + D\Omega_q)^2}{\rho V\Z(1+\Z)}\left[\frac{1}{\ku^2+\omega^2}-\frac{1}{\ku^2(1+\Z)^2+\omega^2}\right]Q_\ab Q_\mn
\\
&-\frac{\left(\mu \Omega_u + D\Omega_q\right)}{\rho V (1+\Z)\Z}
\left[\frac{1}{\ku^2+\omega^2}+\frac{\Z-1}{\ku^2(1+\Z)^2+\omega^2}
    \right]
\left[\mu \left(Q_\ab v_\mn + v_\ab Q_\mn \right) \atop
    +D(Q_\ab \dot Q_\mn + \dot Q_\ab Q_\mn )
    \right]
\\
&-\frac{\left(\mu \Omega_u + D\Omega_q\right)}{\rho V (1+\Z)\Z \ku}
\left[\frac{\ii\omega}{\ku^2+\omega^2}-\frac{\ii\omega}{\ku^2(1+\Z)^2+\omega^2}
    \right]
\left[\mu \left(Q_\ab v_\mn - v_\ab Q_\mn \right) \atop 
    +D(Q_\ab \dot Q_\mn - \dot Q_\ab Q_\mn )
    \right]
\\
&+\frac{2}{\rho V \ku (1+\Z)}\left[\frac{1+\Z}{\ku^2+\omega^2}
    -\frac{1}{ \ku^2(1+\Z)^2+\omega^2  }
    \right]\left[
    \mu^2 v_\ab v_\mn + D^2 \dot Q_\ab \dot Q_\mn +\atop + \mu D \left(v_\ab \dot Q_\mn + \dot Q_\ab v_\mn \right)
    \right]
\end{align}

\section{Noise in hydrodynamic equation}
\label{hydrodynamic-noise}
In the previous sections we calculated the correlation functions of the macroscopic observables. 
On the other hand, we want to have a (stochastic) differential equation that models these correlations. 
In \ref{sm-with-noise} we introduced a random noise to the hydrodynamic equations; recall \cref{constitutive-general-noisy} below:
\begin{align}\label{baz}
    \left(1 + \frac{1}{\ku} \frac{\d}{\d t}\right) \hat\F = (1-\hat\phi) \A \bs{\Omega}_{\ix} + \frac{\hat\phi}{\ku} \A\J + \hat\bxi(t)
\end{align}
where quantities with $\hat{~}$ fluctuate. Here, $\hat\bxi(t)$ is the noise term that generates the fluctuations of the macroscopic observables.
In this section, we derive the spectrum of the noise $\hat\bxi(t)$ that reproduces the observed fluctuations of $\hat\F$ and $\hat\phi$.

We may derive the spectrum of the noise directly from its mes mesoscopic expression \eqref{meso-noise-def}. This is more complicated that the approach we take here, but it is equivalent.

As we mentioned above, we fix the spectrum of the noise $\hat\bxi(t)$ such that the fluctuation spectrum of  $\hat\F$ and $\hat\phi$ are given by \eqref{spectra-sm}.
We will express the spectral density of the noise in terms of the spectral densities of $\hat\F$ and $\hat\phi$.

We will work with zero-mean variables
\begin{align}
\delta \hat\F(t) = \hat\F(t) - \bar{\F}
    \qquad
    \hat\phi(t) - \bar{\phi}+ \delta \hat\phi(t)
\end{align}
because the autocorrelation functions take simpler forms, for example: $C_{\F\phi}(t) = \la \delta\hat\F(t) \delta\hat\phi(0)\ra$, etc. In terms of these variables, \cref{baz} becomes:
\begin{align}
 \left(1 + \frac{1}{\ku} \frac{\d}{\d t}\right) \delta\hat\F =& 
 -\delta\hat\phi\frac{1}{\ku}\A\left(\k-\J\right) + \bxi(t)
 \label{zug}
\end{align}
Take the Fourier transform of this and solve for the noise $\hat\bxi_\omega$:
\begin{align}
    \hat\bxi_\omega= \frac{1}{\ku}\left((\ku+\ii\,\omega)\; \delta\hat\F_\omega + \A (\k-\J)\; \delta\hat\phi_\omega\right)
\end{align}
Next, consider the autocorrelation of the fourier transformed noise:
\begin{align}
    \la \hat\bxi_\omega\hat\bxi_{\omega'} \ra =
    \frac{1}{\ku^2}\Big(&(\ku+\ii\,\omega)(\ku+\ii\,\omega')\la\delta\hat\F_\omega \delta\hat\F_{\omega'}\tr{}\ra
    +\A(\k-\J)(\k-\J)\tr{} \A\tr{} \la\delta\hat\phi_\omega\delta\hat\phi_{\omega'}\ra
    \\
    &+(\ku+\ii\,\omega)\la \delta\hat\F_\omega\delta\hat\phi_{\omega'}\ra (\k-\J)\tr{} \A\tr{}
    +(\ku+\ii\,\omega') \A(\k-\J)\la \delta\hat\phi_{\omega}\delta\hat\F_{\omega'}\ra\tr{} \Big)
\end{align}
Using the \WK{} theorem (see \cref{app-wiener}), we relate the above averages to the corresponding spectral densities:
\begin{align}
    S_{\bxi}(\omega) =
    \frac{1}{\ku^2}\Big(&(\ku^2+\omega^2)
    \;S_{\F}(\omega)
    +\A(\k-\J)\tr{(\k-\J)} \tr{\A} \;S_{\phi}(\omega)
    \\
    &+(\ku+\ii\,\omega)\;S_{\F\phi}(\omega) \tr{(\k-\J)} \tr{\A}
    +(\ku-\ii\,\omega) \A(\k-\J)\,S_{\phi\F}\tr{(\omega)} \Big)
\end{align}
Substituting the spectral densities from \cref{spectra-sm}, we obtain:
\begin{align}
    S_{\bxi}(\omega) =& \frac{2}{N\ku^2(1+\Z)}\A\left(
    \kk
    +\frac{\Z}{\ku}\J\tr{\J}
    +\frac{1}{2\ku}\left(\J\tr{\k}+{\k}\tr{\J}
    \right)\right)\tr{\A}
\end{align}
Thus, the autocorrelation function of the noise $\bxi(t)$ is:
\begin{align}
    \la \bxi(t)\bxi(t')\tr{}\ra =
    \frac{2}{N\ku^2(1+\Z)}\A\left(
    \kk
    +\frac{\Z}{\ku}\J\J\tr{}
    +\frac{1}{2\ku}\left(\J{\k}\tr{}+{\k}\J\tr{}
    \right)\right)\A\tr{}
    \delta(t-t')
\end{align}
Note that the noise amplitude is always positive definite:
\begin{align}
    \left(
    \kk
    +\frac{\Z}{\ku}\J\J\tr{}
    +\frac{1}{\ku}\left(\J{\k}\tr{}+{\k}\J\tr{}
    \right)\right) = \int (\ix+\tfrac{1}{\ku}\J)(\ix+\tfrac{1}{\ku}\J)\tr{}\kb(\ix)\d \ix
\end{align}
Finally, we infer the autocorrelation of noise in the stress. Writing $\bxi = (\bxi^{\sigma}, \bxi^{H})$, we have:
\begin{align}
    \la \xi^\sigma_\ab(t) \xi^\sigma_\mn (t') \ra = \Lambda^{\sigma}_\abmn \frac{1}{V}\delta (t-t')
\end{align}
where
\begin{align}\label{xi-sigma-general}
    \Lambda^{\sigma}_\abmn =& 4\eta \kBT ~\Tiso_\abmn + \frac{2}{\rho \ku^3(1+\Z)}\times \Big\{
    \\
    &~\ku^2\;\left[
    \rho\kBT\mu\Omega_0\Tiso_\abmn 
    + \mu^2 \Omega^{uu}_\abmn
    +\mu D \left(\Omega^{uq}_\abmn+\Omega^{qu}_\abmn\right)
    +D^2\Omega^{qq}_\abmn  \right]
    \\
    &+\ku \left(\mu \Omega_u + D\Omega_q\right)\left[\mu \left(Q_\ab v_\mn + v_\ab Q_\mn \right)
    +D(Q_\ab \dot Q_\mn + \dot Q_\ab Q_\mn )
    \right]
    \\
    &+\Z\; \left[
    \mu^2 v_\ab v_\mn + D^2 \dot Q_\ab \dot Q_\mn + \mu D \left(v_\ab \dot Q_\mn + \dot Q_\ab v_\mn \right)
    \right]\quad \Big\}
\end{align}

where 
\begin{align}
    \Omega^{uu}_\abmn =& \int \Omega(\u,\q) u_\ab u_\mn \;g(\u,\q) \d\u\, \d\q =\sum_{i=0}^2 \Omega_{uu}^{(i)}\T^{(i)}_\abmn
    \\
    \Omega^{uq}_\abmn =& \int \Omega(\u,\q) u_\ab q_\mn \;g(\u,\q) \d\u\, \d\q 
    =\sum_{i=0}^4 \Omega_{uq}^{(i)}\T^{(i)}_\abmn
    \\
    \Omega^{qu}_\abmn =& \int \Omega(\u,\q) q_\ab u_\mn \;g(\u,\q) \d\u\, \d\q =\Omega^{uq}_{\mn\ab}
    \\
    \Omega^{qq}_\abmn =& \int \Omega(\u,\q) q_\ab q_\mn \;g(\u,\q) \d\u\, \d\q
    =\sum_{i=0}^2 \Omega_{qq}^{(i)}\T^{(i)}_\abmn
\end{align}
\section{Passive, equilibrium case}
\label{sec-passive-reduction}

In the absence of activity, where $\k \propto \bs{\Omega}_{\ix} = 0$, and under equilibrium conditions with no fluxes ($\J=0$), the noise correlation is given by:
\begin{align}
    \la \bxi(t)\bxi(t')\tr{}\ra =
    \frac{2}{N\ku^2(1+\Z)}\A\kk\A\tr{}
    \;\delta(t-t')
\end{align}
From \cref{k2-def} and \cref{kakd-ix}, $\kk$ is defined as:
\begin{align}
    \kk =& \int \ix\ix\tr{} \kb(\ix)\,\d\ix
    \quad= \int \ix\ix\tr{} \ku \,\ee^{-\beta \left(\epsilon_0+\frac{1}{\rho}f(\ix)\right)}\,\d\ix
    \quad =\ku\Z\rho T\A^{-1}
\end{align}
Thus, the noise correlation simplifies to:
\begin{align}
    \la \bxi(t)\bxi(t')\tr{}\ra =2\frac{\bar{\phi}}{\ku}\A\, T \frac{1}{V}\delta(t-t')
\end{align}
The noise in stress, considering the definition of $\A$ from \cref{A-def}, is:
\begin{align}
    \la \bxi^\sigma_\ab(t)\bxi^\sigma_\mn(t')\tr{}\ra =&4\frac{\bar{\phi}\mu}{2\ku}  \, T \frac{1}{V}\delta(t-t')\;\Tiso_\abmn
    \\
    =&4\eta  \, T \frac{1}{V}\delta(t-t')\;\Tiso_\abmn
\end{align}
This result aligns with our previous equilibrium findings in \cref{rutio} and the Fluctuation-Dissipation Theorem (see \cref{app-FDT-passive}).

\section{Why is this so complicated?}
The above results are essentially the same expressions compactly written in \cref{CY} and \cref{SY} in main text. Note that these expressions contain a different exponential decaying function, which comes from matrix exponential $\ee^{-M t}$ in \cref{CY}. Ultimately, the white spectrum of the noise is more convincing from the calculation in the main text.

The approach presented in this section was our initial method for deriving the results. Although more complex, it is robust and could potentially allow for the calculation of higher moments of the stochastic process, which we do not cover in this paper.



\end{document}